\documentclass[fleqn,usenatbib]{mnras}

\usepackage{times,txfonts}

\usepackage[T1]{fontenc}
\usepackage{ae,aecompl}

\usepackage{xcolor}
\usepackage{graphicx,amssymb}
\usepackage[english]{babel}
\usepackage{xspace}
\usepackage{lscape}
\usepackage{breqn}

\newcommand{\kms}{km\,s$^{-1}$\xspace}
\newcommand{\mum}{${\mu}$m\xspace}
\newcommand{\htwo}{H$_{\rm 2}$\xspace}
\newcommand{\hi}{H\,{\sc i}\xspace}
\newcommand{\zsun}{$Z_{\odot}$\xspace}
\newcommand{\mhi}{M(H{\sc i})\xspace}
\newcommand{\mhtwo}{M(H$_2$)\xspace}
\newcommand{\mstar}{M$_{\star}$\xspace}
\newcommand{\mustar}{${\mu}_{\star}$\xspace}
\newcommand{\msun}{M$_{\odot}$\xspace}

\newcommand{\arcm}{$^{\prime}$\xspace}
\newcommand{\arcs}{$^{\prime\prime}$\xspace}
\newcommand{\bd}{{\sc Bluedisk}\xspace}
\newcommand{\sighi}{${\Sigma}_{\rm HI}$\xspace}
\newcommand{\sightwo}{${\Sigma}_{\rm H_2}$\xspace}
\newcommand{\sigsfr}{${\Sigma}_{\rm SFR}$\xspace}

\newcommand{\siggas}{${\Sigma}_{\rm gas}$\xspace}

\title[Molecular gas in the \bd galaxies]{
The \bd Survey:
molecular gas distribution and scaling relations
in the context of galaxy evolution}

\author[D.~Cormier et al.]{
D.~Cormier$^{1}$\thanks{Email: diane.cormier@zah.uni-heidelberg.de},
F.~Bigiel$^{1}$,
J.~Wang$^{2}$,
J.~Pety$^{3}$,
A.~Usero$^{4}$,
S.~Roychowdhury$^{5}$,
\newauthor
~D.~Carton$^{6}$,
J.~M.~van der Hulst$^{7}$,
G.~I.~G.~J{\'o}zsa$^{8}$,
M.~Gonzalez Garc{\'i}a$^{9}$ and
\newauthor
~A.~Saintonge$^{10}$ \vspace{5pt} \\ 
$^{1}$Zentrum f\"ur Astronomie der Universit\"at Heidelberg,
Institut f\"ur theoretische Astrophysik,
Albert-Ueberle-Str. 2, 69120 Heidelberg, \\ ~Germany\\
$^{2}$CSIRO Astronomy and Space Science, Australia Telescope National Facility, PO Box 76, Epping, NSW 1710, Australia\\
$^{3}$Institut de Radioastronomie Millim{\'e}trique, 300 Rue de la Piscine, F-38406 Saint Martin d'H{\`e}res, France\\
$^{4}$Observatorio Astron{\'o}mico Nacional (IGN), C/ Alfonso XII 3, Madrid E-28014, Spain\\
$^{5}$Max-Planck-Institut f{\"u}r Astrophysik, Karl-Schwarzschild-Str. 1, D-85748 Garching, Germany\\
$^{6}$Leiden Observatory, Leiden University, PO Box 9513, 2300 RA, Leiden, The Netherlands\\
$^{7}$Kapteyn Astronomical Institute, University of Groningen, PO Box 800, 9700 AV, Groningen, The Netherlands\\
$^{8}$SKA South Africa, 3rd Floor, The Park, Park Road, Pinelands, 7405, South Africa\\
$^{9}$Instituto de Astrof{\'i}sica de Andaluc{\'i}a IAA-CSIC, Glorieta de la Astronom{\'i}a s/n, E-18008, Granada, Spain\\
$^{10}$Department of Physics \& Astronomy, University College London, Gower Place, London WC1E 6BT, UK
}

\begin{document}

\pagerange{\pageref{firstpage}--\pageref{lastpage}}
\date{Accepted 2016 August 18. Received 2016 August 18; in original form 2016 June 10}
\pubyear{2016}
\maketitle
\label{firstpage}

\begin{abstract}
One of the key goals of the \bd survey is to characterize the impact
of gas accretion in disc galaxies in the context of galaxy evolution.
It contains 50 disc galaxies in the stellar mass range $10^{10}-10^{11}$\,\msun,
of which half are bluer and more \hi-rich galaxies than their
\hi-normal (control) counterparts.
In this paper, we investigate how ongoing disc growth affects the molecular
gas distribution and the star-formation efficiency in these galaxies.
We present $^{12}$CO observations from the IRAM 30-m telescope 
in 26 galaxies of the \bd survey. 
We compare the amount and spatial distribution of the molecular gas 
to key quantities such as atomic gas, stellar mass and surface density, 
star-formation rate and metallicity. We analyse the star-formation rate
per unit gas (SFR/\hi and SFR/\htwo) and relate all those parameters
to general galaxy properties (\hi-rich/control disc, morphology, etc.). 
We find that the \hi-rich galaxies have similar \htwo masses as
the control galaxies. In their centres, \hi-rich galaxies have lower
\htwo/\hi ratios and marginally shorter molecular gas depletion times.
However, the main differences between the two samples occur in
the outer parts of the discs, with the \hi-rich galaxies having slightly
smaller CO discs (relative to the optical radius $R_{25}$) and steeper
CO and metallicity gradients than the control galaxies. The ongoing
accretion of \hi at large radii has thus not led to an appreciable growth
of the CO discs in our sample.
Based on depletion times, we estimate that this gas will contribute
to star formation on time-scales of at least 5\,Gyr.
\end{abstract}


\section{Introduction}
\label{sect:intro}
At all redshifts, the star-formation activity and the amount 
of stars in most galaxies are related and define what is known 
as the main sequence of galaxies \citep[e.g.,][]{noeske-2007,elbaz-2007}. 
However, for present-day galaxies to maintain their star formation  
at the observed level, they need to be replenished in gas 
\citep[e.g.,][]{kennicutt-1983,erb-2008,hopkins-2008,bauermeister-2010}. 
There is growing evidence of large atomic gas reservoirs in the 
outer parts of galaxies \citep[e.g.,][]{bigiel-2010a,bigiel-2010b}, due to 
external accretion or galactic fountains \citep[e.g.,][]{sancisi-2008,catinella-2010}. 
Galaxies with significant excess atomic gas feature bluer, younger, 
and more metal-poor outer discs \citep{wang-2010,moran-2010,moran-2012}. 
This supports an ``inside-out'' picture of galactic disc formation, 
in which the inner part forms first and infalling cool gas in the 
outer part then contributes to the build-up of the disc. 
How specifically this gas influences star formation, disc growth,
and galaxy evolution remains unclear. 
In this frame, knowledge of the molecular gas content is necessary 
to better understand the link between the accreted atomic gas and 
star formation occurring predominantly in the centre of the disc. 
Surveys of the atomic and molecular reservoirs in nearby, stellar-mass
selected galaxies such as GASS \citep{catinella-2010} and COLD\,GASS
\citep{saintonge-2011a} have shed light on the global gas content
of massive galaxies. \cite{catinella-2010} showed that most galaxies
lie on a tight plane linking the atomic gas content with the UV/optical
colours and stellar mass surface densities.
For galaxies with an excess of atomic gas, i.e. above that plane,
\cite{saintonge-2011a} do not find differences in their molecular gas
mass. However, only global measurements were considered. 
With spatially resolved observations, the \bd survey aims to go
one step forward compared to previous studies.

\begin{table*}\small
\begin{minipage}{\textwidth}
  \caption{Properties of our selected galaxy sample.}
\begin{center}
\begin{tabular}{clcccccccc}
 \hline\hline
SDSS Name & Galaxy ID & Dist. & Incl. & P.A. & $D_{25}$ & $NUV-r$ & $\log$ \mstar & $\log$ \mhi & Morphology \\
 & & [Mpc] & [$^{\circ}$] & [$^{\circ}$] & [$^{\prime}$] & [mag] & [$\log$ M$_{\odot}$] & [$\log$ M$_{\odot}$] & \\
\hline
\multicolumn{10}{c}{\bd Galaxies observed with HERA}\\
\hline
J081422.02+391504.8 & ID1$^{HI-rich}$ 	& 121 	& 34 & 100 & 1.27 & 2.37	& 10.4 & 10.1 	& SBc face-on  \\ 
J082846.72+403957.0 & ID2$^{HI-rich}$ 	& 107	& 66 & 111 & 1.36 & 3.02	& 10.6 & 9.9 	& Sb  \\ 
J083706.51+412722.8 & ID3$^{control}$	& 128 	& 37 & 73 & 0.82 & 2.44	& 10.4 & 9.9 	& Sc  \\
J083833.99+304755.2 & ID4$^{HI-rich}$	& 112 	& 62 & 109 & 1.48 & 2.36	& 10.6 & 10.3 	& Sc, lopsided  \\
J084916.39+360711.3 & ID5$^{HI-rich}$	& 110 	& 21 & -8 & 1.09 & 2.01	& 10.3 & 10.2 	& Sab face-on  \\
J084922.57+364237.5 & ID6$^{HI-rich}$	& 109	& 52 & 61 & 1.73 & 2.94	& 10.8 & 10.3 	& SBb, one arm  \\ 
J090842.68+444838.0 & ID8$^{HI-rich}$	& 117	& 48 & 54 & 1.10 & 2.19	& 10.3 & 10.2 	& Sc, clumpy  \\ 
J091458.31+512139.8 & ID9$^{control}$	& 120 	& 63 & 105 & 1.22 & 2.93	& 10.8 & 9.9 	& SB, clumpy  \\
J093225.04+572858.4 & ID10$^{control}$	& 129 	& 37 & 25 & 1.00 & 2.43	& 10.9 & 10.0 	& SBa \\
J101054.38+455701.3 & ID11$^{control}$	& 105 	& 56 & -4 & 1.23 & 2.74	& 10.6 & 9.7 	& Sb \\
J090842.68+444838.0 & ID14$^{HI-rich}$	& 104	& 61 & 159 & 1.46 & 3.25	& 10.8 & 10.1 	& Sa  \\ 
J114859.46+350057.7 & ID15$^{HI-rich}$	& 93 		& 51 & 22 & 1.86 & 2.43	& 10.8 & 10.4 	& Sab, two arms \\
J125203.54+514048.2 & ID16$^{HI-rich}$	& 119 	& 40 & 19 & 0.90 & 2.01	& 10.3 & 10.0 	& Sc, clumpy \\
J130713.58+580806.1 & ID17$^{HI-rich}$	& 120 	& 44 & 17 & 1.02 & 2.85	& 10.7 & 10.3 	& Sb face-on \\
J143759.94+400622.3 & ID20$^{HI-rich}$	& 114 	& 52 & 2 & 1.29 & 1.55	& 10.2 & 10.1 	& Sc, clumpy \\
\hline
\multicolumn{9}{c}{\hspace{2cm}\bd Galaxies observed with EMIR} & Pair \\
\hline
J080240.65+343117.2 & ID27$^{control}$		& 126 	& 39 & -77 & 0.62 & 2.90	& 10.3 & 9.3 	& ID3$^{control}$ \\
J082914.91+553122.8 & ID29$^{excluded}$ 	& 112 	& 64 & 95 & 1.25 & 2.90	& 10.5 & 10.4 	& ID4$^{HI-rich}$ \\
J091645.74+454844.1 & ID31$^{excluded}$ 	& 114 	& 52 & -38 & 0.75 & 2.68	& 10.2 & 10.1 	& ID20$^{HI-rich}$ \\
J092609.43+491836.7 & ID33$^{control}$ 	& 118 	& 58 & -62 & 1.12 & 3.99	& 10.7 & 9.2 	& ID9$^{control}$ \\
J095740.85+451531.2 & ID35$^{HI-rich}$ 	& 106 	& 71 & -62 & 1.28 & 3.32	& 10.6 & 9.9 	& ID15$^{HI-rich}$ \\
J101511.43+564019.5 & ID37$^{control}$ 	& 114	 & 31 & -80 & 1.05 & 2.81	& 10.4 & 9.7 	& ID5$^{HI-rich}$\\
J111415.25+340915.8 & ID40$^{control}$ 	& 119 	& 39 & 103 & 0.73 & 2.83	& 10.4 & 9.7 	& ID1$^{HI-rich}$ \\
J122139.99+405056.4 & ID41$^{control}$ 	& 130 	& 31 & -23 & 0.91 & 3.45	& 10.8 & -- 	& ID10$^{control}$ \\
J133329.90+403146.8 & ID43$^{control}$ 	& 118 	& 48 & 162 & 0.76 & 2.74	& 10.3 & 9.6 	& ID16$^{HI-rich}$ \\
J134100.24+422553.1 & ID44$^{control}$ 	& 122 	& 51 & 40 & 1.03 & 3.34	& 10.8 & 9.3 	& ID17$^{HI-rich}$ \\
J161731.80+311140.1 & ID47$^{HI-rich}$ 	& 105 	& 54 & 199 & 1.10 & 2.91	& 10.5 & 10.0 	& ID11$^{control}$ \\
 \hline\hline
\end{tabular}
\end{center}
\end{minipage}
\begin{minipage}{\textwidth}
Notes. Parameters taken from \cite{wang-2013}: 
name, identification number, luminosity distance, diameter 
of the $g$-band 25\,mag/arcsec$^{-2}$ isophote, inclination, 
position angle of the major axis, stellar mass, \hi mass, and morphology. 
The position angle is measured from SDSS data, w.r.t. west.
The distance is the luminosity distance, calculated from the redshift
assuming H$_0$=70, $\Omega_M$=0.3, $\Lambda_0$=0.7, q$_0$=-0.55.
The upper script next to the ID indicates which group the galaxy 
belongs (\hi-rich, control, excluded; where excluded galaxies are multi-source systems, 
see \citealt{wang-2014}).  
\end{minipage}
\label{table:general}
\end{table*}

The main science objective of the \bd survey is to search
for accretion signatures and therefore to characterize the gas distribution
and kinematics of \hi-rich galaxies.
The \bd survey is a multi-wavelength campaign of 50 nearby
galaxies with similar stellar masses, of which about half are \hi-rich
galaxies and half are \hi-normal or ``control'' galaxies.
The ancillary data of that survey comprise UV data from GALEX, 
optical broadband data and spectroscopy from the SDSS, WHT 
spectroscopy \citep[PI Brinchmann;][]{carton-2015} at a resolution
of $\sim$1.5\arcs, and WSRT \hi 21-cm data \citep[PI Kauffmann;][]{wang-2013,wang-2014} 
at a resolution of $\sim$15\arcs.
\hi observations have revealed that the discs of the \hi-rich
sample are more extended relative to the optical size and generally
more clumpy than in the control sample \citep{wang-2013}.
The \hi gas is not kinematically disturbed, giving preference to
rather continuous accretion. The excess \hi is located in the
outer parts of the disc (beyond $R_{25}$, defined as the radius
of the $g$-band 25\,mag/arcsec$^{-2}$ isophote) for half of the 
\hi-rich galaxies and in the centre for the other half \citep{wang-2014}.
In the outer parts, the \hi surface density profiles of the \hi-rich
and control galaxies are similar though.
Moreover, all galaxies display gas-phase metallicity drops,
which are steeper for the \hi-rich sample and occur at larger stellar
radii for the control sample \citep{carton-2015}.

Observations of the cold, star-forming molecular gas are necessary 
to study the evolution of the star formation properties and disc growth 
in those galaxies.
In this paper, we present CO observations with the IRAM-30m telescope
of $26$ galaxies from the \bd survey, listed in Table~\ref{table:general}.
The main goal is to investigate whether the \hi-rich galaxies have different
molecular gas properties than the control galaxies, and what are
the implications for the evolution of those galaxies.
To do this, we compare the amount and distribution of atomic and
molecular gas to the star-formation rate between the two samples,
and we relate those quantities to galaxy morphology, metallicity,
and stellar densities.  
We describe the observing strategy, data reduction, and further 
processing in Sections~\ref{sect:data} and \ref{sect:analysis}. 
Our results are presented in terms of integrated, resolved, and
radial properties in Section~\ref{sect:results}, and discussed in
Section~\ref{sect:discussion}.
We summarize our findings in Section~\ref{sect:summary}.

\section{Sample selection}
\subsection{\bd sample}
Selection of the \bd sample is described in detail in \cite{wang-2013}.
We summarize the main characteristics. Galaxies of the \bd survey
are selected from the SDSS DR7 MPA/JHU catalog and
required to be massive ($10< \log$ \mstar/\msun$<11$), nearby
($0.01<z<0.03$), northern ($dec.>30^{\circ}$) and extended
enough on the sky to be resolved ($D_{25}>50$\arcs).
The \hi-rich sample consists of galaxies that are above the
fundamental plane\footnote{$\log({\rm M(HI)/M_{\star}})+0.234\times\log(\mu_{\star})+0.342\times(NUV-r)-2.329=0$}
defined by \cite{catinella-2013}. The \hi-rich
galaxies have $\sim3\times$ higher \mhi than the median
at the same \mstar. The control sample lies close to that plane,
with galaxies closely matched to the \hi-rich sample in stellar
mass (\mstar), stellar surface density ($\mu_{\star}$), redshift
and inclination, but with normal \hi content.

\subsection{CO follow-up}
Due to sensitivity limitations, not all of the \bd galaxies could be
observed with IRAM. We decided to target a sub-sample of $26$
galaxies (Table~\ref{table:general}), mapping the brightest ones
with the HEterodyne Receiver Array (HERA) and observing with
single pointings the others with the Eight Mixer Receiver (EMIR).
The target selection was done by identifying the most \hi-rich or
highest-SFR galaxies and their paired galaxies, while covering 
a range of parameters (\mhi, \mstar, SFR) close to that covered
by the full \bd sample (Fig.~\ref{fig:coldg1}).
This is driven largely by observability considerations (reducing
the number of receiver tunings/overheads while keeping a sufficiently
large number of galaxies).

Of the 26 galaxies in our sub-sample, 15 belong to the \hi-rich sample
and 11 belong to the control sample. Two \hi-rich galaxies (ID29 and ID31)
are not isolated but multi-source systems. We include them in the
Figures (1 and 2) as black symbols but exclude them from correlations
and hence label them as {\em excluded} in Table~\ref{table:general}.
Note that ID41 was also labeled as excluded in \cite{wang-2013} because
initially no \hi data were taken for this galaxy. Since we have obtained
new WSRT \hi observations for this galaxy, we re-include it as a control
galaxy.

\section{Observations}
\label{sect:data}

\subsection{HERA observations of CO(2-1)}
\subsubsection{Observing details}
We mapped 15 \bd galaxies in the $^{12}$CO(2-1) line 
at 230.54\,GHz with the HERA instrument. 
The spatial resolution achieved with HERA, which has a 
HPBW of 11\arcs, is $\sim$6.5\,kpc at the median distance
of our sample (120\,Mpc). 
HERA is a heterodyne receiver consisting of 3$\times$3 pixels 
separated by 24\arcs, with two orthogonal polarizations. 
Observations were performed in the on-the-fly mapping mode 
to map an area of about 1.2\arcm$\times$1.2\arcm, slightly larger 
than the optical disc size of our galaxies. 
To make maps as uniform as possible and fill the gaps between 
the detectors, we used the oversampling observing
mode\footnote{\url{http://www.iram.es/IRAMES/otherDocuments/manuals/HERA_manual_v20.pdf}} 
and performed up and down scans separated by 12\arcs along 
two alternated perpendicular directions oriented 45$^{\circ}$ from 
the major axis. The pattern is repeated as many times as required 
to achieve the desired rms. 
We used the backend WILMA, which covers a total bandwidth 
of 8\,GHz per polarization ($\sim$11\,000\,\kms at the observed 
frequency 225\,GHz), with velocity resolution 2.6\,\kms. 
Observations were carried out between October 2012 and March 2014, 
under good winter conditions (mean zenith opacity of $0.08$ at 225\,GHz), 
as part of the programs {073-12} (PI Bigiel; 35\,hrs) and {205-13} (PI Cormier; 50\,hrs). 
The average system temperature was 340\,K. 
The observing details are given in Table~\ref{table:obsdetails}.

\begin{table}\small
  \caption{Details of the IRAM 30-m observations.} 
\begin{center}
\begin{tabular}{l l c}
    \hline\hline
     \vspace{-8pt}\\
    \multicolumn{1}{l}{Galaxy ID} & 
    \multicolumn{1}{l}{Observing Dates} &
    \multicolumn{1}{c}{Noise [mK]}  \\
    \hline
    \multicolumn{3}{c}{HERA observations} \\
    \hline
	{ID1}			& 2012 Dec. 1, 2014 Mar. 11 		& 3.9 \\
	{ID2}			& 2012 Dec. 2				 	& 6.1 \\
	{ID3}			& 2014 Mar. 8, 9			 	& 4.5 \\
	{ID4}			& 2012 Dec. 3, 2014 Mar. 10, 11 	& 3.6 \\
	{ID5}			& 2012 Oct. 15 					& 10 \\
	{ID6}			& 2012 Dec. 1			 		& 8.1 \\
	{ID8}			& 2012 Dec. 2			 		& 4.9 \\
	{ID9}			& 2014 Jan. 14, Mar. 7, 8 			& 3.1 \\
	{ID10}		& 2012 Dec. 2 					& 4.5 \\
	{ID11}		& 2012 Dec. 3 					& 6.5 \\
	{ID14}		& 2012 Dec. 3				 	& 6.6 \\
	{ID15}		& 2012 Dec. 2 					& 5.7 \\
	{ID16}		& 2014 Mar. 8 					& 4.0 \\
	{ID17}		& 2012 Dec. 2, 2014 Mar. 8 		& 2.6 \\
	{ID20}		& 2012 Dec. 2, 2014 Mar. 11 		& 4.1 \\
    \hline
    \multicolumn{3}{c}{EMIR observations} \\
    \hline
	{ID27}		& 2014 Mar. 9, 10			 	& 2.3, 3.0 \\
	{ID29}		& 2014 Feb. 8	 				& 1.4, 2.6 \\
	{ID31}		& 2014 Mar. 9			 		& 2.7, 3.9 \\
	{ID33}		& 2014 Feb. 8		 			& 1.4, 2.2 \\
	{ID35}		& 2014 Jan. 12 					& 2.3, 3.3 \\
	{ID37}		& 2014 Jan. 12 					& 1.8, 3.0 \\
	{ID40}		& 2014 Mar. 9			 		& 2.7, 3.5 \\
	{ID41}		& 2014 Jan. 10 					& 1.6, 3.1 \\
	{ID43}		& 2014 Jan. 12					& 1.9, 3.1 \\
	{ID44}		& 2014 Jan. 12			 		& 2.5, 3.8 \\
	{ID47}		& 2014 Mar. 9, 10 				& 1.6, 2.0 \\ 
    \hline \hline
\end{tabular}
\end{center}
    \vspace{-8pt}
Notes. The noise level (in units of $T_{\rm mb}$)
for the HERA observations is taken as the median value
of the error maps (see Section~\ref{sect:mapmaking}). 
For the EMIR observations, the noise level in both 
the CO(1-0) and CO(2-1) spectra are indicated (left 
and right values, respectively), at a spectral resolution
of 15.6\,\kms. 
  \label{table:obsdetails}
\end{table}

\subsubsection{Data reduction of the spectral cubes}
The data reduction was performed using the Continuum and 
Line Analysis Single-dish Software (CLASS), which is part of the 
{GILDAS} package\footnote{\url{http://www.iram.fr/IRAMFR/GILDAS}}. 
The steps are as follows, and executed for each galaxy. 
We first read the data in, selected the part of the spectra around 
the expected line centre (about $\pm$600\,\kms), and ignored bad 
spectra (strange baseline, fringes) as well as data from the unstable 
pixels 4 and, for the 2014 observations, 9 of the second receiver. 
The data were shifted to rest-frame using the velocity from
optical data converted to the radio definition, and put in units of main
beam temperature using a beam efficiency value extrapolated
from the measured value of 0.58 at 230\,GHz\footnote{\url{http://www.iram.es/IRAMES/mainWiki/Iram30mEfficiencies}}
to our observed frequencies with the Ruze's equation and a forward
efficiency of 0.94.
We then applied a baseline correction of order 0 (masking the inner
$\pm$200\,\kms where the line is expected) and excluded 
spectra with overall noise above 3 times the theoretical noise. 
Spectral cubes were produced by projecting the remaining spectra 
on a grid of pixel size 2\arcs, with final spatial resolution 13.5\arcs 
(convolved with a Gaussian kernel) and spectral resolution 15.6\,\kms. 
We clipped the very edges of the map where the noise increased 
due to smaller number of scans from our mapping strategy.
For this, we clipped pixels with time weights below a
value of 0.03 typically.

To further improve the baseline correction, we defined spectral
windows to mask the CO line based on the \hi data. \hi moment maps
from \cite{wang-2013} with robust weighting of 6 and tapering
of 30\arcs, which optimize the sensitivity, were used.
Our first assumption was that CO emission is expected inside 
the \hi envelope, hence our windows were defined at positions
where \hi is detected at $\ge$ 5-$\sigma$ level in the moment
zero maps. We used the \hi first and second order moment
maps to determine the centre and width of the CO windows.
Then, we iterated the process and adapted the windows by hand 
(where the CO line is clearly detected) to ensure that our windows 
are as narrow as possible while encompassing all the CO emission. 
Since the CO line profiles are observed to be wider at the galaxy
centre than in the disc, the windows were made wider at the centre
and around the minor axis and linearly decreasing in size going to
larger radii. Beyond 0.7\,$R_{25}$ typically, where CO is
not clearly detected, we kept the size of the window fixed to $\sim0.7$
times the window size at the galaxy center.

Once our windows were finalized, we masked all velocity
channels inside each window and fit a polynomial of order~1 to correct
for the baseline.

\subsubsection{Intensity and error maps}
\label{sect:mapmaking}
We created moment maps after masking all channels outside 
of the windows. At the mean distance of our sample, we probe 
linear scales of 6.5\,kpc.
The line widths are large, from 80\,\kms to 240\,\kms at
the centre of the disc, and the CO profiles are not Gaussian.
Hence we prefer to use moment maps rather then perform a line fit. 
The intensity maps and averaged spectra within a circular
aperture of diameter 22\arcs on the galaxy 
centres are shown in Appendix~\ref{append:hera}. Line widths are 
smallest for ID1, ID5, and ID10, and larger than 200\,\kms for 
the other galaxies. 

We also generated maps reflecting the statistical error on the intensities. 
For each spectrum, the error is calculated as the standard deviation 
on the baseline (channels outside of the window), multiplied by the 
channel width times the square root of the number of channels inside the window. 
This is the error per window and it is independent of the channel size.

To verify the accuracy of the described method and that we 
are not biasing the results when including manual steps, we also 
produced moment maps for which the windows were defined: 
(1)~centre from \hi moment~1 maps and width where \hi is emitting 
above a 3-$\sigma$ level (i.e. without further refinement); and 
(2)~centre from \hi moment~1 maps and a fixed width. 
The three methods generally agree within the errors, but the
manual method tends to produce better signal-to-noise maps. 
With a fixed window, some of the CO emission is sometimes
missed, and based on the \hi only, the windows often include
noisy channels which degrade the results.

\subsection{EMIR observations of CO(1-0) and CO(2-1)}
\subsubsection{Observing details}
We used the EMIR instrument to simultaneously observe the
$^{12}$CO(1-0) and $^{12}$CO(2-1) lines in 11 \bd galaxies.
The rest frequencies are 115.27\,GHz and 230.54\,GHz, and
the beam HPBW are 22\arcs ($\sim$13\,kpc) and 11\arcs
($\sim$6.5\,kpc) respectively. 
We made pointed observations in ONOFF wobbler switching mode 
with a wobbler throw between 55-120\arcs (80\arcs in general) 
depending on the source. The WILMA backend was connected. 
Observations were carried out between January and March 2014, 
under good winter conditions (mean zenith opacity of $0.16$ at 225\,GHz), 
as part of the program {205-13}. 
The average system temperature was 340\,K. 

\subsubsection{Data reduction}
The data were reduced with CLASS. 
For each galaxy, we extracted spectra in a bandwidth of 1\,GHz
around the observed line frequency. The spectra were shifted to
rest-frame using the radial velocity, and put in units of main beam
temperature using beam efficiency values extrapolated
from the measured values of 0.78 at 115\,GHz and 0.58 at 230\,GHz
to our observed frequencies with the Ruze's equation and forward
efficiencies of 0.92 and 0.94, respectively. The spectra were then
averaged and rebinned to a spectral resolution of 15.6\,\kms.
A baseline correction of order~3 was applied.
We measured the rms of the data as the standard deviation on 
the continuum, i.e. outside of a $\pm200$\,\kms window around 
the expected line centre (see Table~\ref{table:obsdetails}). 

The spectra are shown in Fig.~\ref{fig:emir}. 
Several spectra have double peaked profiles, others have asymmetric 
profiles, especially ID\,31 which is an interacting system.
The lines are detected in all galaxies, except CO(2-1) in ID33 
and a marginal detection in ID41. 
The CO intensities are reported in Table~\ref{table:masses}. 
They were derived by integrating the signal inside a window 
of size $\sim$250\,\kms depending on the source.

\begin{table*}\small
\begin{minipage}{\textwidth}
  \caption{CO intensities, gas masses, and star-formation rates.}
\begin{center}
\begin{tabular}{l c c c c c c c c}
 \hline\hline
Galaxy ID & $I_{\rm CO(1-0)}$ & $I_{\rm CO(2-1)}$ & $M_{\rm H_2,~22}$ & $M_{\rm H_2,~total}$ 
	& $M_{\rm HI,~22}$ & $M_{\rm HI,~total}$ & ${\rm SFR_{22}}$ & ${\rm SFR_{total}}$\\
 & [K\,\kms] & [K\,\kms] & [log \msun] & [log \msun] & [log \msun] & [log \msun] & [\msun\,yr$^{-1}$] & [\msun\,yr$^{-1}$] \\
\hline
\multicolumn{9}{c}{\bd Galaxies observed with HERA}\\
\hline
ID1   & $ - $   & $ \{1.37~(0.20)\}$ &    $8.95~(7.84) $   & $9.27~(8.73)$ &    $8.90~(7.71) $   & $10.10~(8.44)$ &    $1.37~(0.01)$   & $3.16~(0.01)$ \\
ID2   & $ - $   & $ \{0.58~(0.33)\}$ &    $8.39~(8.06) $   & $8.66^*$ &    $9.13~(7.67) $   & $9.95~(8.25)$ &    $0.22~(0.01)$   & $0.65~(0.01)$ \\
ID3   & $ - $   & $ \{1.77~(0.35)\}$ &    $9.14~(8.31) $   & $9.35~(7.98)$ &    $9.13~(7.82) $   & $9.91~(8.36)$ &    $1.41~(0.01)$   & $3.39~(0.04)$ \\
ID4   & $ - $   & $ \{2.70~(0.22)\}$ &    $9.20~(7.75) $   & $9.50~(8.62)$ &    $9.27~(7.77) $   & $10.27~(8.52)$ &    $1.39~(0.01)$   & $4.74~(0.03)$ \\
ID5   & $ - $   & $ \{3.57~(0.60)\}$ &    $9.29~(8.42) $   & $9.63~(9.09)$ &    $8.87~(7.65) $   & $10.21~(8.58)$ &    $2.42~(0.01)$   & $5.62~(0.01)$ \\
ID6   & $ - $   & $ \{2.00~(0.60)\}$ &    $9.06~(8.52) $   & $9.44^*$ &    $8.83~(7.67) $   & $10.33~(8.51)$ &    $0.36~(0.01)$   & $0.88~(0.06)$ \\
ID8   & $ - $   & $ \{1.89~(0.31)\}$ &    $9.14~(8.49) $   & $9.40^*$ &    $8.88~(7.74) $   & $10.21~(8.57)$ &    $0.45~(0.01)$   & $1.97~(0.01)$ \\
ID9   & $ - $   & $ \{2.97~(0.20)\}$ &    $9.31~(8.07) $   & $9.52~(8.50)$ &    $9.20~(7.89) $   & $9.93~(8.33)$ &    $2.03~(0.01)$   & $5.38~(0.01)$ \\
ID10   & $ - $   & $ \{3.84~(0.24)\}$ &    $9.49~(8.37) $   & $9.76~(8.77)$ &    $9.01~(7.83) $   & $10.05~(8.46)$ &    $3.18~(0.01)$   & $8.31~(0.02)$ \\
ID11   & $ - $   & $ \{3.11~(0.41)\}$ &    $9.20~(8.23) $   & $9.49~(8.89)$ &    $8.88~(7.87) $   & $9.73~(8.11)$ &    $0.84~(0.01)$   & $2.52~(0.01)$ \\
ID14   & $ - $   & $ \{1.53~(0.50)\}$ &    $8.93~(8.50) $   & $9.24^*$ &    $8.99~(7.65) $   & $10.15~(8.47)$ &    $0.49~(0.01)$   & $1.84~(0.01)$ \\
ID15   & $ - $   & $ \{3.06~(0.42)\}$ &    $9.08~(7.96) $   & $9.50~(9.06)$ &    $8.66~(7.50) $   & $10.40~(8.58)$ &    $0.74~(0.01)$   & $3.83~(0.04)$ \\
ID16   & $ - $   & $ \{2.12~(0.23)\}$ &    $9.13~(7.82) $   & $9.30~(8.51)$ &    $9.22~(7.93) $   & $10.02~(8.50)$ &    $2.32~(0.01)$   & $5.50~(0.01)$ \\
ID17   & $ - $   & $ \{1.01~(0.16)\}$ &    $8.83~(7.89) $   & $9.05~(8.45)$ &    $8.84~(7.98) $   & $10.23~(8.82)$ &    $0.78~(0.01)$   & $2.05~(0.01)$ \\
ID20   & $ - $   & $ \{1.98~(0.24)\}$ &    $9.12~(8.35) $   & $9.33~(8.67)$ &    $9.21~(7.76) $   & $10.22~(8.50)$ &    $1.15~(0.01)$   & $3.66~(0.02)$ \\
\hline
\multicolumn{9}{c}{\bd Galaxies observed with EMIR}\\
\hline
ID27   & $2.57~(0.18)$   & $1.90~(0.23)$ &    $9.36~(8.21) $   & $9.48^*$ &    $8.87~(7.70) $   & $9.36~(7.96) $ &    $1.05~(0.01)$   & $2.25~(0.01)$ \\
ID29   & $2.43~(0.14)$   & $3.48~(0.27)$ &    $9.23~(7.98) $   & $9.48^*$ &    $9.17~(7.91) $   & $10.23~(8.65) $ &    $0.58~(0.01)$   & $1.62~(0.03)$ \\
ID31   & $4.94~(0.20)$   & $7.65^{(a)}~(0.30)$ &    $9.55~(8.17) $   & $9.71^*$ &    $9.21~(7.74) $   & $9.89~(8.39) $ &    $2.63~(0.01)$   & $6.60~(0.01)$ \\
ID33   & $0.94~(0.15)$   & $0.25^{(b)}~(0.25)$ &    $8.85~(8.05) $   & $9.09^*$ &    $8.59~(7.57) $   & $9.33~(7.68) $ &    $0.27~(0.01)$   & $0.64~(0.03)$ \\
ID35   & $2.95~(0.23)$   & $3.15~(0.33)$ &    $9.26~(8.16) $   & $9.50^*$ &    $8.96~(7.61) $   & $9.92~(8.37) $ &    $0.46~(0.01)$   & $1.00~(0.01)$ \\
ID37   & $2.94~(0.16)$   & $3.06~(0.26)$ &    $9.32~(8.07) $   & $9.56^*$ &    $8.99~(7.74) $   & $9.72~(8.10) $ &    $0.67~(0.01)$   & $1.85~(0.01)$ \\
ID40   & $3.13~(0.23)$   & $3.78~(0.31)$ &    $9.39~(8.26) $   & $9.55^*$ &    $8.81~(7.66) $   & $9.69~(8.31) $ &    $0.71~(0.01)$   & $1.57~(0.03)$ \\
ID41   & $1.51~(0.16)$   & $1.43~(0.31)$ &    $9.15~(8.17) $   & $9.37^*$ &    $8.59~(7.69)$	& $9.43~(8.01)$ &    $0.43~(0.01)$   & $1.25~(0.01)$ \\
ID43   & $2.53~(0.20)$   & $3.36~(0.32)$ &    $9.29~(8.18) $   & $9.45^*$ &    $8.77~(7.69) $   & $9.61~(8.18) $ &    $0.60~(0.01)$   & $1.35~(0.01)$ \\
ID44   & $2.56~(0.23)$   & $4.14~(0.34)$ &    $9.33~(8.28) $   & $9.55^*$ &    $8.77~(7.71) $   & $9.40~(7.75) $ &    $1.42~(0.01)$   & $3.41~(0.02)$ \\
ID47   & $1.66~(0.15)$   & $1.76~(0.14)$ &    $9.00~(7.97) $   & $9.24^*$ &    $8.65~(7.62) $   & $10.01~(8.48) $ &    $0.25~(0.01)$   & $1.02~(0.01)$ \\
 \hline\hline
\end{tabular}
\end{center}
\label{table:masses}
\end{minipage}
\begin{minipage}{\textwidth}
Notes. The EMIR beam sizes are 22\arcs and 
11\arcs for CO(1-0) and CO(2-1), respectively. 
EMIR intensities are given per beam. 
HERA intensities are averages within a circle of diameter 
$\sim$22\arcs positioned on the galaxy center. 
Uncertainties are indicated in parenthesis and correspond to statistical errors. 
Total masses are measured out to $R = 0.7\times R_{25}$ for $M_{\rm H_2,~total}$ 
and $R\simeq3\times R_{25}$ for ${\rm SFR_{total}}$ and $M_{\rm HI,~total}$. 
$(a)$~the CO(2-1) line shape is more asymmetric than that of CO(1-0). 
We may be missing extended emission due to the different beam sizes. 
$(b)$~1-$\sigma$ limit on the CO(2-1) intensity. 
$(*)$~mass extrapolated from the central measurement (see Section~\ref{sect:total}).
\end{minipage}
\end{table*}

\subsection{Sources of uncertainty}
%
Besides statistical errors, there are other systematic sources 
of uncertainty to be aware of. IRAM flux calibration uncertainties
are on the order of 10\% and pointing accuracy is about 2\arcs.
Uncertainties on the quantities listed in Table~\ref{table:masses}
only include statistical errors.

\subsection{WSRT observations of H\,{\sc i} 21-cm}
To complement the \hi 21-cm data set of the \bd survey published 
in \cite{wang-2013}, we observed \hi in the galaxy ID41 with the WSRT.
ID41 was not observed in the initial survey due to time constraints.
The data were obtained in February 2014, with an on-source 
integration time of 13\,h, as part of the program R14A028 (PI Bigiel). 
The reduction steps are identical to those performed in \cite{wang-2013}
and the \hi masses are reported in Table~\ref{table:masses}. 
The beam size is 16$\times$20\,arcsec$^2$, the velocity resolution
is 12.4\,kms and the noise level is 0.33\,Jy\,beam$^{-1}$.

\section{Analysis}
\label{sect:analysis}
In this study, we work with integrated results for each galaxy and
radial profiles for galaxies mapped in CO. We describe how those
measurements are made in the following.

\subsection{Surface density maps}
\label{sect:surfmaps}
To compare ISM tracers to star formation properties, we create 
surface density maps of \htwo, \hi, and SFR (denoted \sightwo,
\sighi, and \sigsfr). 
The CO(2-1) intensities are converted to \htwo surface densities 
assuming an intensity ratio of CO(2-1)/CO(1-0) of 0.8 and a Galactic 
conversion factor 
$\alpha_{\rm CO}=4.38$~\msun\,pc$^{-2}$~(K\,km\,s$^{-1}$)$^{-1}$, 
which includes helium, both appropriate for Milky-Way type galaxies 
\citep[e.g.,][]{leroy-2013}.
We compare the assumed CO line ratio to that measured with EMIR 
in Section~\ref{sect:icoratio} and we discuss effects of using a 
metallicity-dependent conversion factor in Section~\ref{sect:coldg}. 
The \hi products are taken from \cite{wang-2013} and we refer to 
that paper for information on the data acquisition and processing 
of the \bd galaxies. We use the same version of the \hi data as in
their analysis, i.e. with robust weighting of 0.4 and no tapering,
which is the best compromise between sensitivity and resolution.
The SFR is measured from a combination of archival WISE 22\mum
\footnote{\url{http://wise2.ipac.caltech.edu/docs/release/allsky/}}
\citep{wright-2010} and GALEX FUV\footnote{\url{http://galex.stsci.edu/GR6/}}
\citep{martin-2005} data.
We first removed a background using areas of the maps where
there is no signal, convolved the maps from their native resolution
(12\arcs for WISE 22\mum, 4.2\arcs for GALEX FUV)
 to a resolution of 13\arcs with a Gaussian kernel, and converted
the WISE fluxes to MIPS 24\mum fluxes using a conversion factor
of $0.86$ (based on the \citealt{chary-2001} templates and respective
response curves). The datasets are then combined using the formula
from \cite{leroy-2008}:
\sigsfr~$= 0.081 \times I_{\rm UV} + 0.0032 \times I_{24}$.
In addition, we compare the gas masses and SFR to metallicity
measurements. Optical spectroscopy was obtained with the WHT and
converted to oxygen abundance using the models of \cite{charlot-2001}
(see \citealt{carton-2015}, for details). Observations consisted of
a slit passing through the centre of each disc and aligned with
the rotation axis of the galaxy.
Note that ID6, ID11, ID15 and ID17 do not have metallicity
measurements at their centre because of signatures of non star-formation
activity (AGN/LINER) in their optical spectra \citep{carton-2015}.
However, the WISE colours based on bands 1, 2 and 3
at shorter wavelengths indicate that our galaxies do not fall in the
AGN wedge but have colours compatible with those of star-forming
galaxies \citep{wright-2010,cutri-2013}. Therefore we do not expect
the derived SFR to be significantly contaminated by AGN activity.

\subsection{Structure in the CO maps}
At a working resolution of $\sim$13\arcs, we detect CO emission
(above 3-$\sigma$) in the \bd galaxies mapped with HERA in roughly
$4$ independent beams. As a consequence, we cannot identify
clear sub-structure, but we do see extended CO emission (see the
maps in Appendix~\ref{append:hera}). CO peaks at the optical centre
of the galaxies and is usually elongated along the major axis. 
The signal-to-noise ratio is low for ID2, ID6, ID8, and ID14.
Their CO intensity maps appear patchy because CO is detected
only in the brightest peaks. We present their profiles but we
exclude them from any quantitative radial analysis. 
CO and \hi emission follow each other well globally. The CO and
\hi peaks are noticeably offset (by at least 10\arcs) in ID5, ID6, ID8,
ID11, ID15, and ID17. This is because the \hi emission is less centrally
concentrated, or, in the cases of ID15 and ID17, more clumpy.
Visual inspection of the CO intensity maps indicates that features
appearing outside of the optical discs are most likely noise rather than
emission from a companion galaxy.
We note for individual galaxies:
\begin{itemize}
\vspace{-1mm}
\setlength\itemsep{0em}
\item {\it ID1:} the knots N and S of the central emission blob 
	are real and coincide with the spiral arms of the galaxy.  
\item {\it ID3:} the two stripes in the NE direction are noise. 
\item {\it ID11:} the extensions of the main peak to the N and E 
	are in agreement with the asymmetric \hi distribution. 
	There seems to be a companion in the optical and \hi deficit to the NE.  
\item {\it ID15:} the emission in the SE is bright and may be connected 
	to the spiral arm. 
\end{itemize}

\subsection{Radial Profiles}
\label{sect:profiles}
%
To analyze the radial behaviour of the different tracers, we produce 
surface density profiles as a function of radius, with step size
$1/2$ of the spatial resolution of the CO data (i.e. 6.5\arcs) and
out to a radius of 1.5\,$R_{25}$.
We caution that this sampling applies to the major axis, but
for highly inclined galaxies, the minor axis remains largely unresolved.
At a given radius, we consider the average surface density within 
a tilted ring of width $\pm0.5$ times the chosen step size. The assumed 
inclinations and position angles are given in Table~\ref{table:general}.
Those are measured from SDSS r-band data.
Error bars on the profile measurements are also taken as the average
within each ring in our error maps.
All profiles are corrected for inclination. 
The radial profiles of \sighi, \sightwo, \sigsfr, metallicity, and their 
associated errors are shown in Fig.~\ref{fig:allprofiles} and discussed 
in Section~\ref{sect:radial}.

\subsection{Integrated results}
\label{sect:total}
\subsubsection{Quantities: total and within 22\arcs}
\label{sect:qtot22}
Most of the control galaxies have CO(1-0) observed only 
in their centre (EMIR beam size of 22\arcs at 115.3\,GHz) 
and we have full coverage of the molecular emission mainly 
for the \hi-rich galaxies which molecular gas discs were mapped
with HERA. Therefore, for all galaxies, we compile 
measurements of molecular gas mass, atomic gas mass, 
and SFR inside an aperture of diameter 22\arcs as well as 
total measurements. For total measurements, we perform
aperture photometry (with weights of 1 inside the
aperture and weights of 0 outside the aperture) on the CO
(if available), \hi and SFR maps. Total quantities are derived
out to a radius of $0.7\times R_{25}$ for the molecular gas
and out to $3\times R_{25}$ for the atomic gas and SFR. 
When doing aperture photometry, the integrated errors 
correspond to the quadratic sum of the errors multiplied
by the square root of the oversampling factor of the beam.

For the galaxies that have CO(1-0) observed only in their 
centre (EMIR sample) or CO(2-1) mapped but detected only in their centre 
(ID2, ID6, ID8, and ID14), we estimate their total molecular 
gas mass by extrapolation of their central measurements. 
We assume that CO is distributed in an exponential disc of 
scale-length $0.2\times R_{25}$ \citep[e.g.,][]{leroy-2008,lisenfeld-2011} 
and we employ the 2D prescription of \cite{boselli-2014a} since 
none of our discs are edge-on (their equations 9 and 10).
This extrapolation corresponds to a correction factor of
the central mass to the total mass of a factor of $\sim$2
(see Table~\ref{table:masses}).
Testing this method on the galaxies that were actually mapped, 
we find that the molecular gas masses measured inside 
a radius of $0.7\times R_{25}$ and those extrapolated from 
their central mass agree within 20\%. This gives confidence 
in the methods.
We also tried to use individual scale-lengths of the stellar disc
from the stellar surface density maps of \cite{carton-2015},
instead of the canonical value of 0.2, but for the mapped
galaxies, the scatter between the measured total mass and
the mass extrapolated from the centre is not improved.
For consistency, we prefer to use the fixed value of 0.2.

\subsubsection{The CO(2-1)/CO(1-0) intensity ratio}
\label{sect:icoratio}
For the EMIR galaxies, we have obtained with central pointings
both CO(1-0) and CO(2-1) in different beam sizes (22\arcs and
11\arcs, respectively). 
The intensity ratio at face value (assuming uniform distribution) 
varies between 0.65 and 1.6, which is globally higher than 
the standard value of $\sim$0.8 observed in nearby disc galaxies 
\citep[e.g.,][]{leroy-2009a}. 
In the case of point-like emission, the ratios have to be corrected 
for the difference in beam solid angles and thus multiplied by $1/4$. 
Note that both scenarios are likely not good assumptions for
the distribution of CO emission in the inner parts of galaxies
which often decreases exponentially with radius.

Given the size of the optical radii of our galaxies, we expect 
CO emission to be marginally extended compared to the 
EMIR beam. In order to compare the intensities of the two 
CO transitions, a more realistic approach would be to correct 
for the aperture difference assuming a radial 
decline of the CO emission. 
Here, we use the HERA data to simulate what would be the
CO(2-1) mean surface brightness as if observed by a single
pointing. We employed two different methods: circular aperture
photometry, and multiplying our map with a Gaussian kernel.
We find the two methods to be equivalent within $\sim$5-15\%
for FWHMs of 22\arcs and 11\arcs respectively. Therefore we
prefer to use the aperture photometry method for simplicity.
The CO(2-1) mean surface brightness within
a diameter of 22\arcs about the centre is $0.60-0.95$ times lower
than the mean surface brightness within a diameter of 11\arcs.
Hence for the EMIR data, we multiply the observed CO(2-1)
intensity (for which the beam size is 11\arcs) by a factor of $0.75$
to estimate what would be the CO(2-1) intensity in 22\arcs, and 
we compare this estimate to the observed CO(1-0) intensity
(for which the beam size is 22\arcs). Doing so, we find CO intensity
ratios between $0.5-1.2$ in the EMIR galaxies, with a median
value of $0.8$. 
If conditions in the CO-emitting clouds are similar in the \hi-rich 
and control galaxies, this validates our choice of a CO(2-1)/CO(1-0) 
intensity ratio of $0.8$ used for the discs mapped with HERA 
where we have no CO(1-0) data.

\section{Results}
\label{sect:results}
\subsection{Comparison to the surveys COLD\,GASS and HRS}
\label{sect:coldg}
\begin{figure}
\centering
 \includegraphics[clip,trim=0 8mm 0 4mm,width=6.7cm]{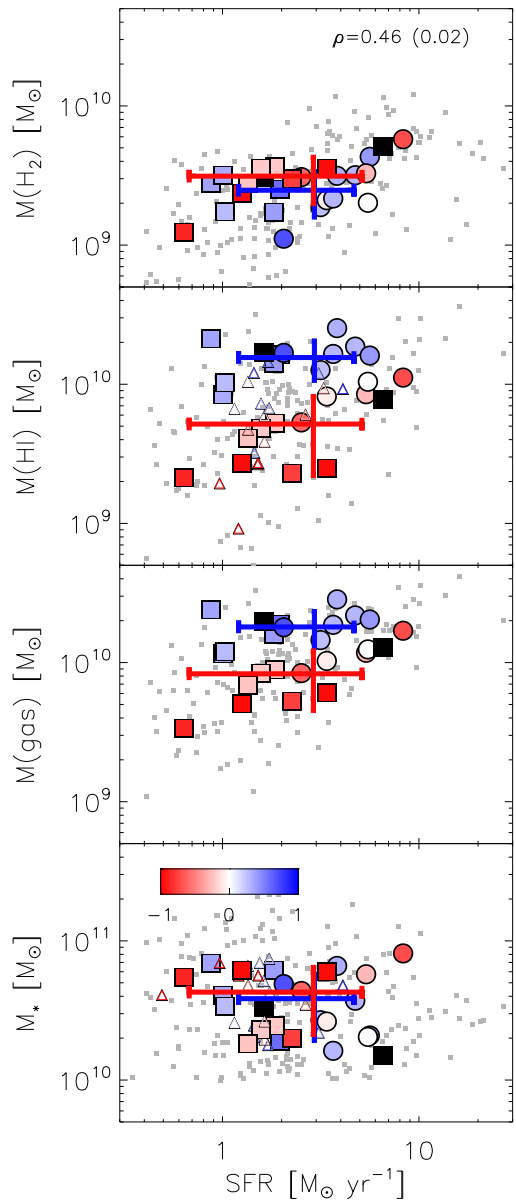}
\caption{ 
 Comparison of stellar and gas masses as a function of SFR.
 Circles are galaxies with measured CO masses (HERA sample) and
 squares are galaxies with CO central measurements extrapolated
 to total masses (EMIR sample and ID2, 6, 8, 14); see Sect.~\ref{sect:qtot22}
 for details.
 Open triangles are galaxies not observed in CO.
 Colour-coding is based on the distance from the plane
 in fig.~3 of \protect\cite{catinella-2013}. 
\hi-rich galaxies are above the plane, control galaxies are
below that plane, and excluded galaxies are in black. 
The mean and standard deviation of each quantity are 
overplotted as crosses for the \hi-rich and control samples. 
The background grey data points are CO-detected galaxies
from the COLD\,GASS 3rd data release \protect\citep{saintonge-2011b}.
}
\label{fig:coldg1}
\end{figure}
\begin{figure*}
\centering
 \includegraphics[clip,trim=2cm 2.2cm 1cm 0,width=17cm]{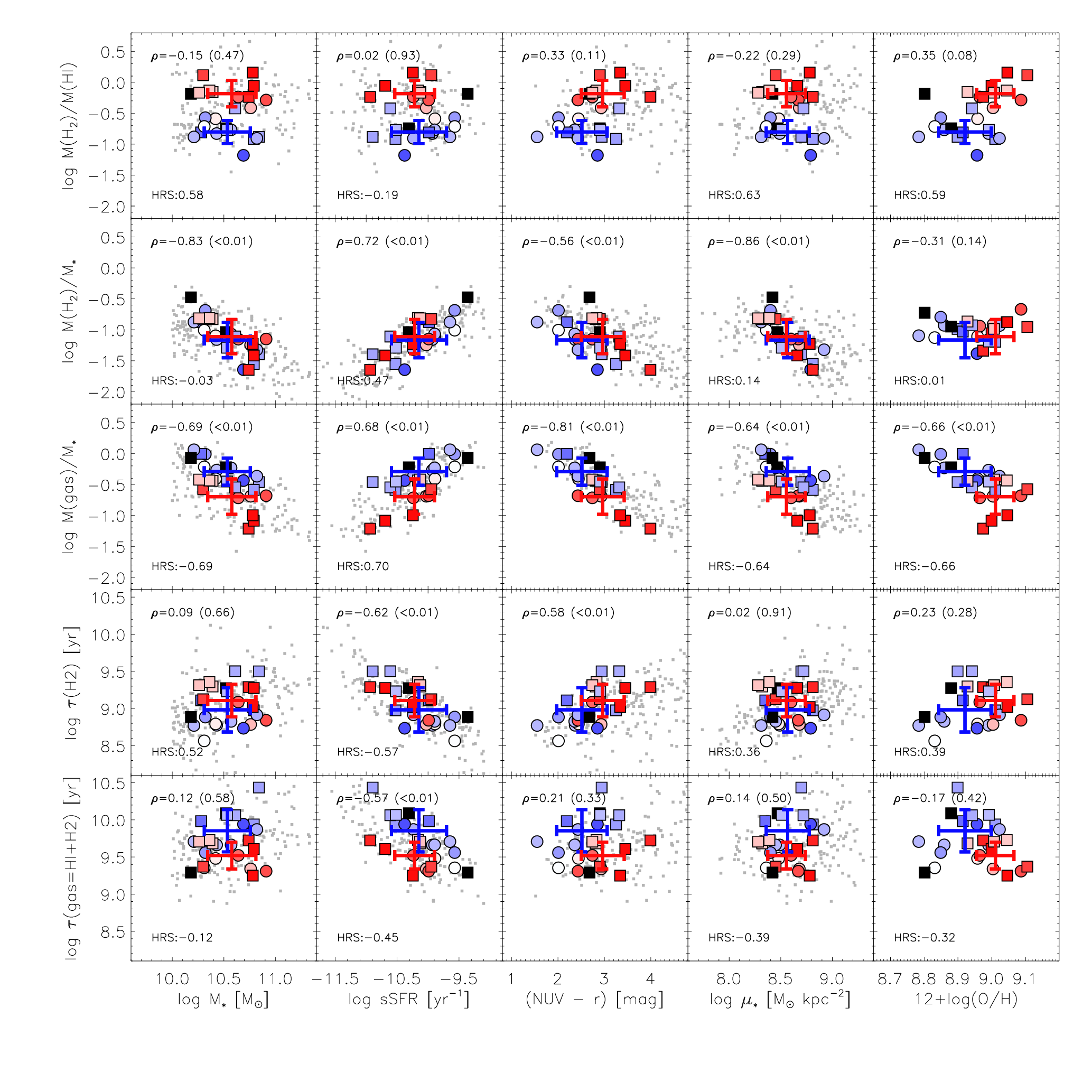}
\caption{ 
Scaling relations for the \bd galaxies.
We show the molecular fraction, \mhtwo/\mhi, the molecular 
and atomic to stellar mass ratios, \mhtwo/\mstar and \mhi/\mstar, 
the molecular gas depletion time, $\tau_{\rm H2}$, 
and total gas depletion time, $\tau_{\rm gas=HI+H_2}$, 
as a function of stellar mass, specific SFR, sSFR, colour, $NUV-r$,
stellar mass surface density, \mustar, and metallicity, 
for total measurements of the \bd galaxies. 
Same colour-coding as in Fig.~\ref{fig:coldg1}. 
The mean and standard deviation of each quantity are 
overplotted as crosses for the \hi-rich and control samples.
We also indicate Pearson correlation coefficients of the 
\bd galaxies as well as coefficients of the HRS galaxies 
for comparison \protect\citep{boselli-2014b}. 
The background grey data points are CO-detected galaxies
 from the COLD\,GASS 3rd data release \protect\citep{saintonge-2011b}.
}
\label{fig:hrs}
\end{figure*}

We compare total gas quantities with general galaxy parameters
for the \bd galaxies and contrast our findings with the results from
COLD\,GASS \citep{saintonge-2011a,saintonge-2011b} and HRS
\citep{boselli-2014a,boselli-2014b} in Figs.~\ref{fig:coldg1} and \ref{fig:hrs}.
Trends between variables (in logarithmic scale) are quantified with
the Pearson correlation coefficient and their significance are indicated
in parenthesis.
COLD\,GASS studied a sample of 215 CO-detected galaxies at
redshifts $0.025<z<0.05$ and stellar masses $10.0<$log(\mstar)$<11.5$,
while the HRS sample consists of 143 CO-detected galaxies at $z<0.005$
and $9.0<$log(\mstar)$<11.0$.

Figure~\ref{fig:coldg1} shows the mass of stars, atomic, molecular
and total gas (atomic + molecular) as a function of SFR for the \bd
galaxies. We find that the \hi-rich galaxies span the same range of SFR
and stellar masses as the control galaxies. They clearly have more \hi,
by a factor of $\sim$4 on average, but they have similar \htwo masses.
Overall, the SFR is not correlated with the stellar mass, nor with the
\hi mass, but is correlated with the \htwo mass (Pearson correlation
coefficient $\rho\simeq0.46$).
Our \bd galaxies fall within the range of parameters covered by the
COLD\,GASS sample, with the \hi-rich galaxies being on the upper
end of the \hi and total gas mass distributions.
Comparing the ranges of parameters covered by the entire \bd sample,
the sub-sample of galaxies that we observed in CO seems representative
of the whole \bd sample. The galaxies mapped with HERA are amongst
the most \hi-rich and actively star-forming, characteristics
to keep in mind for the interpretation
of our resolved analysis in the context of galaxy evolution.

Figure~\ref{fig:hrs} shows the molecular fraction, \mhtwo/\mhi, 
the molecular and total gas mass to stellar mass ratios, \mhtwo/\mstar 
and M(gas)/\mstar, the molecular gas depletion time, $\tau_{\rm H2}$, 
and the total gas depletion time, $\tau_{\rm gas=HI+H_2}$, as a
function of stellar mass, specific star-formation rate, sSFR, colour,
$NUV-r$, stellar mass surface density, \mustar, and average metallicity. 
The depletion time is defined as the mass of gas divided by the SFR.
The average values of \mhtwo/\mstar and M(gas)/\mstar are 
$\sim$0.1 and 0.4 respectively, with the \hi-rich galaxies having
$\sim$3 times larger total gas masses than the control galaxies. 
The mean value of \mhtwo/\mhi is $\sim$40\% in both the \bd
and COLD\,GASS samples \citep{saintonge-2011a}. 
However, we find that the \hi-rich galaxies have systematically
$4$ times lower \mhtwo/\mhi ratios compared to the control galaxies,
independently of global galaxy parameters. Our molecular gas
fractions are not as well correlated with galaxy parameters as
for the HRS survey. This may be a sample effect as the \hi-rich
galaxies were selected specifically for their large \hi masses. 
The trend of increasing molecular fraction with metallicity that
we observe in the \bd could point to an underestimation of the
molecular gas mass from CO in the lower metallicity galaxies
\citep[e.g.,][]{bolatto-2013,cormier-2014}. The mean metallicities
of the \hi-rich and control samples are higher than solar
and differ only by 0.1\,dex though. Correcting masses
with the metallicity-dependent prescription of $X_{\rm CO}$
from \cite{bolatto-2013} (their equation 31) would increase
global \htwo masses by up to 10\% in the \hi-rich galaxies.
Hence the offset in the \mhtwo/\mhi values is unlikely to be due
to $X_{\rm CO}$ conversion factor effects only given the
small range of metallicities spanned by our galaxies.

Generally, the trends that appear are similar to those found 
by \cite{boselli-2014b}. The strongest correlations are found 
for \mhtwo/\mstar, decreasing with \mstar and \mustar and 
increasing with sSFR; and for M(gas)/\mstar, decreasing with 
\mstar, \mustar, $NUV-r$, and metallicity and increasing with sSFR. 
Interestingly, the \hi-rich and control galaxies seem to follow 
parallel trends in some of those plots. 
Our correlations with \mhtwo/\mstar are more significant 
than those found for the HRS galaxies and closer to what 
\cite{saintonge-2011a} found. Again, this is probably 
a selection effect as our galaxies do not span a range of 
galaxy properties as wide as the HRS survey. 
We also find that the molecular depletion time, $\tau_{\rm H2}$, 
is not constant. It varies between 0.4 and 4\,Gyr, which is 
compatible with previous studies in the stellar mass range 
probed. It is marginally higher in the control galaxies than in
the \hi-rich galaxies, by a factor of $1.4$ on average, because
they are slightly more molecular-rich (Fig.~\ref{fig:coldg1}).
Yet, total gas depletion times are clearly higher in the \hi-rich
galaxies since \hi dominates their gas budget. 

To summarize, the main characteristics of the \hi-rich galaxies
in comparison to the control galaxies are: larger \hi masses,
bluer colours and lower average metallicities, for similar stellar
masses, molecular masses and SFR.
Within the accuracy of our methods, the bluer colours
in our galaxies seem to be related to a separation of the \hi-rich/control
samples in metallicity rather than a separation in SFR.
The low \htwo/\hi ratios indicate that no \htwo formation out
of accreted \hi gas has occurred yet.

We refer the reader to Section~\ref{sect:discussion} for a 
discussion of those results.
In the remainder of this section, we focus the analysis 
on the \bd galaxies mapped with HERA for which we have 
spatial information on the physical quantities. 

\begin{figure*}
\centering
 \includegraphics[clip,trim=0 17mm 1.3mm 0.5mm,width=6cm]{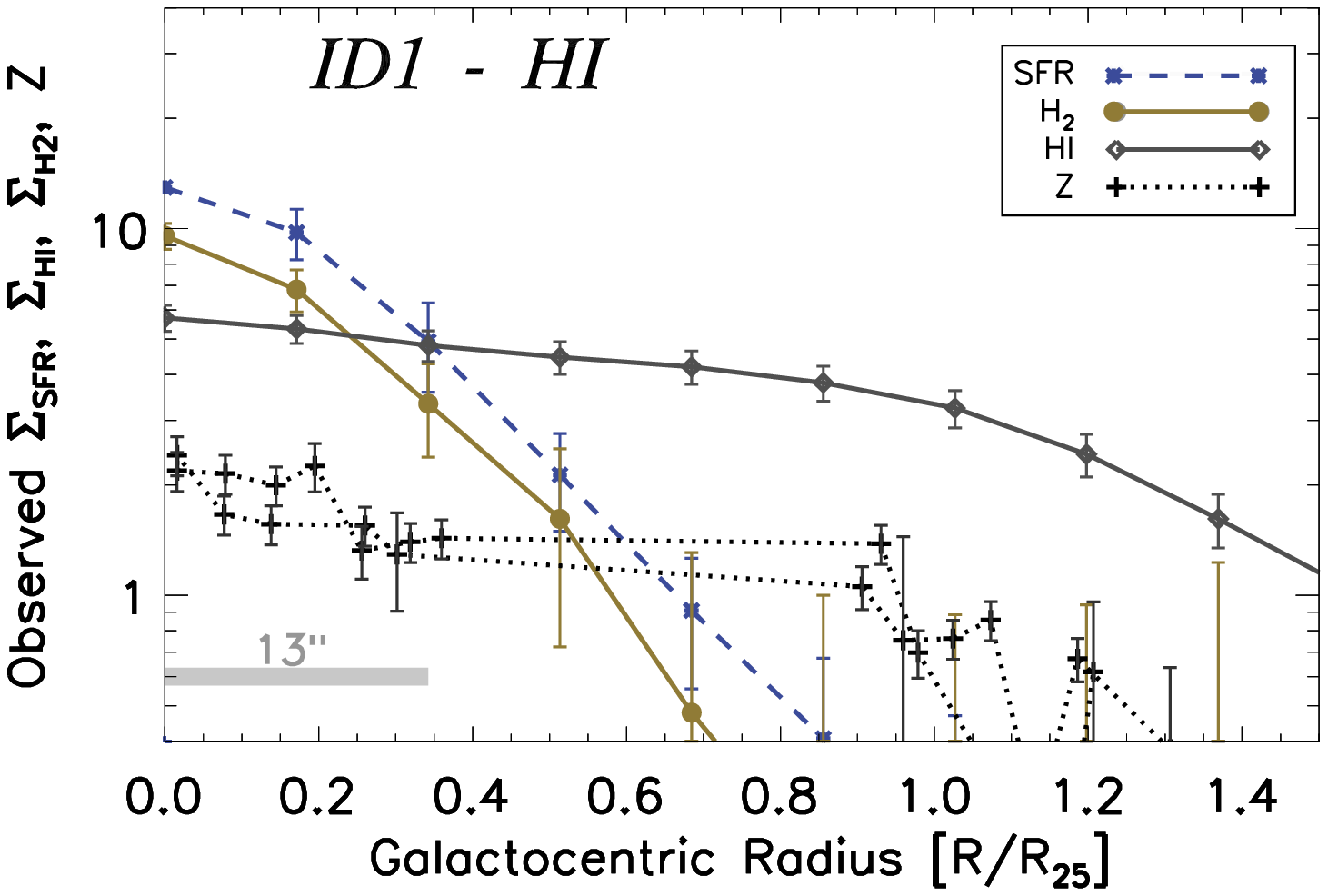}
 \includegraphics[clip,trim=19.5mm 17mm 1.3mm 0.5mm,width=5.2cm]{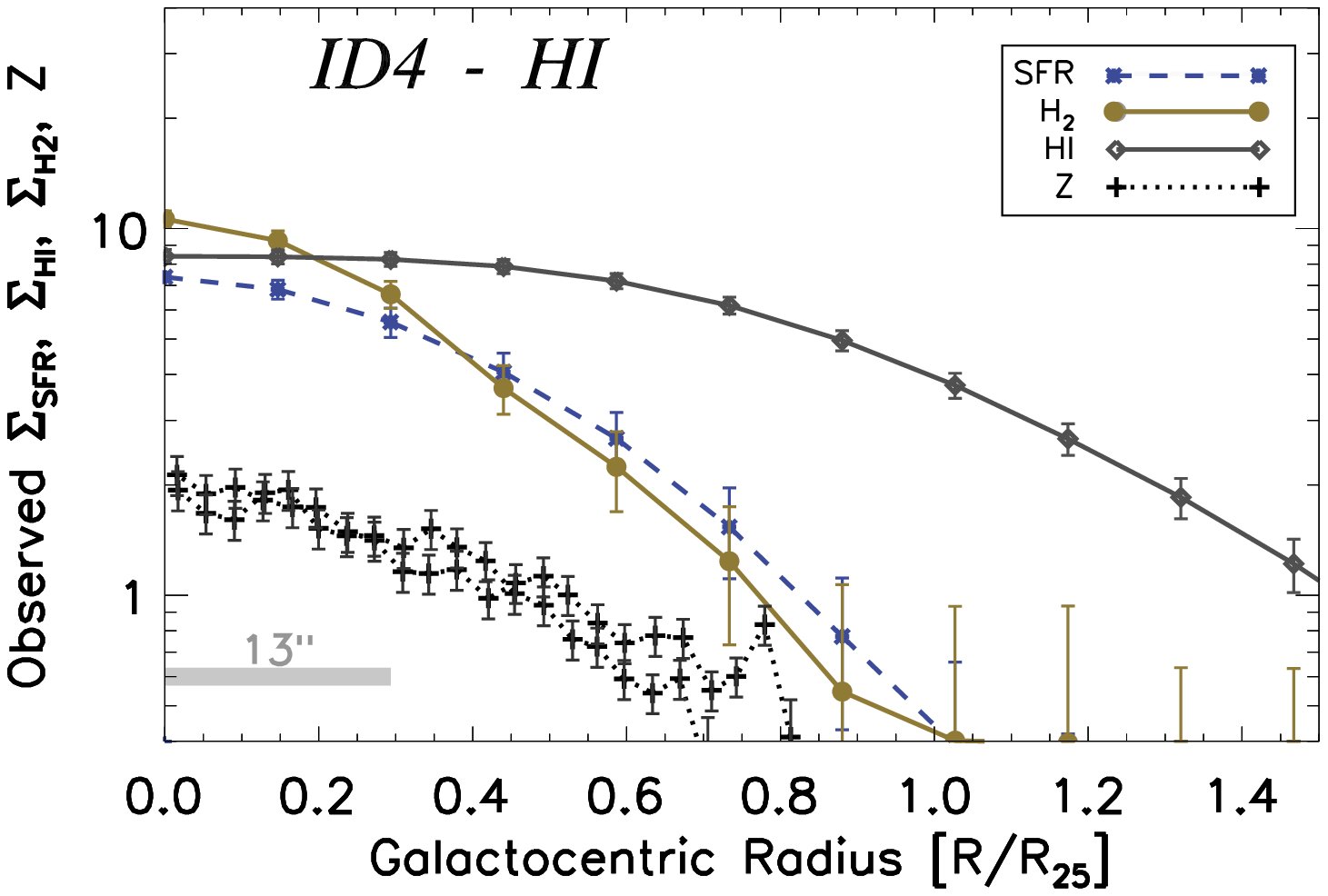}
 \includegraphics[clip,trim=19.5mm 17mm 1.3mm 0.5mm,width=5.2cm]{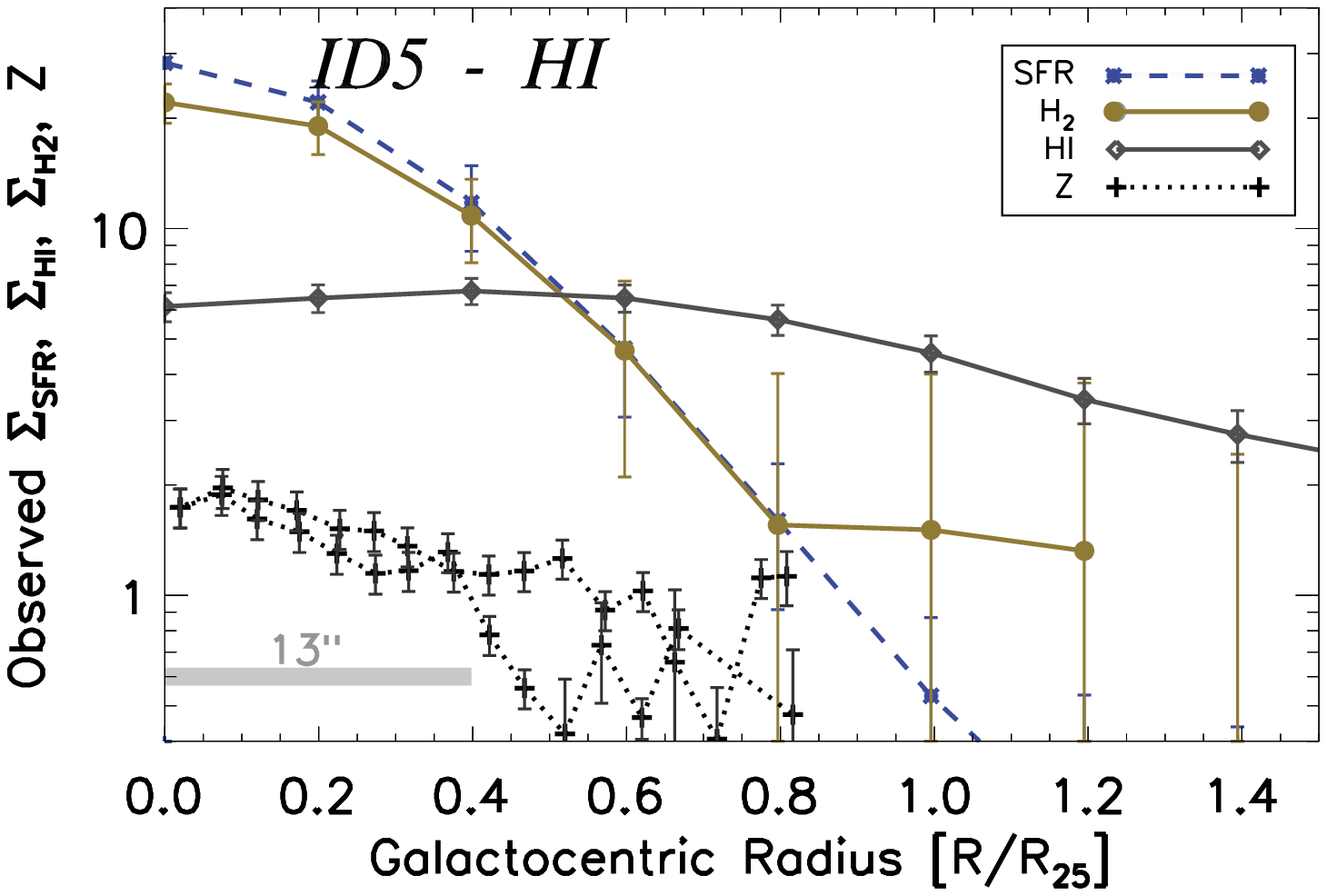}
 \includegraphics[clip,trim=0 17mm 1.3mm 0.5mm,width=6cm]{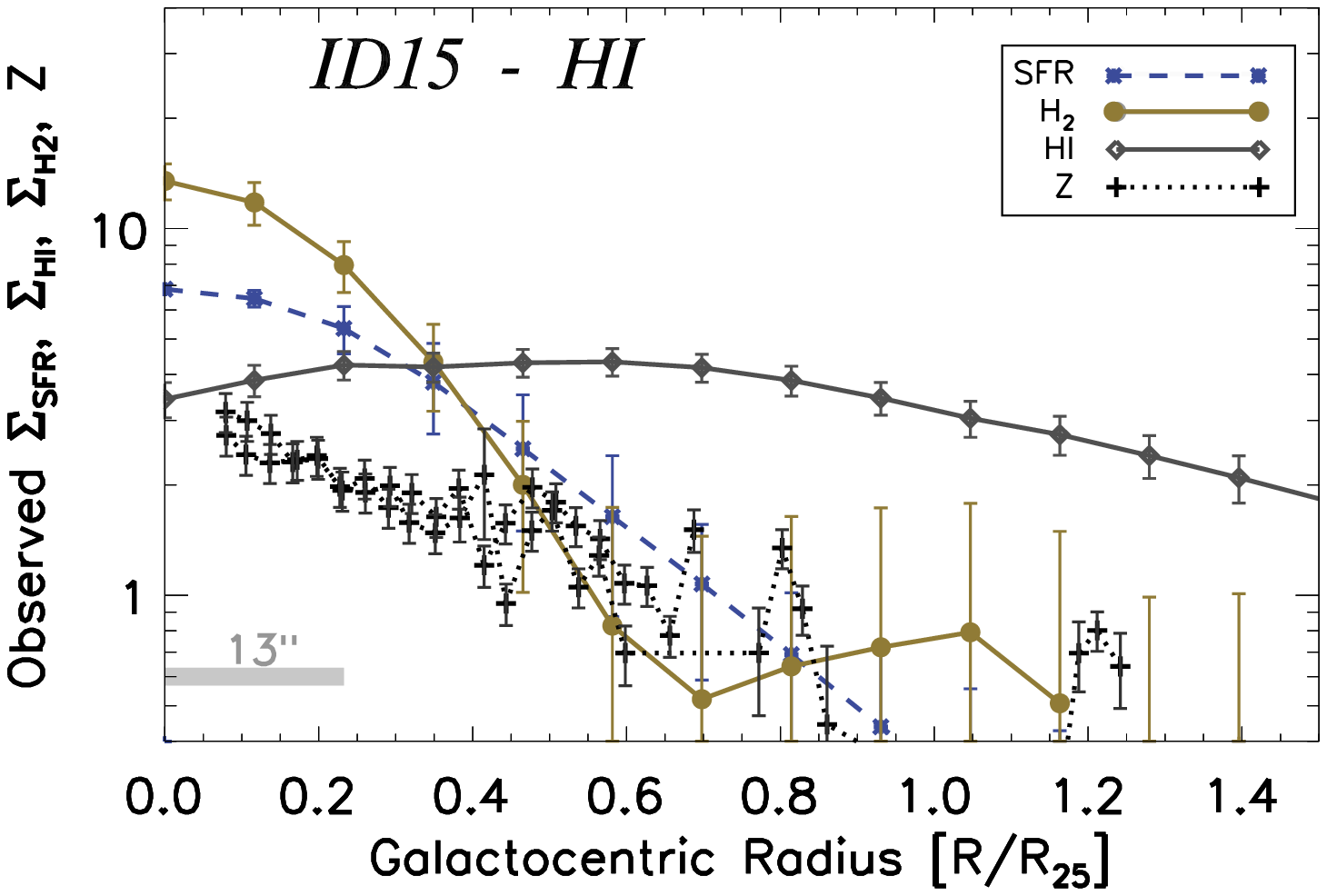}
 \includegraphics[clip,trim=19.5mm 17mm 1.3mm 0.5mm,width=5.2cm]{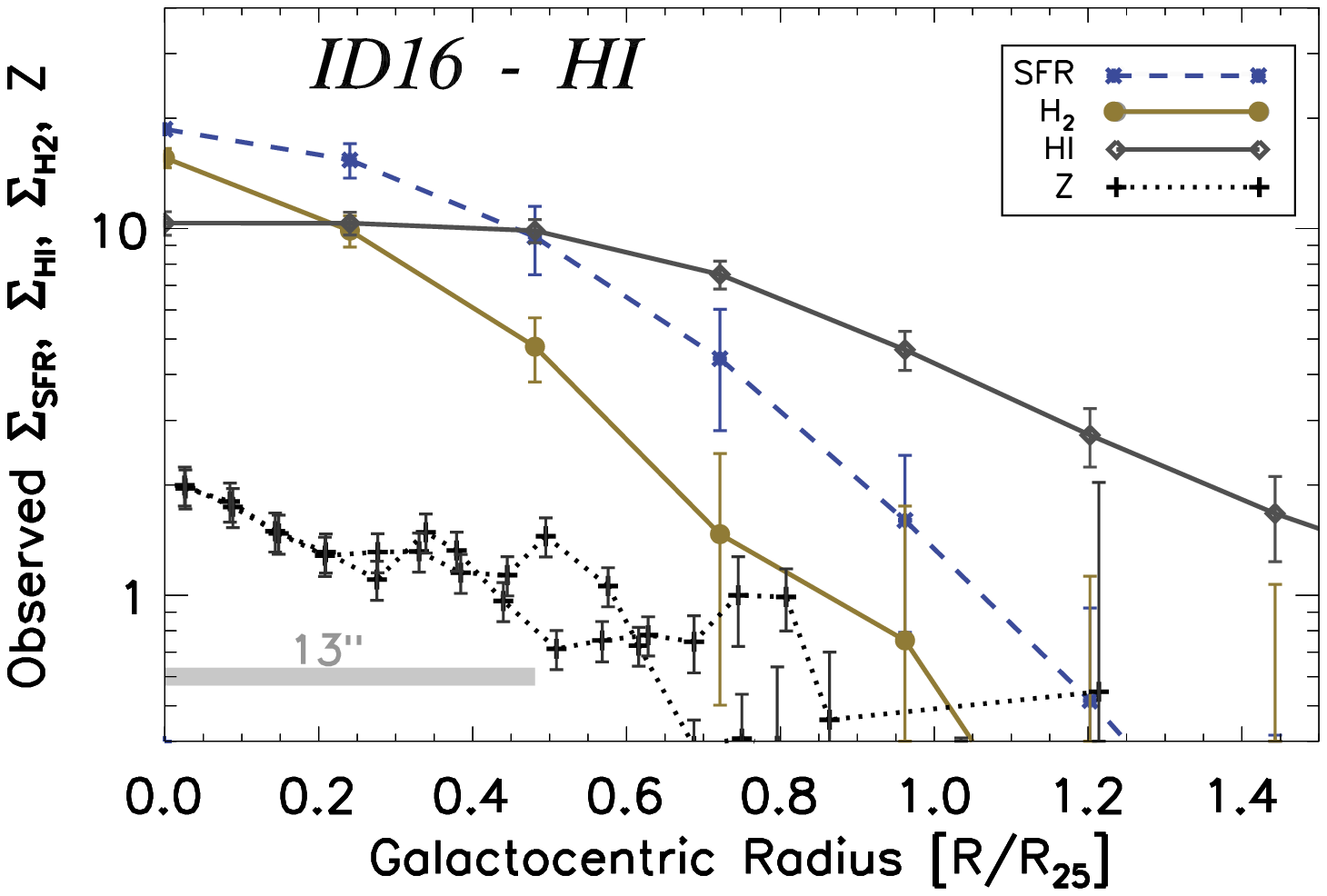}
 \includegraphics[clip,trim=19.5mm 17mm 1.3mm 0.5mm,width=5.2cm]{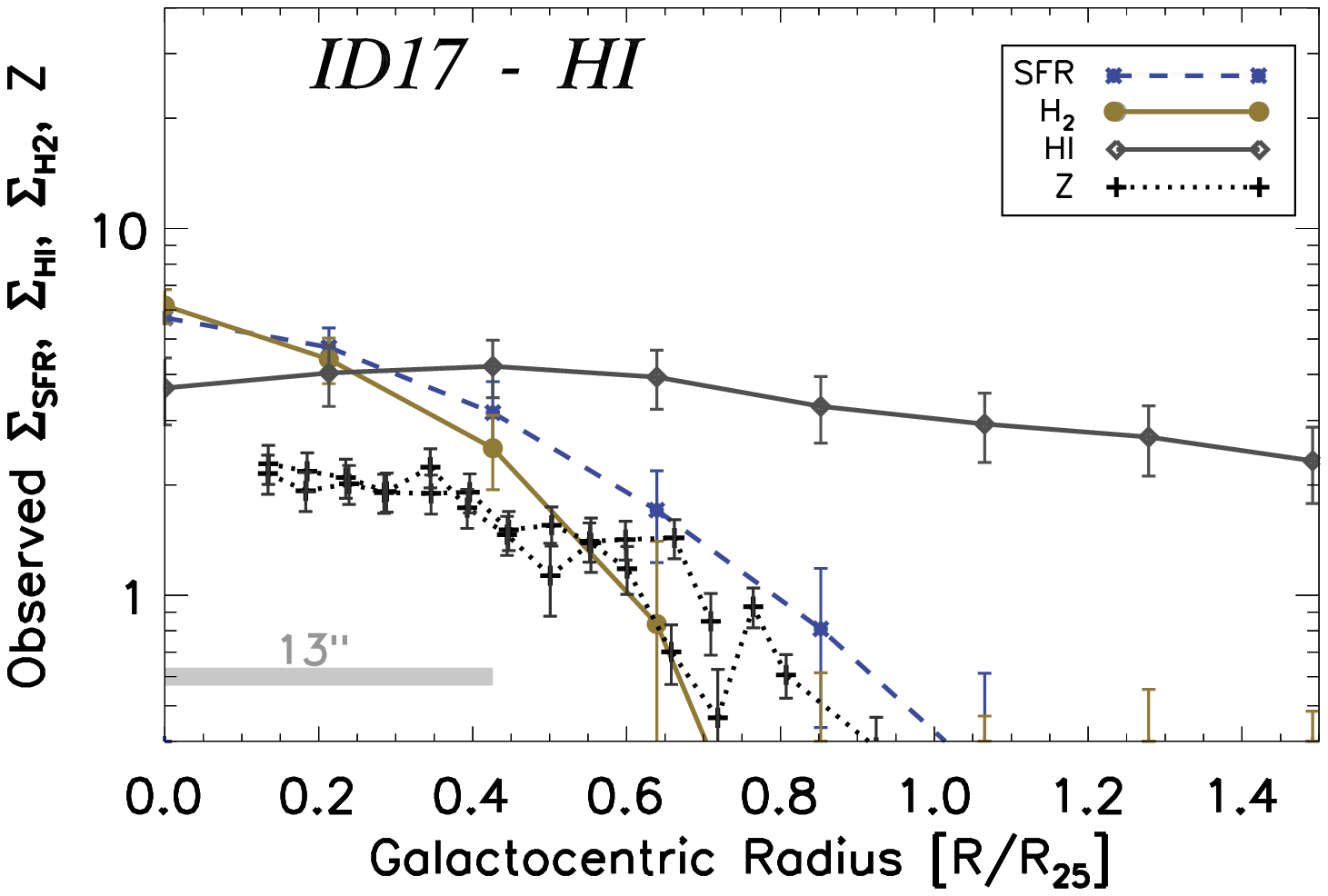}
 \includegraphics[clip,trim=0 17mm 1.3mm 0.5mm,width=6cm]{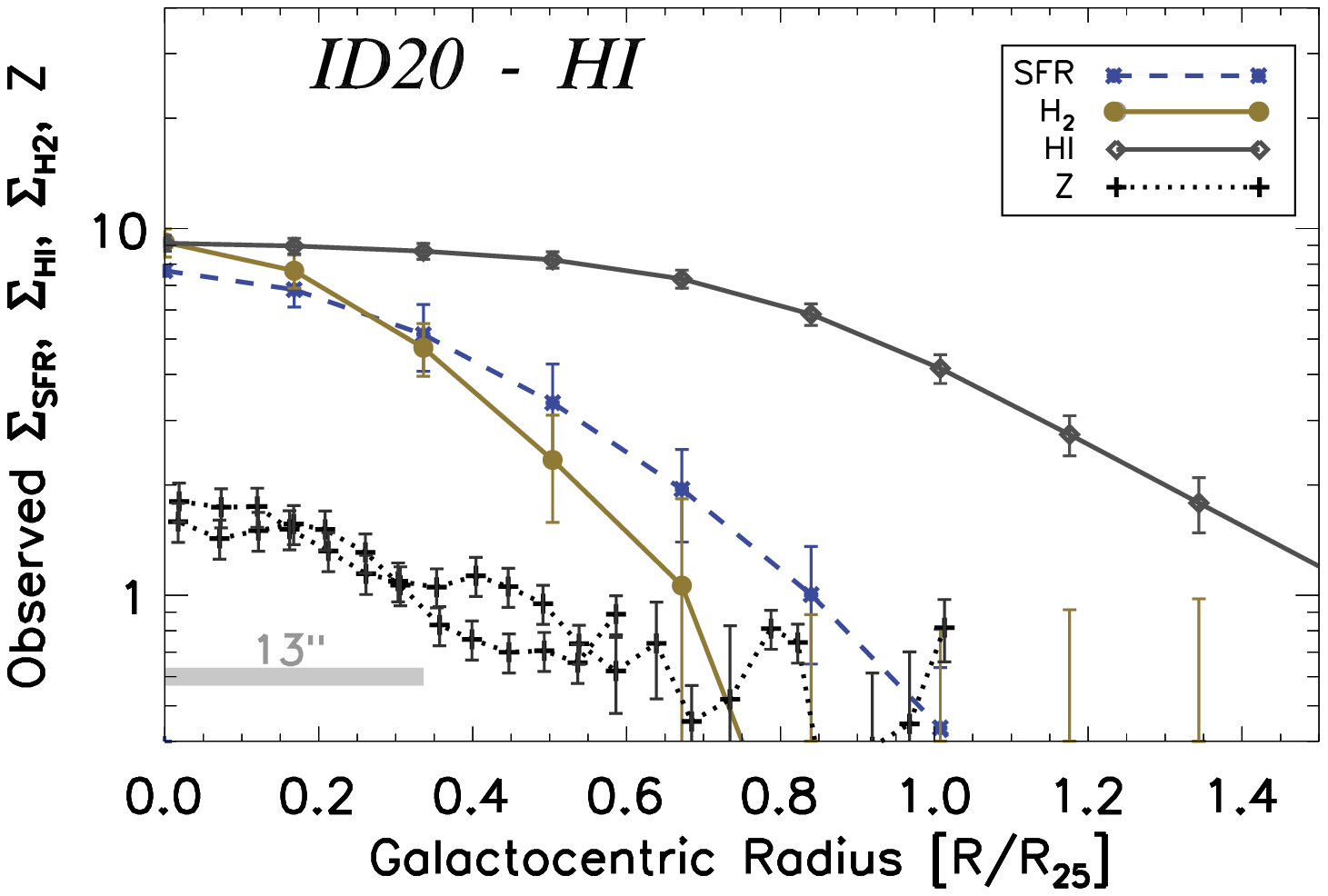} 
 \includegraphics[clip,trim=19.5mm 17mm 1.3mm 0.5mm,width=5.2cm]{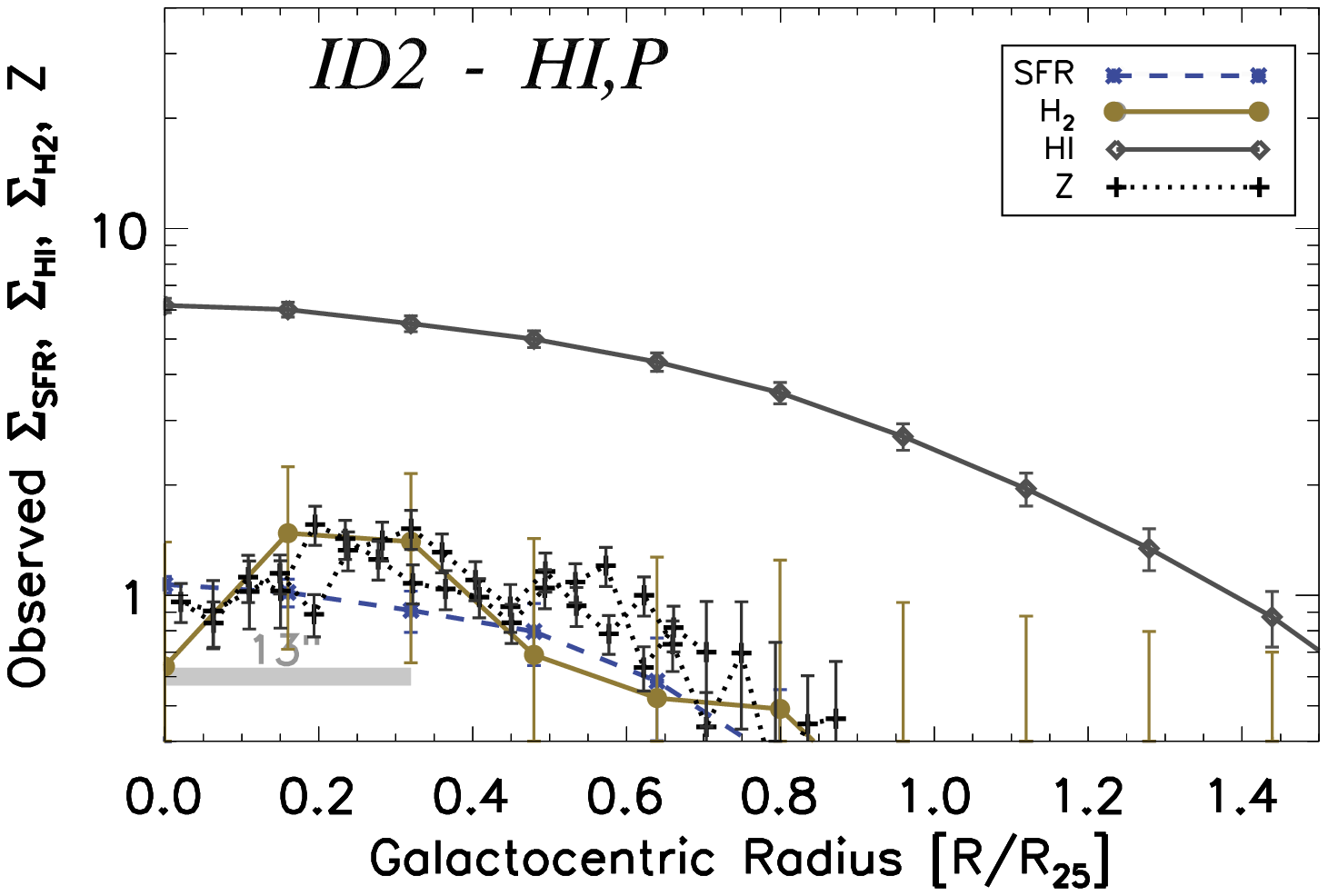}
 \includegraphics[clip,trim=19.5mm 17mm 1.3mm 0.5mm,width=5.2cm]{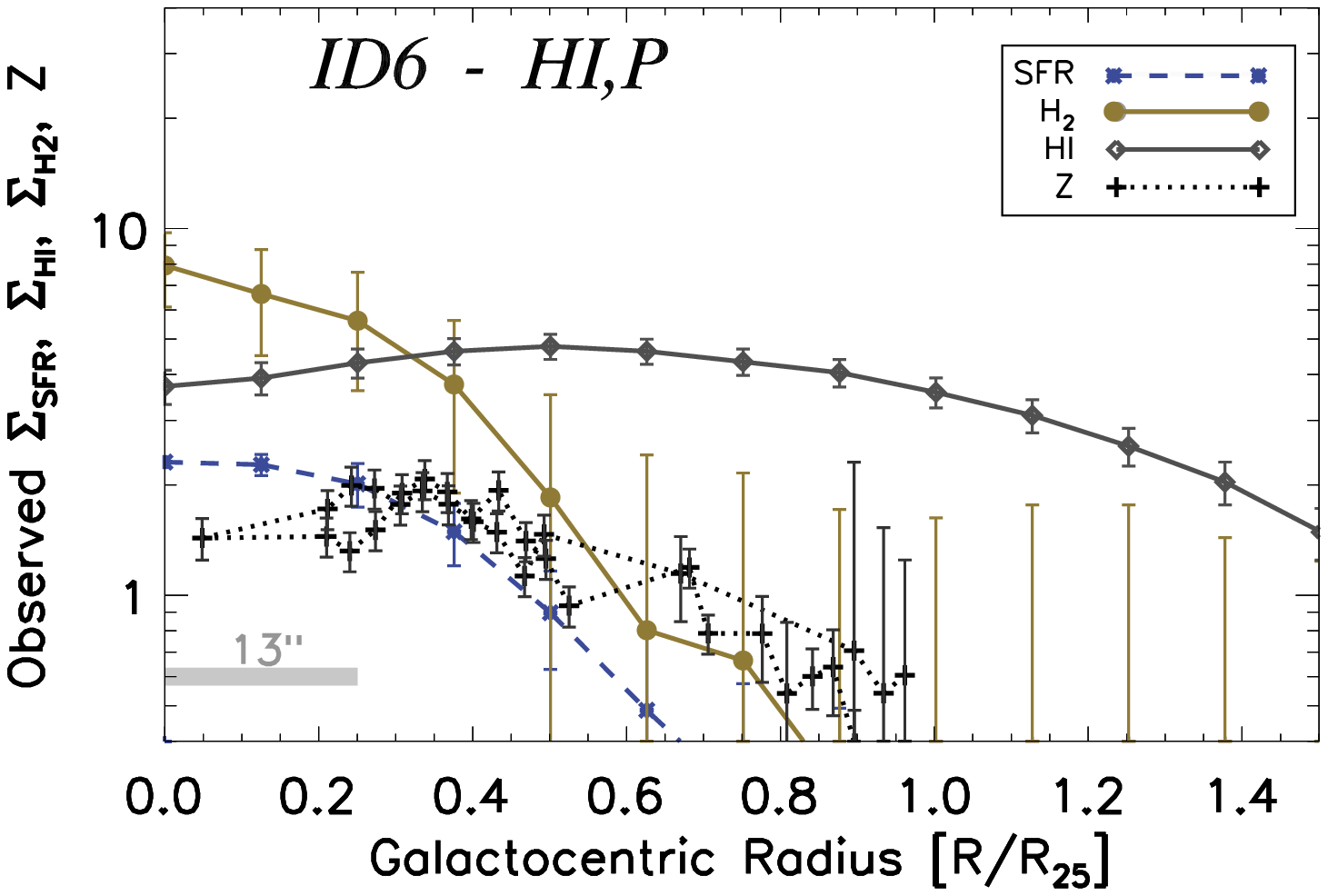}
 \includegraphics[clip,trim=0 17mm 1.3mm 0.5mm,width=6cm]{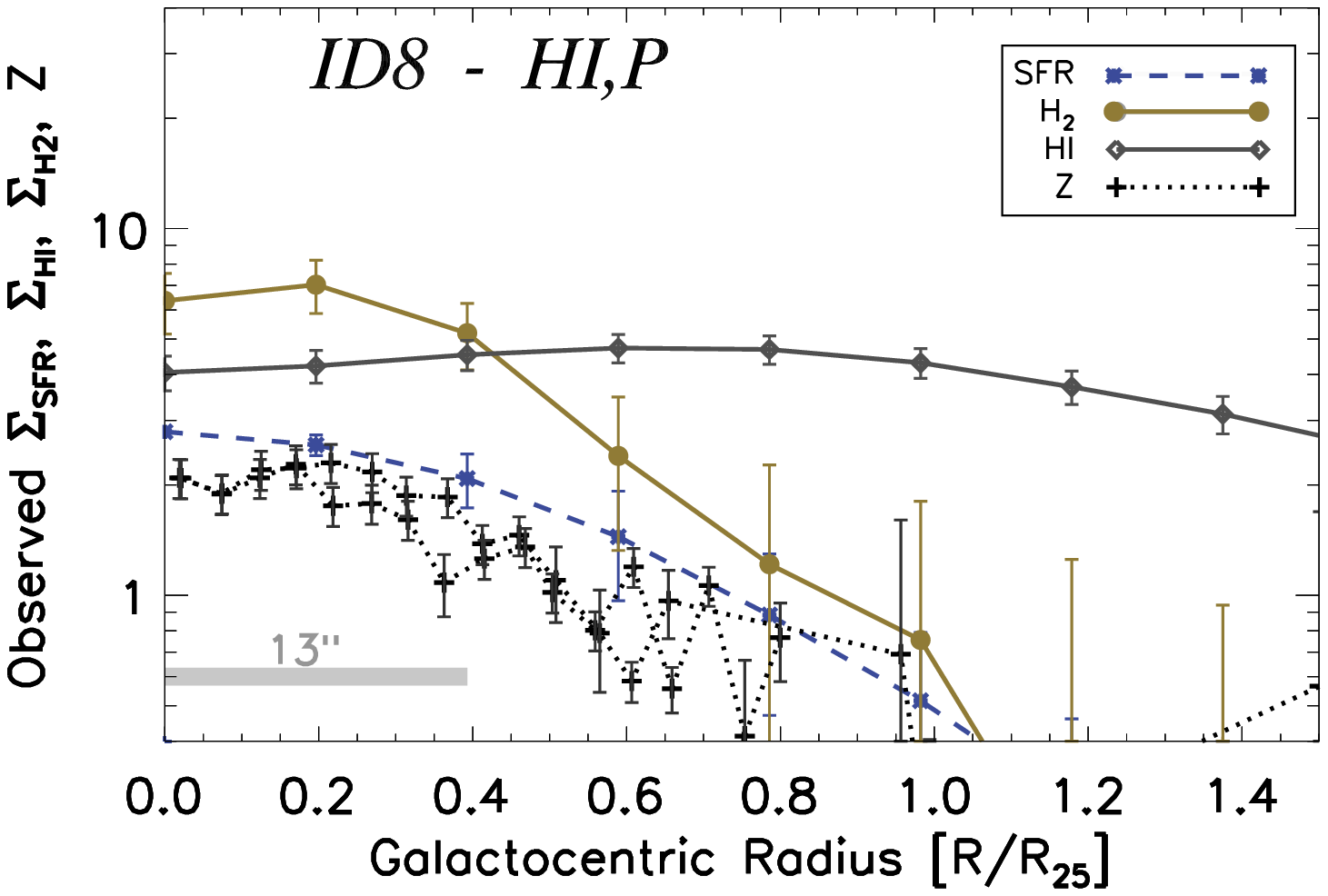}
 \includegraphics[clip,trim=19.5mm 17mm 1.3mm 0.5mm,width=5.2cm]{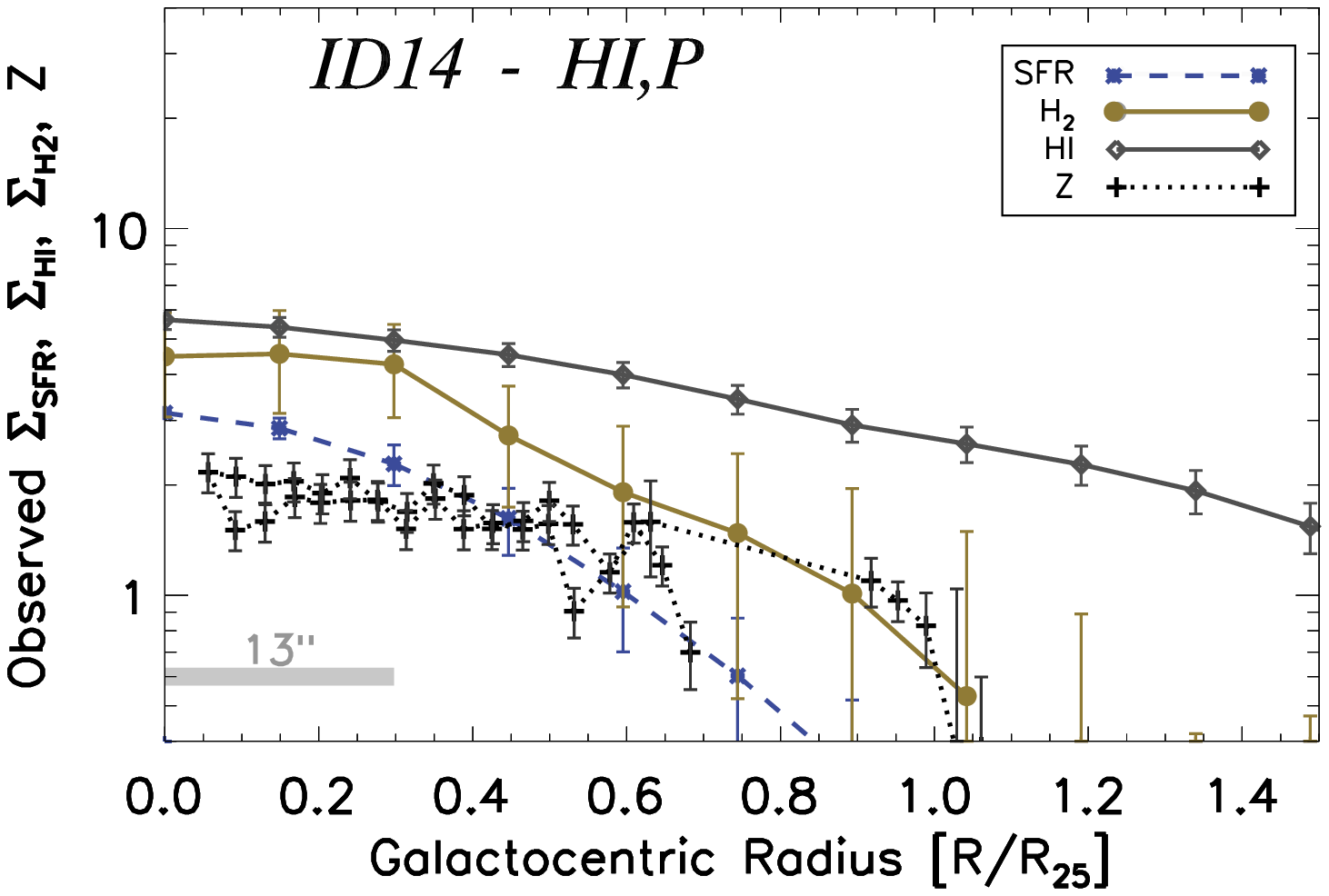}
 \includegraphics[clip,trim=19.5mm 17mm 1.3mm 0.5mm,width=5.2cm]{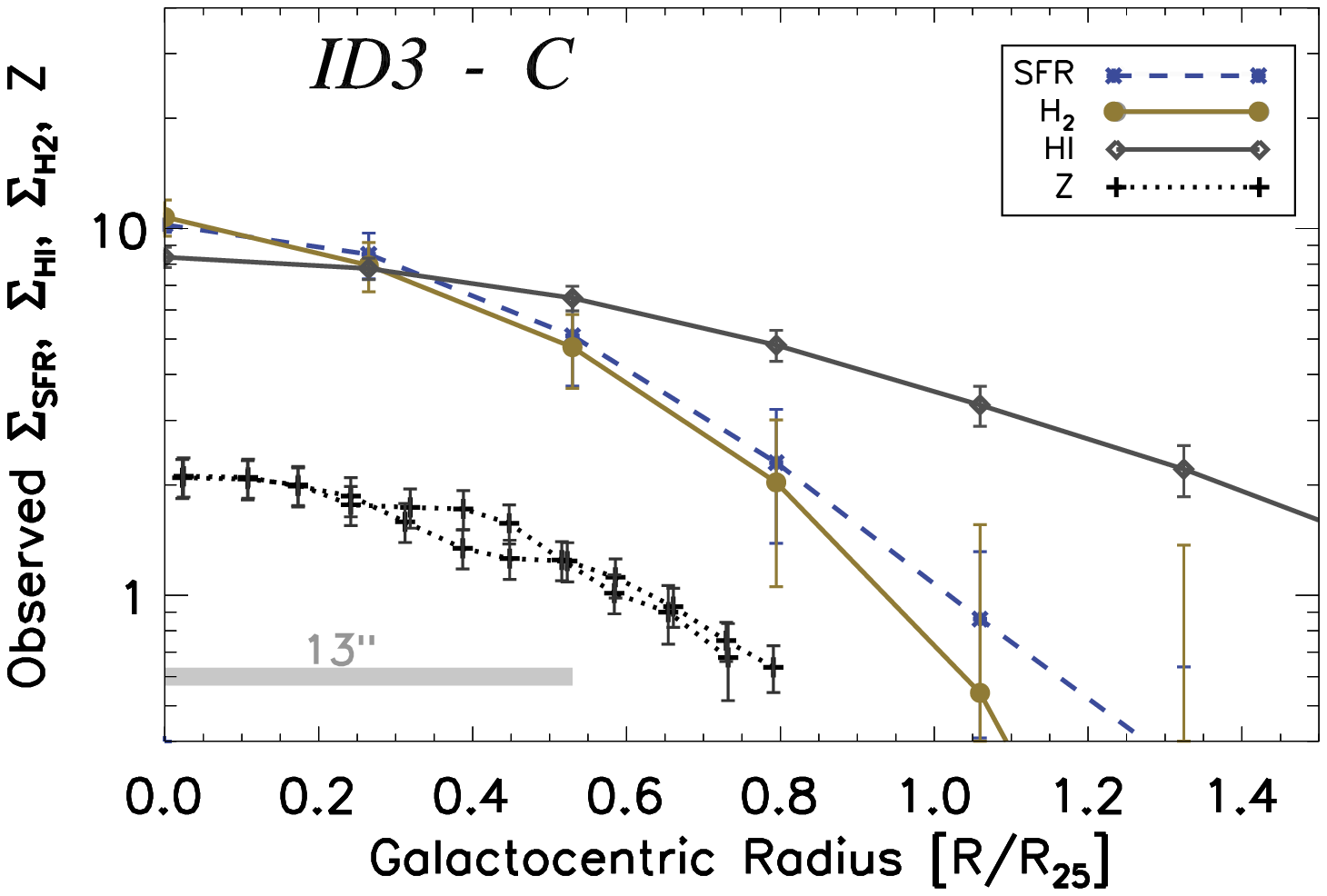}
 \includegraphics[clip,trim=0 0 1.3mm 0.5mm,width=6cm]{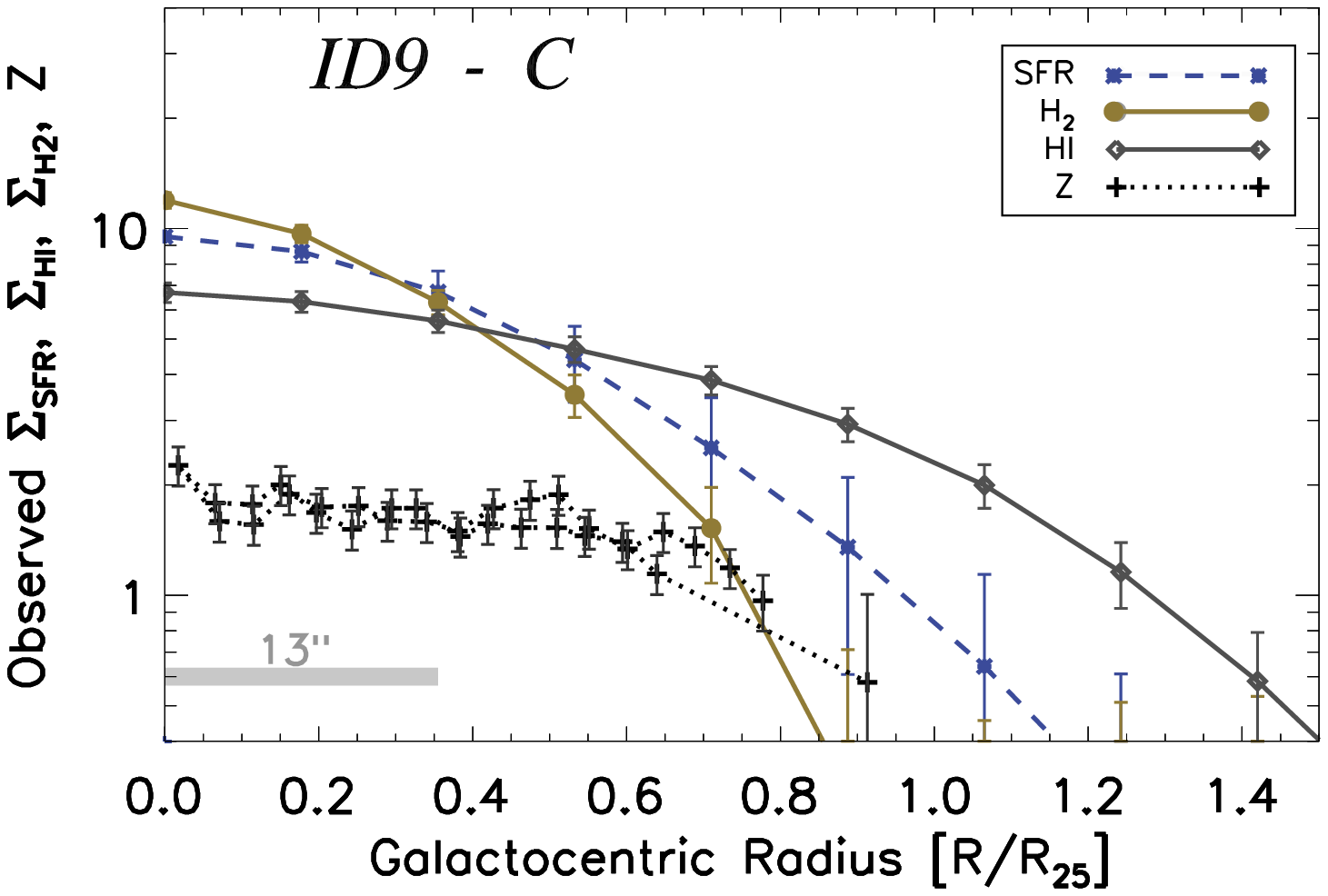}
 \includegraphics[clip,trim=19.5mm 0 1.3mm 0.5mm,width=5.2cm]{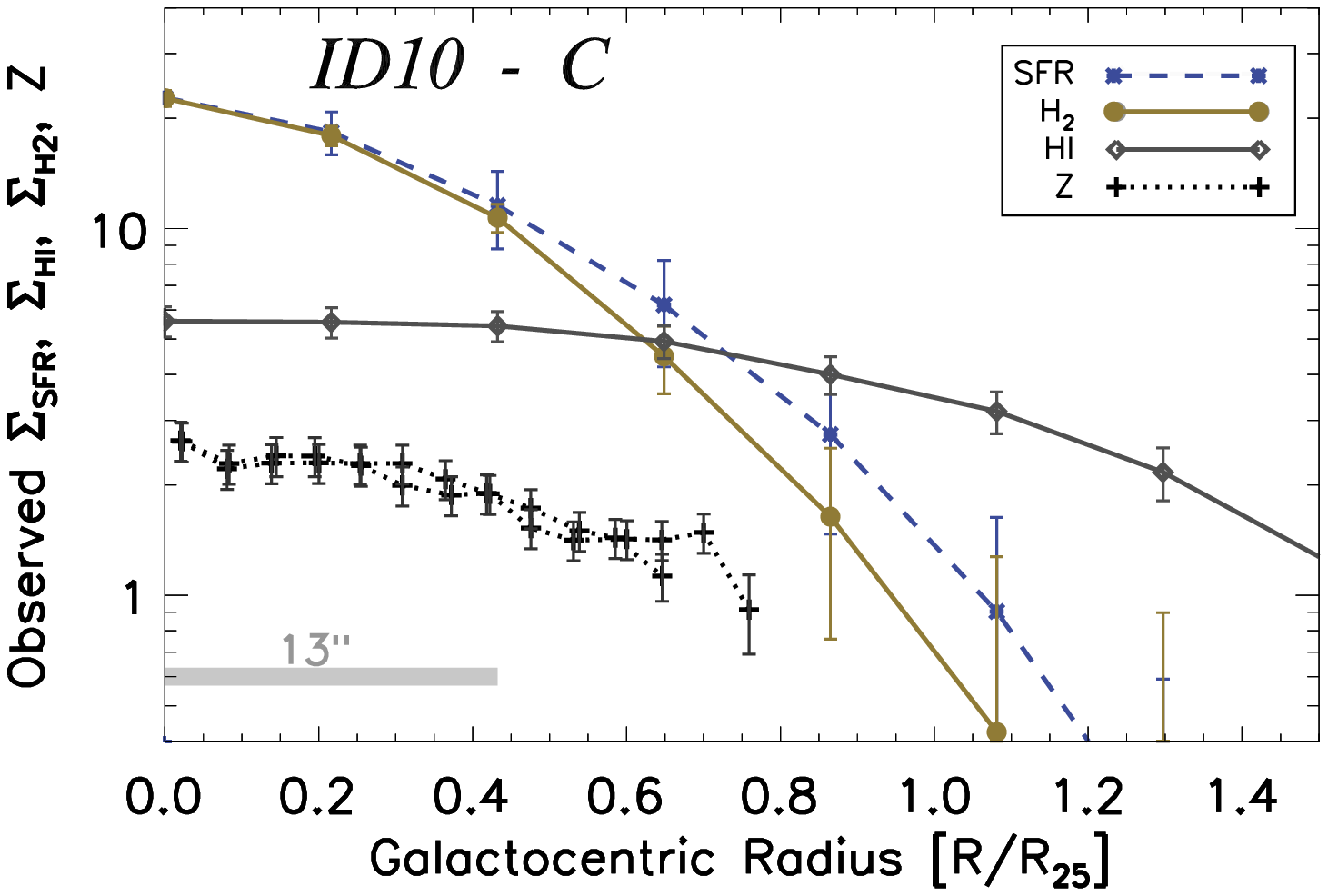}
 \includegraphics[clip,trim=19.5mm 0 1.3mm 0.5mm,width=5.2cm]{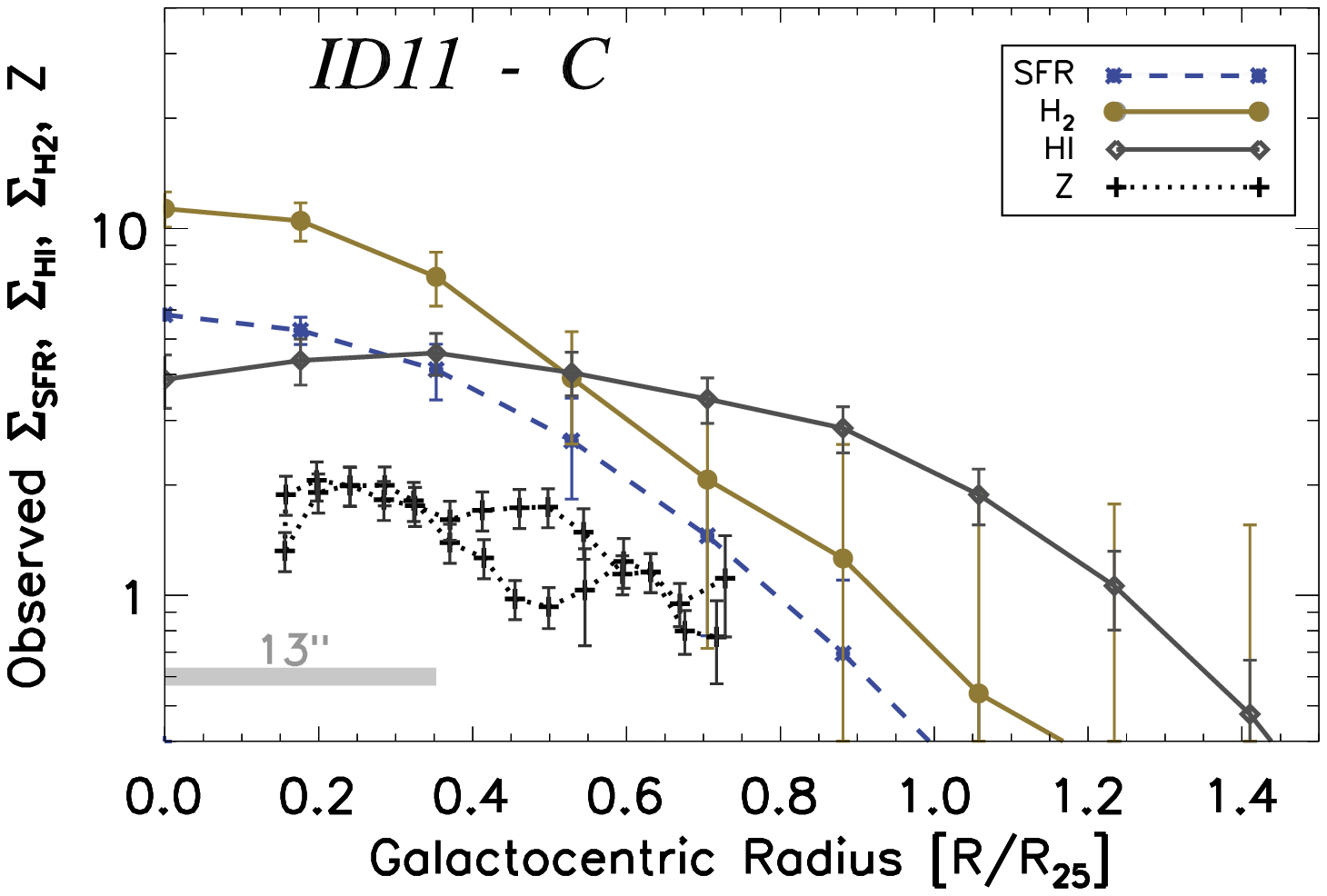} 
\caption{ 
 Surface density profiles, corrected for inclination, of
 \sightwo, \sighi (in units of \msun pc$^{-2}$), and
 \sigsfr (in units of $10^{-3}$\,\msun yr$^{-1}$\,kpc$^{-2}$) as
 a function of radius (normalized to $R_{\rm 25}$) for each galaxy
 mapped with IRAM/HERA.
 Horizontal grey bars indicate the angular resolution.
 The dotted lines are metallicity profiles from \citealt{carton-2015},
 with two values for each radius because the slit of the telescope
 passes through the center, and normalized to the solar metallicity
 (\zsun$=12+\log(O/H)=8.82$).
 Profiles are shown in the following order:
 \hi-rich discs of good quality (HI),
 \hi-rich discs of poor quality (HI,P),
 and control discs (C).
 Average behaviours are shown in Fig.~\ref{fig:onefig}.
 }
\label{fig:allprofiles}
\end{figure*}
\begin{figure}
\centering
 \includegraphics[clip,width=7cm]{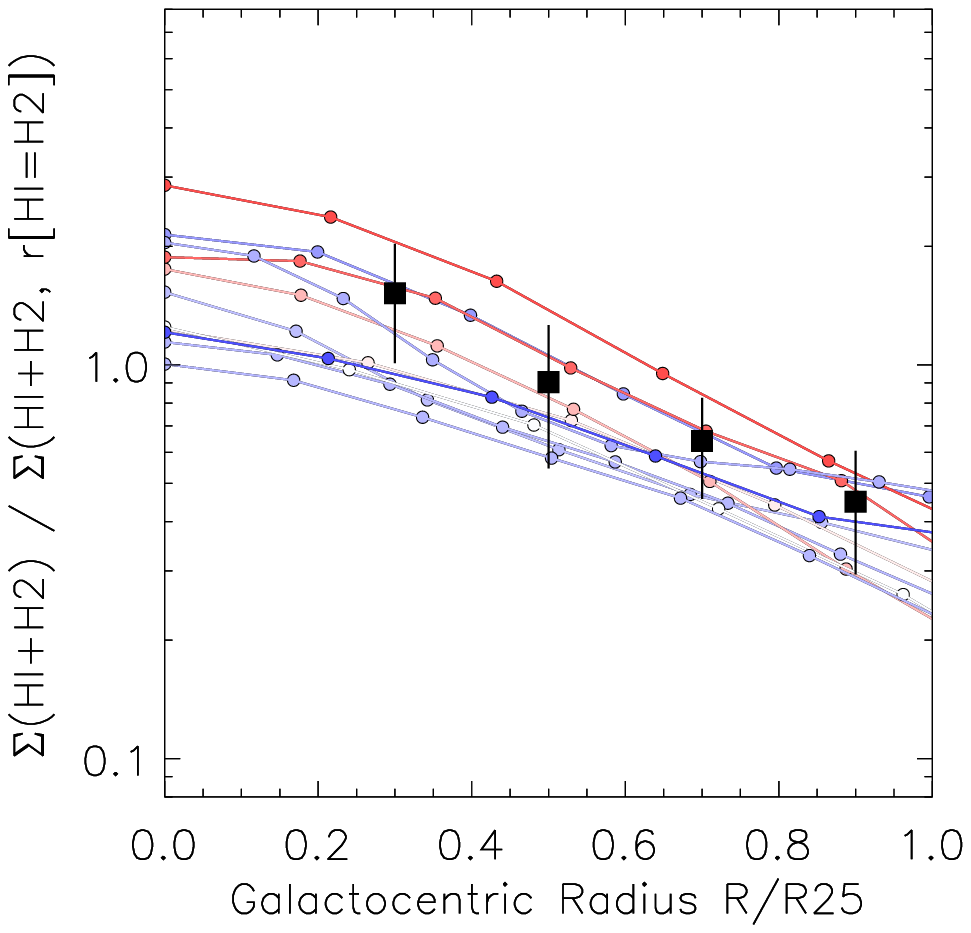}
 \vspace{-10pt}
\caption{ 
Total gas profiles (\hi+ \htwo) of the 11 \bd galaxies with reliable profiles,
normalized to the surface density at the transition radius (where \sighi= \sightwo).
Same colour-coding as in Fig.~\ref{fig:coldg1}. 
The mean values found by \protect\cite{bigiel-2012} in nearby, late-type discs 
are overplotted as black squares.
}
\label{fig:uniprof}
\end{figure}

\subsection{Radial behaviours}
\label{sect:radial}
Our goal in this section is to analyze radial behaviours,
specifically of the gas and SFR distributions, as a function
of distance to the disc centre. 

\begin{figure*}
\centering
 \includegraphics[clip,trim= 6mm 16mm 0 1.8mm,width=8.5cm,height=3.7cm]{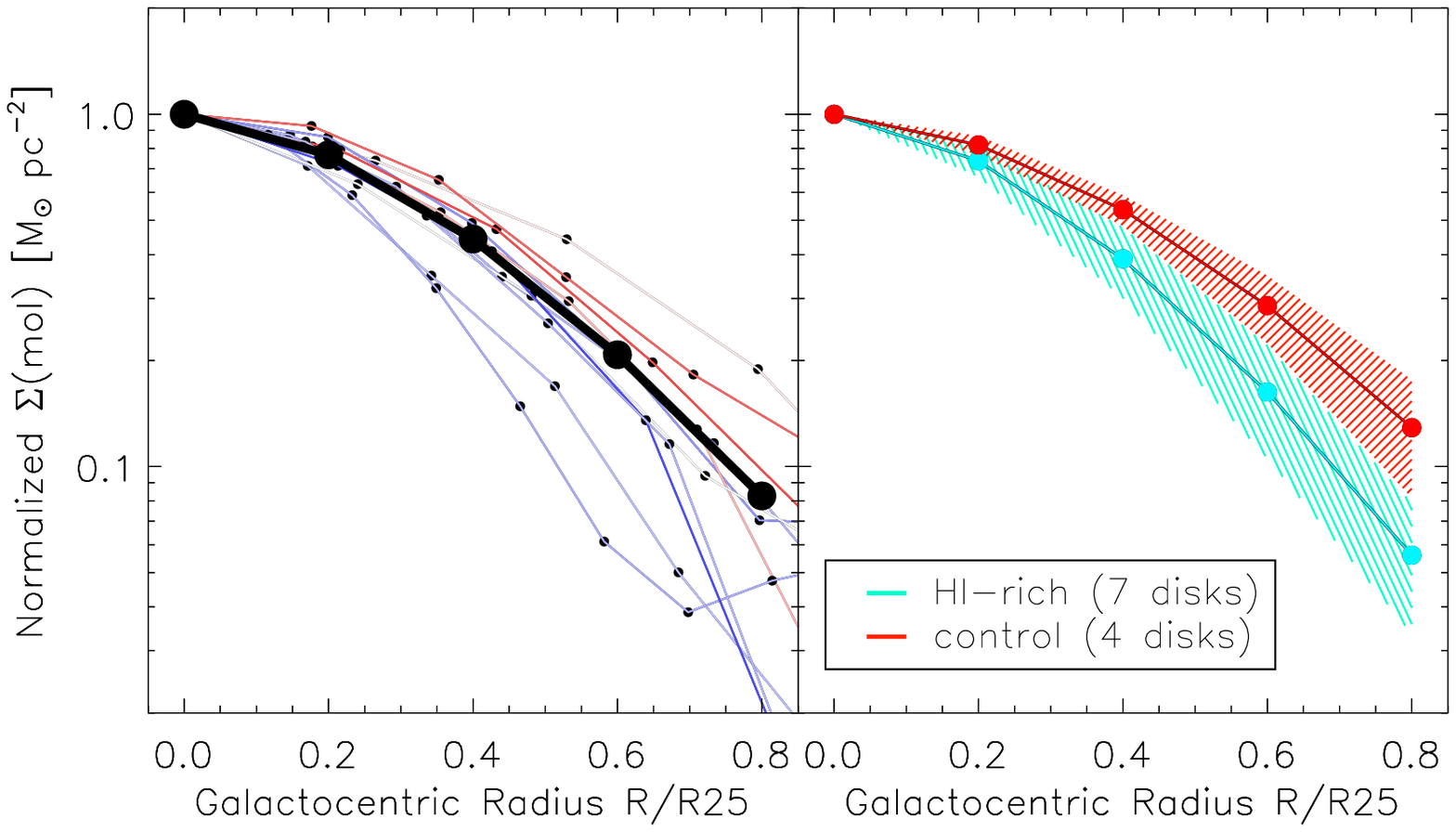}\hspace{5mm}
 \includegraphics[clip,trim= 6mm 16mm 0 1.8mm,width=8.5cm,height=3.7cm]{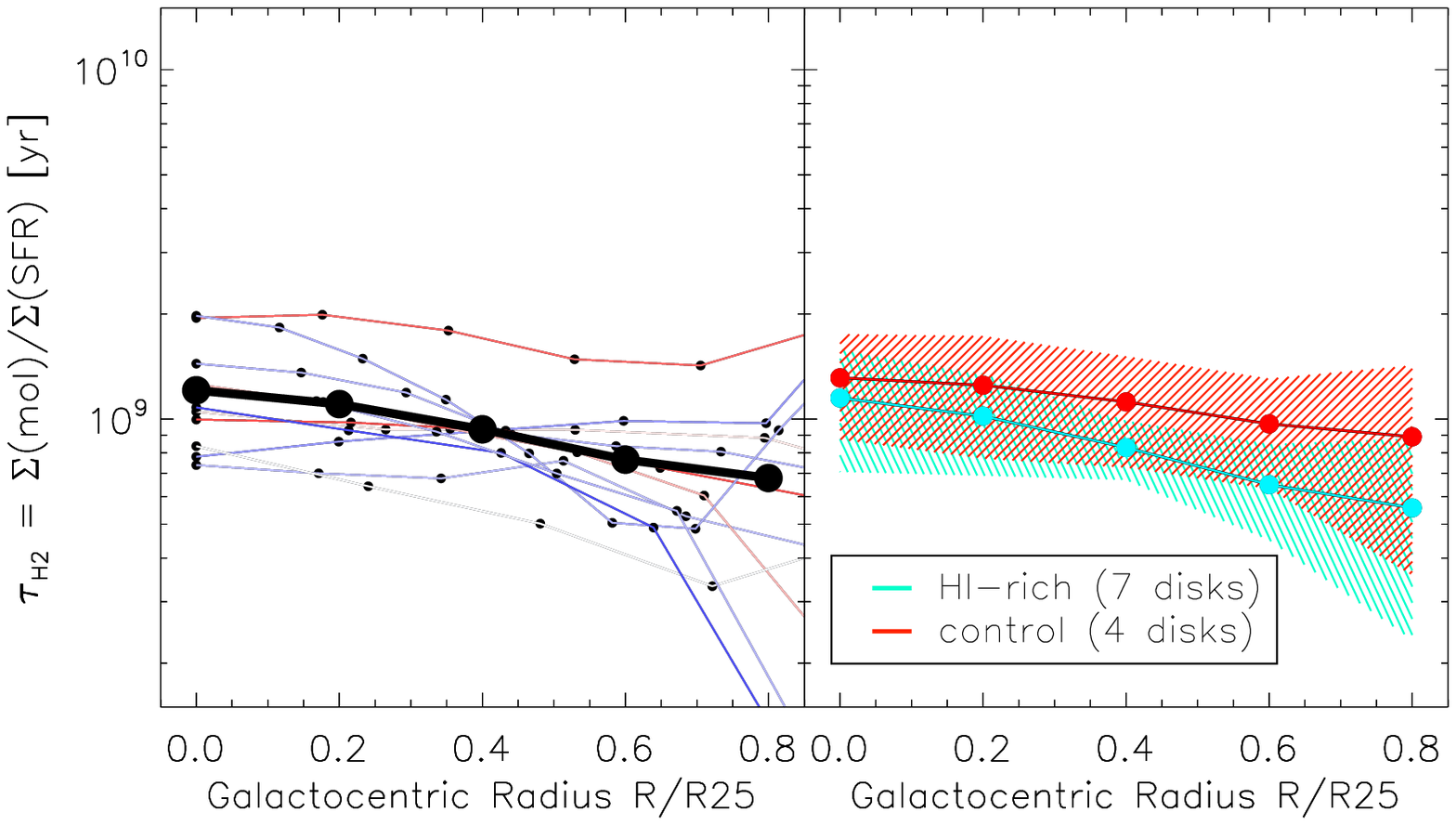}
 \includegraphics[clip,trim= 6mm 16mm 0 1.8mm,width=8.5cm,height=3.7cm]{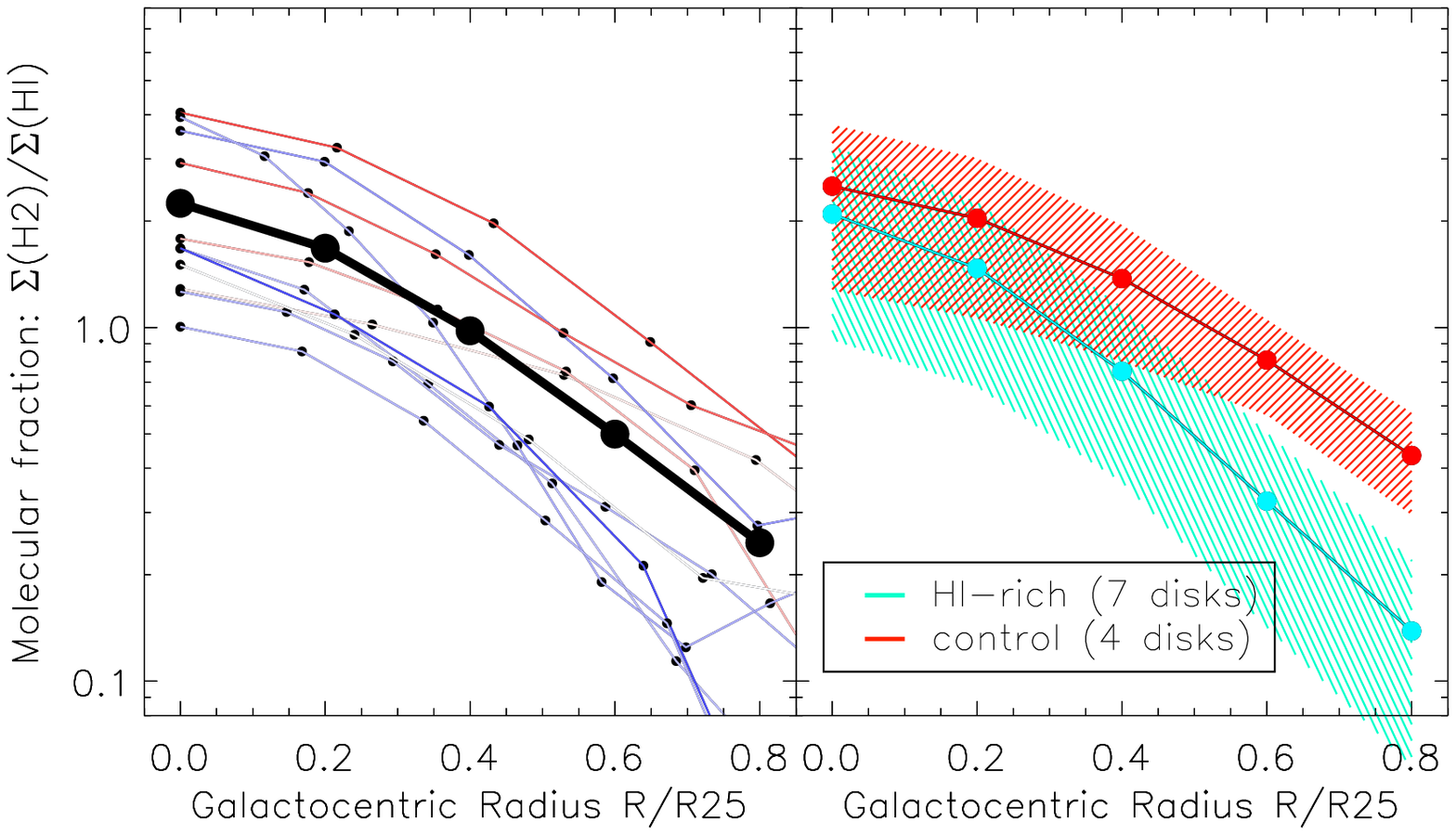}\hspace{5mm}
 \includegraphics[clip,trim= 6mm 16mm 0 1.8mm,width=8.5cm,height=3.7cm]{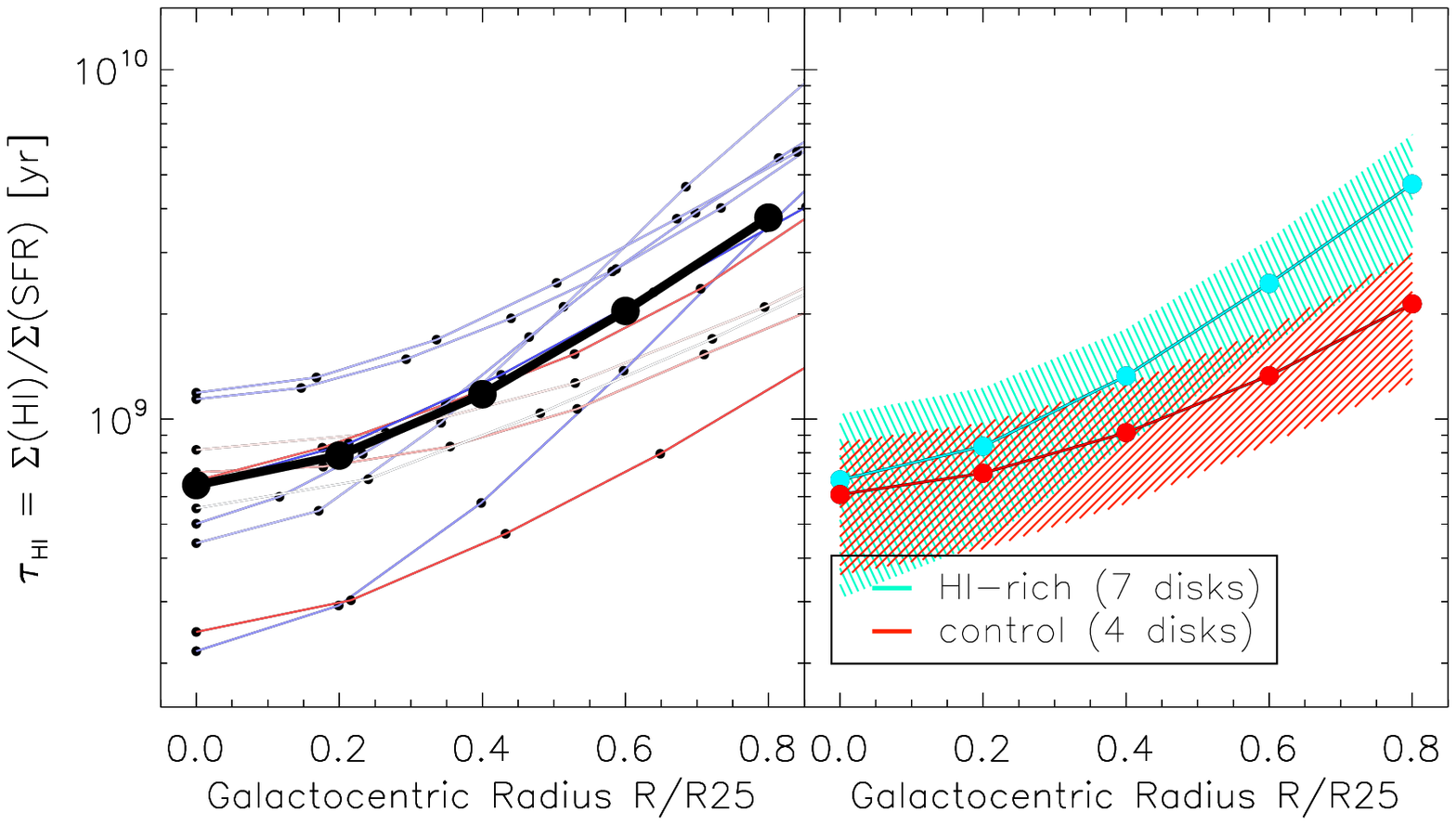}
 \includegraphics[clip,trim= 6mm 0 0 1.8mm,width=8.5cm,height=4.5cm]{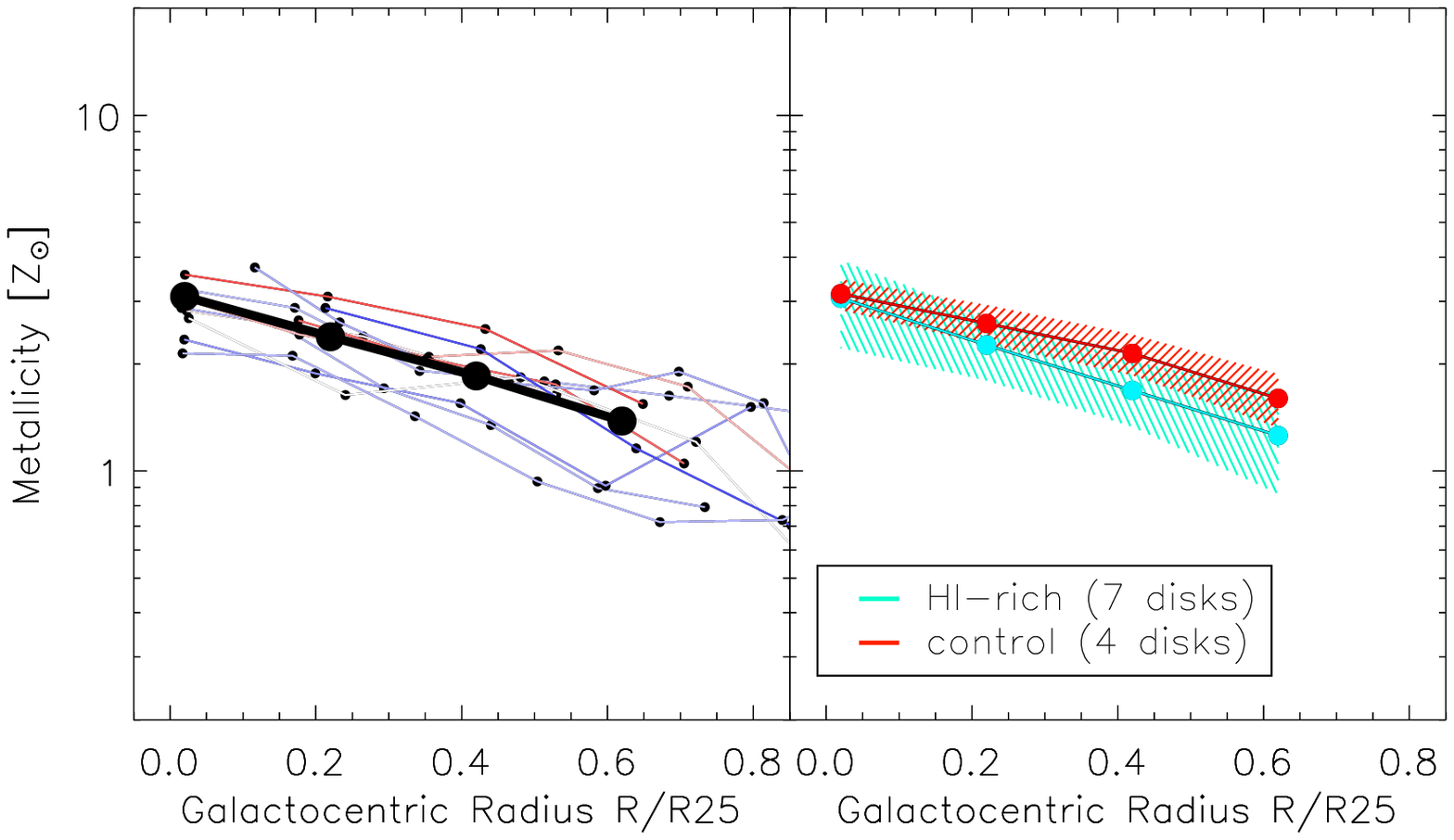}\hspace{5mm}
 \includegraphics[clip,trim= 6mm 0 0 1.8mm,width=8.5cm,height=4.5cm]{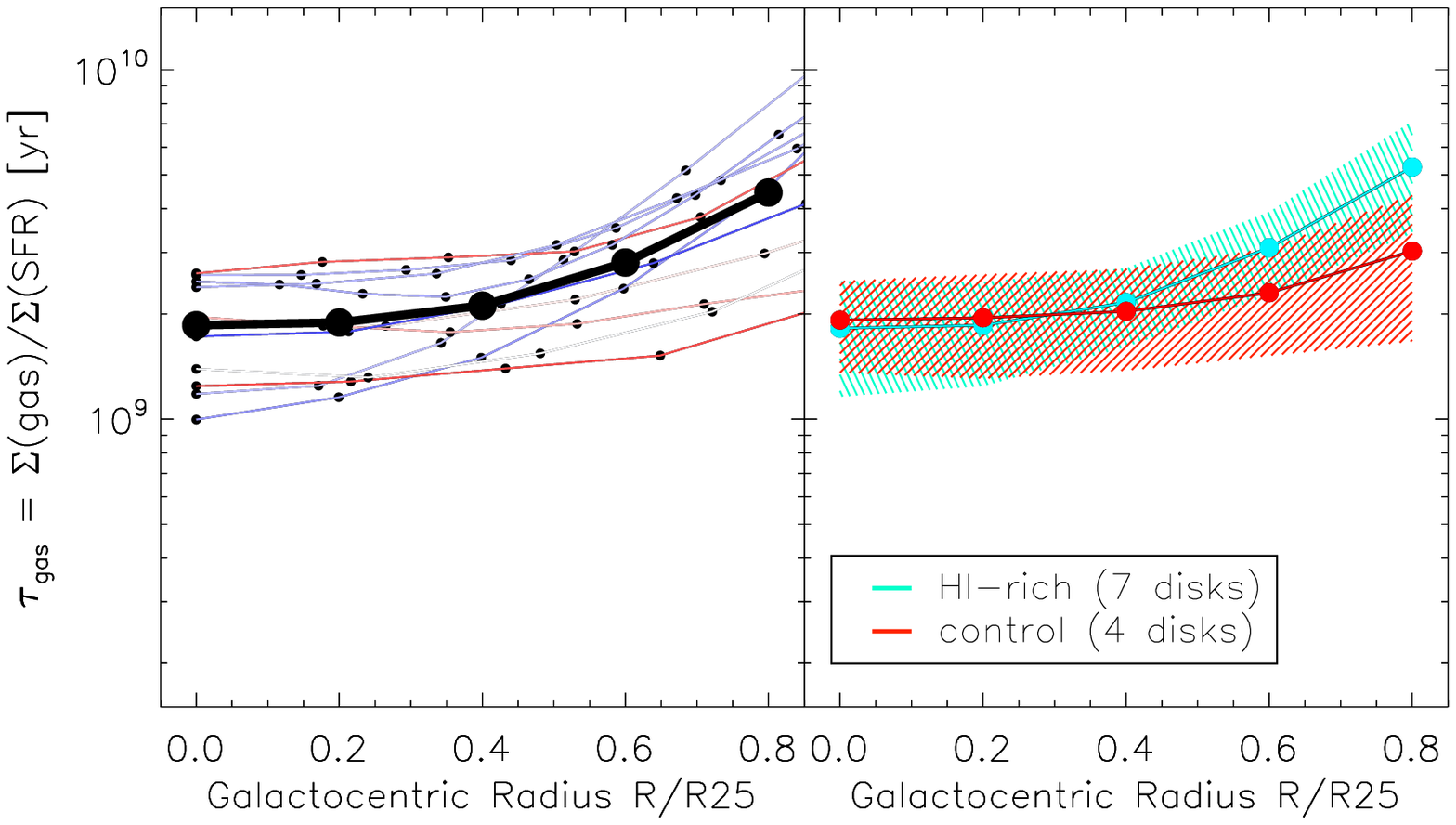}
\caption{ 
 From top to bottom: ({\it left} column)
 normalized \sightwo, \sightwo/\sighi, and metallicity as a function of radius;
 ({\it right} column) molecular, atomic, and total gas star-formation
 depletion times as a function of radius. 
 For each quantity, panels on the left show profiles
 of all the mapped galaxies colour-coded upon 
 the \hi-rich/control classification. 
 The black profile represents the mean profile of the sample. 
 Panels on the right show mean profiles and their dispersion
 for galaxies classified as \hi-rich (cyan)/control (red).
}
\label{fig:onefig}
\end{figure*}

Figure~\ref{fig:allprofiles} shows radial profiles for each disc.
\sightwo and \sigsfr generally behave similarly 
with radius and decline more rapidly than \sighi, as observed 
in the nearby spiral galaxies of HERACLES \citep{bigiel-2008}. 
\sightwo is particularly elevated compared to \sigsfr in the centre
of ID11 and ID15.
Focusing on the centres of the galaxies, only ID2 seems \hi-dominated,
ID5, ID10, ID11, and ID15 are clearly \htwo-dominated, while
the other discs have similar values (within a factor of 2) of
\sightwo and \sighi.
The transition between the \htwo-dominated and \hi-dominated 
regimes always occurs at radii $R<0.7\,R_{25}$, where 
\siggas$>10\,\rm{M_{\odot}\,pc^{-2}}$.
Those are roughly compatible with transition radii of nearby, late-type
discs ($\sim$0.5$\times$$R_{25}$; \citealt{bigiel-2012}). The transition
radius in the \bd galaxies is smallest for the very \hi-rich galaxies
and largest for the most active galaxy (ID10).
Metallicity profiles show fluctuations on small scales. However,
the limited spatial resolution of the gas data precludes us from
measuring fluctuations on those scales.

Following \cite{bigiel-2012}, we investigate whether the \bd galaxies
have total gas profiles similar to the average {\em universal} profile
of well-resolved late-type nearby disk galaxies. Exponential fits to each
profile of Fig.~\ref{fig:uniprof} in the radius range $0.2-1.0 \times R_{25}$
give a median scale-length of $0.58$ with dispersion $0.13$.
This compares to the scale-length of $0.61$ in the nearby discs studied
by \cite{bigiel-2012}, although the scatter in the slopes of our profiles
is quite large. \cite{wang-2014} normalized the x-axis to R1
instead of $R_{25}$, where R1 is the radius where the \hi column
density reaches 1\,\msun\,pc$^{-2}$.
Doing so does not seem to reduce the scatter in our data.
Several \hi-rich galaxies have shallower total gas profiles
and lie below the median relation of \cite{bigiel-2012} because
of their flatter \hi profiles and smaller transition radii.
A universality in the profiles of the outer \hi disc has also been
highlighted by \cite{wang-2014,wang-2016} to explain the
\hi mass-size relation of galaxies.
However, this profile universality may be weaker when considering
the total gas for the \bd galaxies as they show a wider range
of total gas profiles at $R < R_{25}$.

Figure~\ref{fig:onefig} shows the molecular surface density, \sightwo,
the molecular fraction, \sightwo/\sighi, the metallicity, and the depletion 
times, \sightwo/\sigsfr, \sighi/\sigsfr, and \siggas/\sigsfr, 
as a function of galactocentric radius. 
At a given radius, the dynamic range of those quantities is about 
a factor of $2-3$ for the normalized molecular surface density and
metallicity, and a factor of $5$ for the depletion times and molecular
fraction. 
Inspecting averages over the sample (thick black lines), the metallicity
decreases by a factor of $\sim$3 and the molecular fraction by
a factor of $\sim$8 out to $R\simeq 0.7\times R_{25}$. The latter
trend is less dramatic than the decrease of more than an order of
magnitude observed in nearby spirals \citep[e.g.,][]{schruba-2011}
because we probe much larger spatial scales (13\arcs$\sim$7.5\,kpc
for the \bd). 
Regarding depletion times, $\tau_{\rm H2}$ decreases by a factor
of $<$2, i.e. \sightwo drops faster than \sigsfr, though we caution that
a systematic change of $X_{\rm CO}$ with radius/metallicity can at least
partly cancel this trend. 
On global scales, \cite{boselli-2014b} and \cite{saintonge-2011b}
find a weak correlation between $\tau_{\rm H2}$ and \mustar.
Re-analysis of the COLD GASS data by \cite{huang-2014} also
showed the lack of significant trend between $\tau_{\rm H2}$ and \mustar.
The weak correlation is mostly driven by lower depletion times for lower
\mustar values and it disappears when \htwo is calculated from a
metallicity-dependent $X_{\rm CO}$ factor \citep{boselli-2014b}.
The atomic depletion time increases by a factor of 5,
and the total gas depletion time increases by a factor of $\sim$2.
The trend seen in \hi is compatible with observations of nearby spirals
\citep{bigiel-2010b}. The trend seen in \htwo is somewhat different
from the molecular gas-rich, late-type discs studied in \cite{leroy-2013}
for which depletion times are constant with radius and slightly lower in
their centres. The discrepancy with our galaxies is probably due to
the large scatter in our data and larger spatial scales probed.

Moreover, several trends appear when binning the profiles by
their type (\hi-rich/control).
For the following discussion, we caution that we base our analysis
on only 11 profiles (the good-quality HERA data). We identify
behaviours as a trend if the average quantity in one category
is outside of the range of radial distributions (i.e. the scatter,
indicated as the hashed regions) in the other category.
First, \sightwo seems to decrease faster with radius for the \hi-rich
galaxies. The trend is weaker but still present when correcting
CO profiles for metallicity variations with radius. The \hi profiles
are flatter, resulting in comparable normalized total gas surface
density profiles for the \hi-rich and control galaxies.
As a consequence, the molecular fraction also decreases faster
with radius for the \hi-rich galaxies. This is interesting because
the \hi-rich galaxies display steeper metallicity gradients across
their stellar discs than the control galaxies overall \citep{carton-2015}.
We further quantify this in Section~\ref{sect:gradients}.

The molecular depletion time is marginally higher in the control
galaxies than in the \hi-rich galaxies by a factor of $\sim$1.4,
as noticed in Fig.~\ref{fig:hrs} on global scales.
The atomic depletion times increase faster for the \hi-rich galaxies,
again due to the \hi profiles being relatively flatter in the outer parts,
and the total gas depletion times are globally constant with radius,
with an increase at larger radii for the \hi-rich galaxies. 

Although not shown in Fig.~\ref{fig:onefig}, we inspected results
by binning our profiles by stellar mass or by presence of a bar in the galaxy.
We found that several of the plots binned by stellar mass resemble
those binned by colour because 3 of the 4 control discs are in
the high stellar mass range (ID9, ID10, ID11). 
The presence of a bar, identified in only three of our galaxies 
mapped in CO, does not seem to influence the radial behaviour 
of the quantities discussed (at the coarse resolution of our data).

Overall, the response from the inner half of the star-forming disc
(within $0.5 \times R_{25}$) of the \hi-rich galaxies to accretion
seems nonexistent or very slow. Only the molecular fraction and
the molecular depletion time are marginally lower, while the atomic
depletion time is marginally higher.
In the outer parts of the disc, we find no evidence for a significantly
elevated SFR or a detectable CO reservoir. Only the metallicity,
decreasing below half solar in some \hi-rich galaxies, possibly
translates in difficulties to detect CO beyond $\sim R_{25}$.

\subsection{Extents and gradients comparison}
\label{sect:gradients}
\begin{figure*}
\centering
 \includegraphics[clip,width=4.35cm]{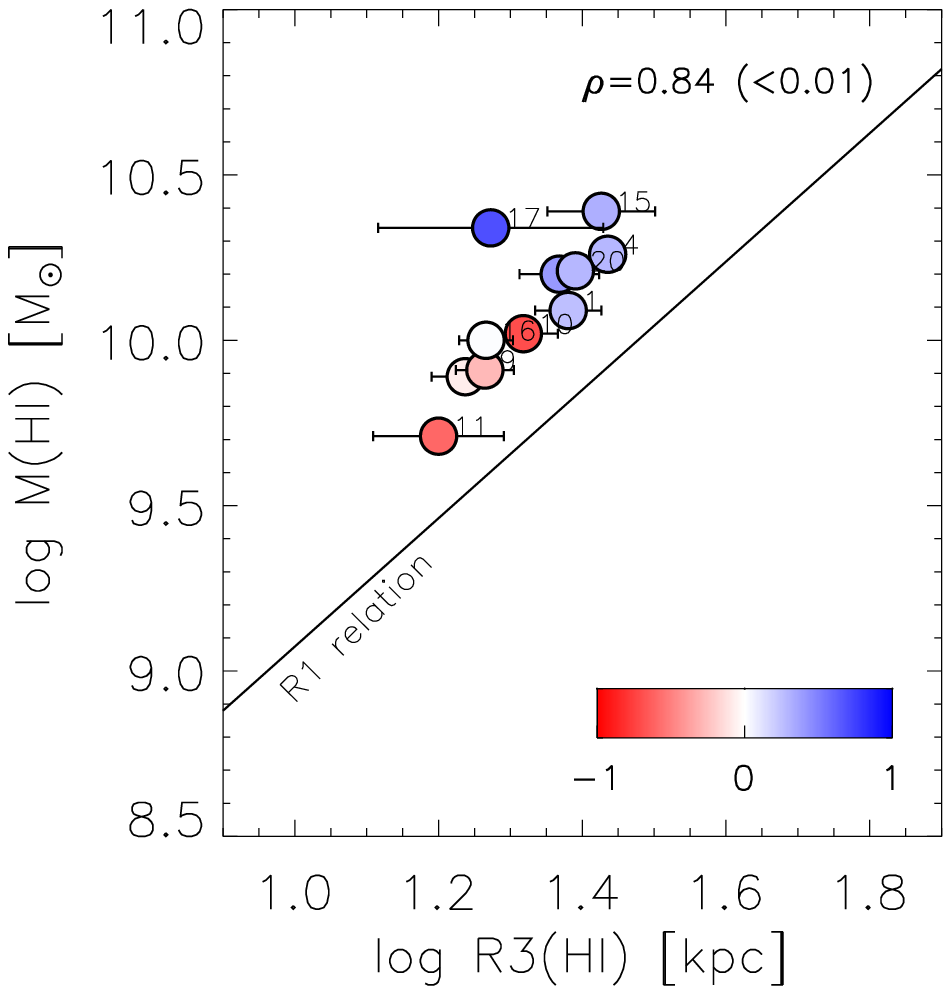}
 \includegraphics[clip,width=4.35cm]{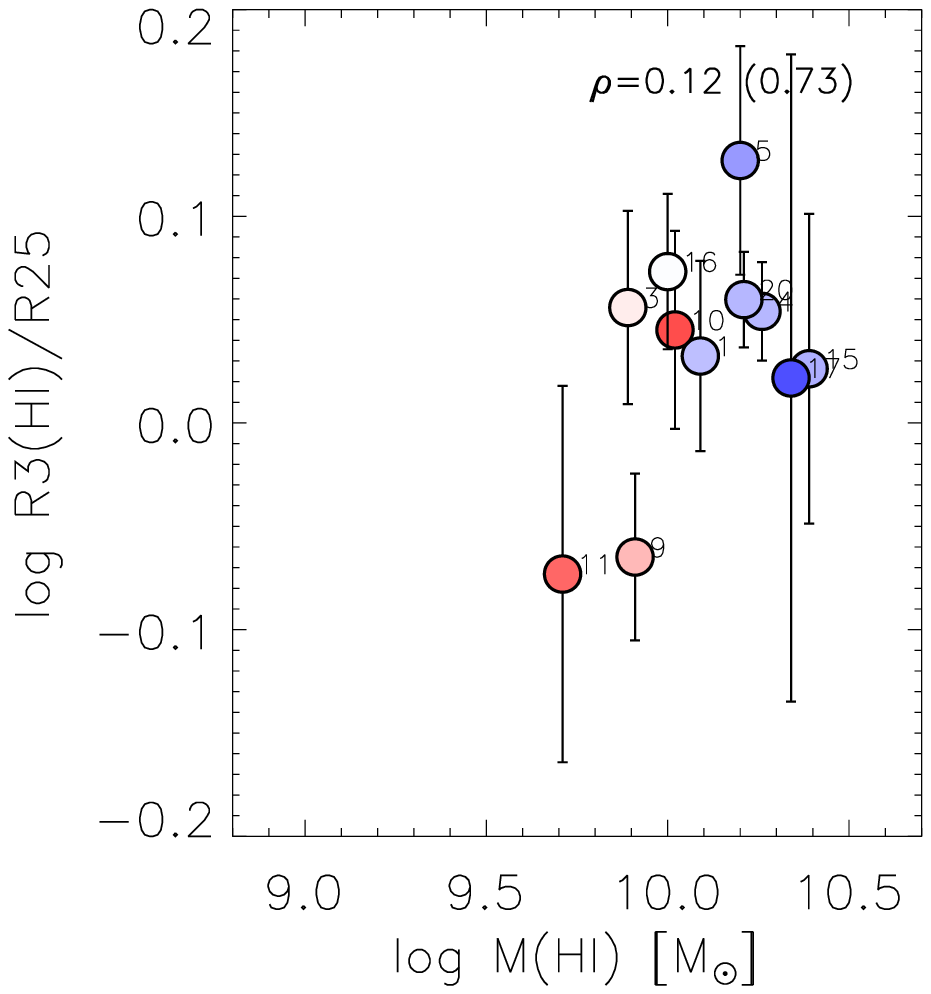}
 \includegraphics[clip,width=4.35cm]{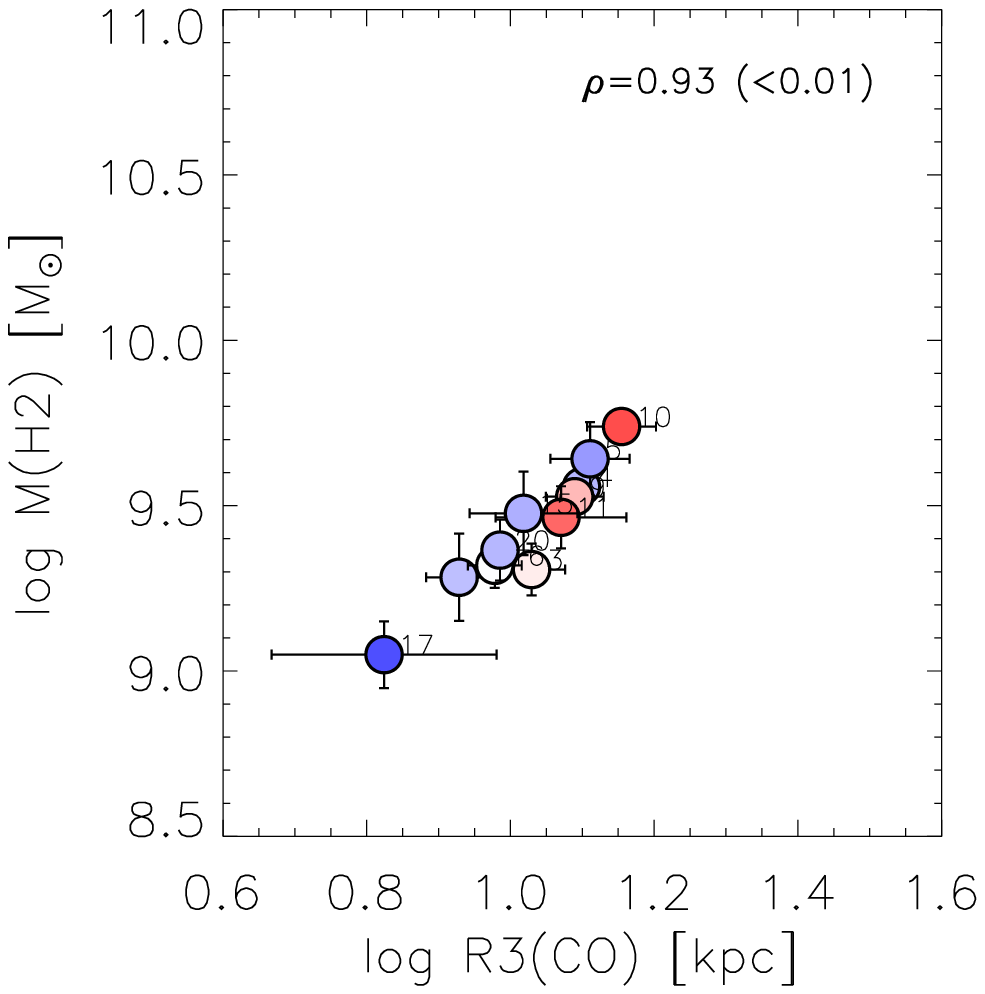} \vspace{2mm}
 \includegraphics[clip,width=4.35cm]{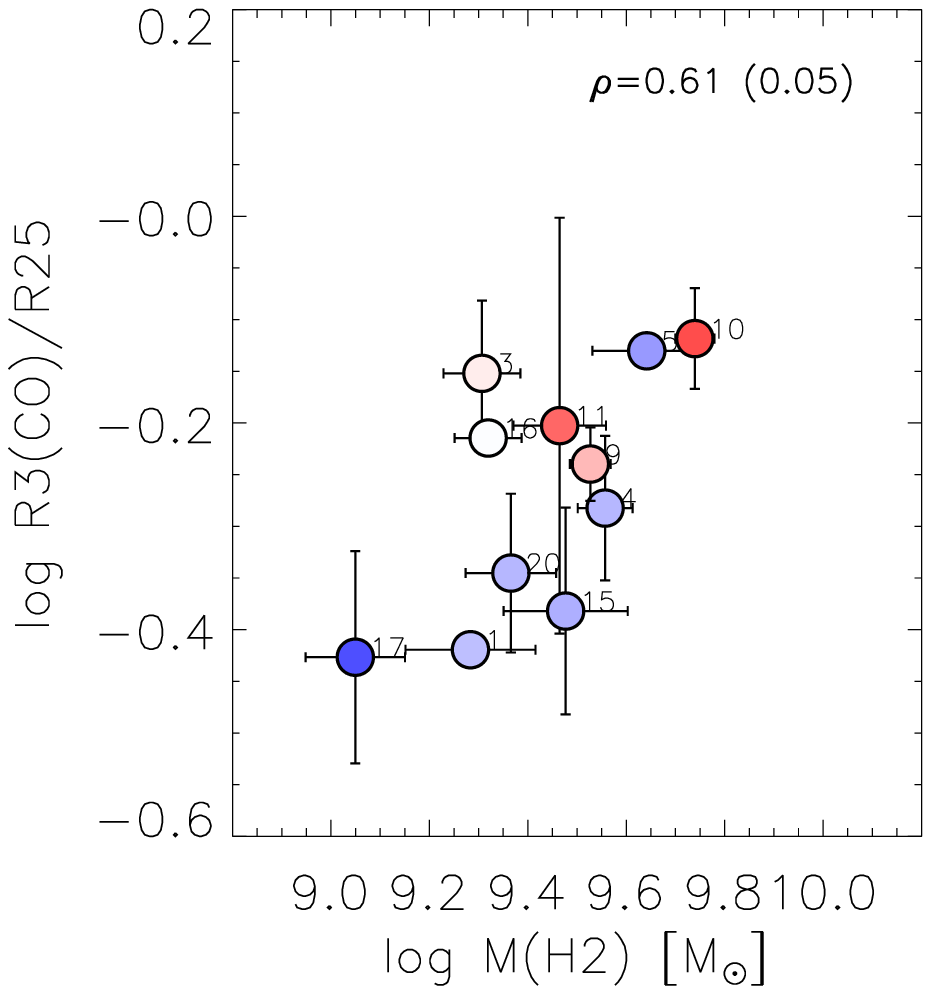}
 \includegraphics[clip,width=4.35cm]{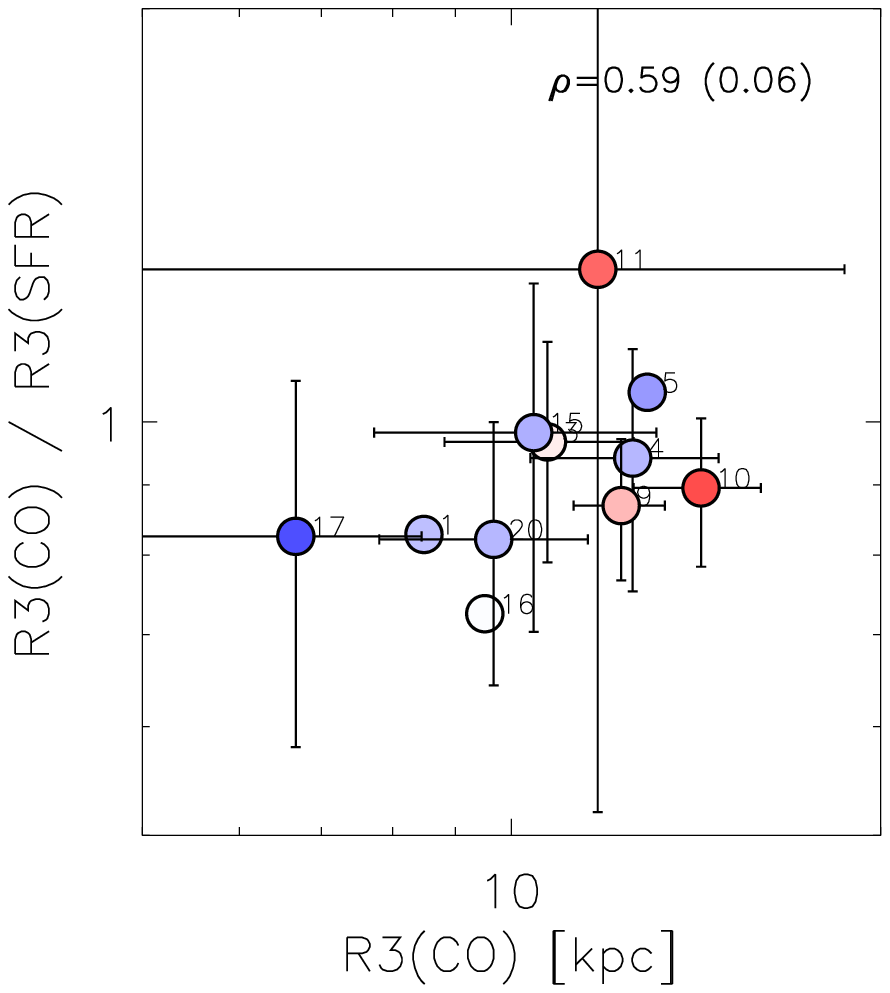}
 \includegraphics[clip,width=4.35cm]{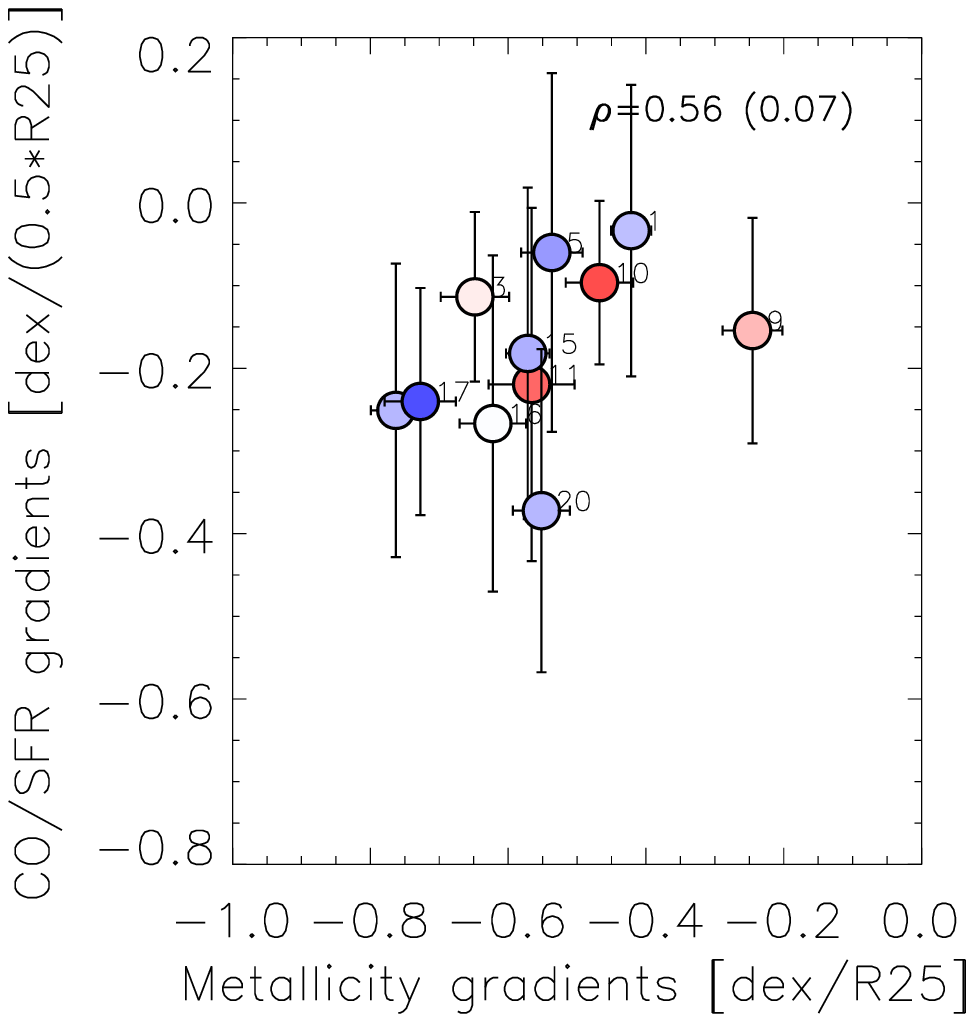}
 \includegraphics[clip,width=4.35cm]{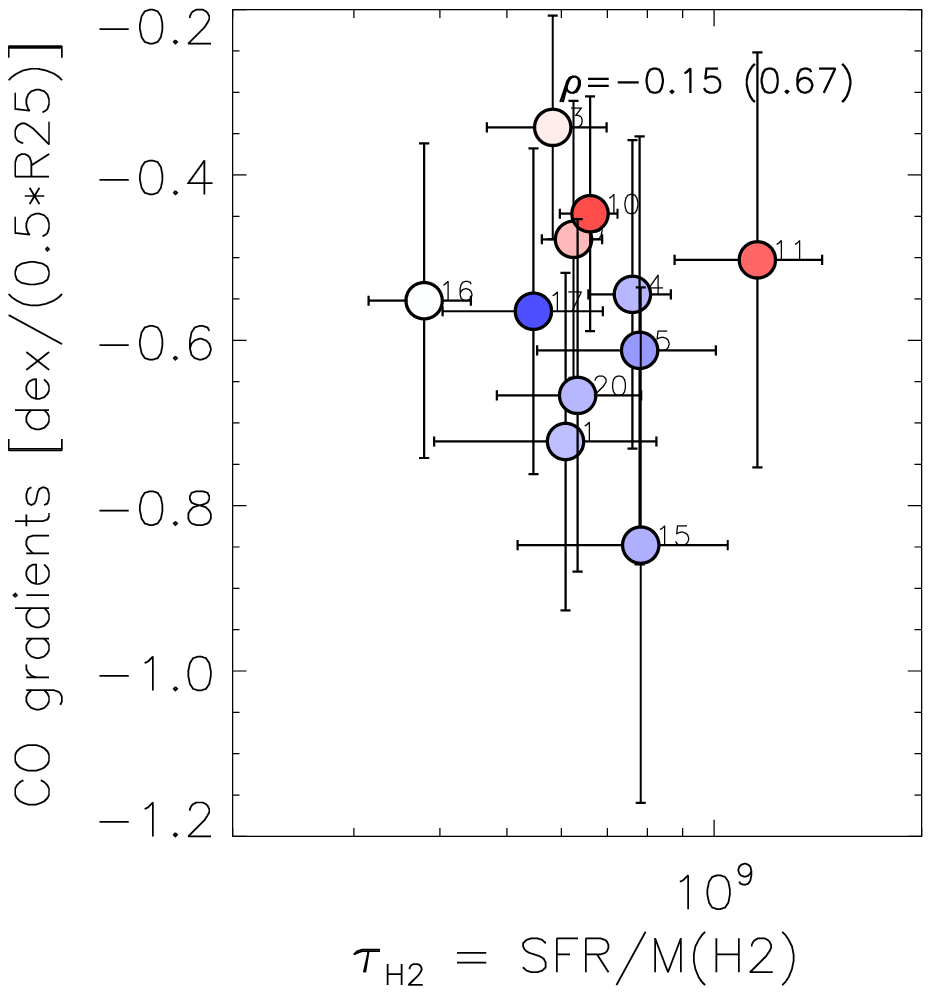}
\caption{ 
Comparison of surface density extents and gradients.
R3 is the radius at which the surface density reaches 3\,\msun pc$^{-2}$ 
for CO and \hi, and $3\times10^{-3}$\,\msun yr$^{-1}$\,kpc$^{-2}$ for SFR. 
The line in the top, left panel is the \mhi-R1 relation found by
\protect\cite{wang-2013}, i.e. measuring sizes where the \hi surface
density reaches 1\,\msun pc$^{-2}$ instead of 3\,\msun pc$^{-2}$ (hence R3$<$R1).
The mean and standard deviation of each quantity are 
overplotted as crosses for the \hi-rich and control samples. 
We also indicate Pearson correlation coefficients. 
Same colour-coding as in Fig.~\ref{fig:coldg1}. 
}
\label{fig:gradients}
\end{figure*}

We measure disc extents as the radius (distance from 
the disc centre) at which the CO surface density profile
reaches 3\,\msun\,pc$^{-2}$, denoted R3(CO). The threshold
is chosen based on data quality; below that value, the CO 
profiles are not reliable. 
For the \hi and SFR profiles, we also choose a threshold of 
3\,\msun\,pc$^{-2}$ and $3\times10^{-3}$\,\msun\,yr$^{-1}$\,kpc$^{-2}$, 
and we denote the corresponding radii R3(\hi) and R3(SFR), 
respectively. 
Uncertainties on those derived quantities are estimated 
by iterating the process $100$ times, varying
data points of the profiles within their errors. 
This gives final uncertainties of $\sim$12\%. 
CO gradients are calculated as the difference in log of the 
surface density at $R=0.5\times R_{25}$ and $R=0$.

Although we work with different threshold values and a smaller 
sample, the two upper, left panels in Fig.~\ref{fig:gradients} 
recover the results from \cite{wang-2013}: \hi-rich galaxies have more 
massive and more extended \hi discs than the control sample.
Our data points lie above the \mhi-R1 relation because R3$<$R1.
In the two upper, right panels, we see that the most massive \htwo 
discs are also the biggest, with a tighter correlation than seen 
in \hi, and, when normalized to $R_{25}$, the CO discs appear 
slightly more extended in the control sample, by $\sim$0.1\,dex. 

The lower, left panel of Fig.~\ref{fig:gradients} shows that 
the CO and SFR sizes are roughly compatible. CO is slightly 
more extended than the SFR in the largest CO discs, i.e. in 
the control discs. 
The two lower, middle panels show the CO/SFR gradients and
CO gradients as a function of metallicity gradients and molecular
depletion time respectively. We find that, globally, the \hi-rich
galaxies have steeper CO gradients, steeper metallicity gradients,
and marginally lower molecular depletion times, as discussed
in Section~\ref{sect:radial}.
Note that even when correcting for metallicity effects (using a higher
$X_{\rm CO}$ factor), weak trends remain present.

\section{Discussion}
\label{sect:discussion}
In this section, we aim to discuss our results on the two
galaxy samples in the context of galaxy evolution. We find
that the \hi-rich galaxies differ from the control galaxies in
being bluer, more gas-rich with lower \htwo/\hi ratios, and
having slightly steeper CO and metallicity gradients and
marginally shorter molecular gas consumption times.

Given the small stellar mass range probed by the \bd galaxies,
variations in the sSFR values reflect variations about the
main sequence of galaxies (plots in Figure~\ref{fig:hrs} look
very similar when normalizing sSFR by the main sequence values,
hence we do not show them).
\cite{genzel-2015} study \mhtwo/\mstar and $\tau_{\rm H2}$
as a function of redshift and offset from the main sequence.
In particular, they find a decrease in $\tau_{\rm H2}$ and an increase
in \mhtwo/\mstar with increasing offset from the main sequence
(to high sSFR values), which is in agreement with the trends reported
for our sample (at $z\simeq0.03$) in Section~\ref{sect:coldg}.
This supports findings at higher redshifts that gas fractions are
higher in galaxies above the main sequence
\citep{magdis-2012,scoville-2016}.

Analysis of global parameters in Section~\ref{sect:coldg}
tend to indicate that the \hi-rich and control galaxies are mostly
offset in the \hi reservoir, and much less in \htwo mass or sSFR.
As found in previous work, \hi is not the main, direct fuel of
the star-formation activity in those galaxies.
The larger \hi reservoir in the \hi-rich galaxies might argue for fresh
accretion \citep{wang-2015} with no time yet to transport the gas
further in. This accreted atomic gas may at some point be
converted into \htwo and then into stars.
The time-scale of this process can be taken as the offset of
the \hi-rich and control galaxies in their $\tau_{\rm gas}$ values,
i.e. $\sim$5\,Gyr.
If the accretion is recent, those galaxies with large \hi envelopes
may not have had time to go back to equilibrium, and therefore
the estimated time-scale is a lower limit. 
Even though the \hi-rich galaxies are gas-rich, their normalised
gas surface densities are below average, especially in the inner
parts of the disc (Fig.~\ref{fig:uniprof}). This may confirm that
the gas has not progressed towards the inner parts yet and
hence the growth of the disc is slow despite evidence for accretion.
This picture is somewhat more secular than in starburst galaxies 
where elevated SFRs are due to efficient conversion of \hi 
to \htwo and stars \citep[e.g.,][]{jaskot-2015}.
The fact that the \hi-rich galaxies in our survey appear blue because
of metallicity rather than elevated SFR (Figs.~\ref{fig:coldg1} and
\ref{fig:hrs}) is consistent with this secular evolution interpretation.
The \hi-rich/control offset may be interpreted as an evolutionary 
sequence and could also partly explain the large dispersions in 
the correlations found for larger galaxy samples. 

Overall, with the \bd survey, we are looking at similar galaxies
(same stellar mass, redshift) but in apparent different moments
of their build-up (different gas budget), where the \hi-rich galaxies
have (recently) accreted atomic gas.
Concerning the duty cycle of gas accretion events, in order to easily
detect fluctuations in the accretion histories, such fluctuations have
to occur on timescales similar to or longer than the timescale to
consume the excess of atomic gas (i.e. several Gyrs).
However, following the evolution of one group to the other is not
trivial as galaxy evolution is a multi-parameter problem and
our sample is small. By analyzing the stellar content of thousands
of galaxies, \cite{dressler-2016} find that,
at a given stellar mass, the same galaxies could have undergone
very different star-formation histories. They conclude that
there is not one typical evolutionary track for a galaxy at a given
mass, in line with our finding here.
We can still speculate on how the \hi-rich galaxies will evolve
with time if they no longer accrete significant amounts of gas.
For them to reach similar \mhi/\mstar levels as the control galaxies,
they have to consume $\sim1\times10^{10}$\,\msun, which
corresponds to an increase in stellar mass of 25\%.
With time (and in a closed box) their metallicities will also increase
and their colours turn redder as the bulk of the stars become older. 
By that time, the main characteristics of the \hi-rich galaxies will
disappear and the \hi-rich galaxies will look like control galaxies.
On even longer time-scales, as those galaxies are isolated, their
SFR will probably decrease due to strangulation (supply of cold
gas halted) rather than removal of external gas \citep[e.g.,][]{peng-2015}.

Within the \hi-rich galaxies, we also find some scatter in the central
\htwo/\hi values. ID4, ID16, ID20 have lower \htwo/\hi ratios than
the other discs.
Interestingly, these three galaxies show a clumpy morphology in
optical images, and they also have an \hi excess in their centres
\citep{wang-2014}. 
Qualitatively, this is consistent with theoretical models by \cite{krumholz-2009}
which suggest a clumping factor of unity for those discs.
Clumpier \htwo distributions would dilute the surface density
of molecular gas relative to \sighi in our beam. Additionally,
in a clumpy medium, the radiation field pervades and is probably
an important parameter in setting the atomic/molecular balance.
Theoretical models also highlight the importance of metallicity
in influencing the \hi-to-\htwo transition \citep{krumholz-2009,sternberg-2014}.
ID4, ID16, ID20 do not show lower central metallicities than the
other \hi-rich galaxies hence we do not expect metallicity to be
responsible for their low \htwo/\hi ratios.
The variations of the \htwo/\hi ratio, lower in the clumpy galaxies
of the \bd survey and higher in the Sa/b galaxies, are also consistent
with studies showing an evolution of the molecular fraction with
morphological type \citep[e.g.,][]{young-1989,obreschkow-2009a}.

Concerning disc extents, \cite{obreschkow-2009c} report, using
the Millennium Simulation, a smooth evolution of the \hi mass
and disc radius with redshift, while molecular masses are higher
and molecular discs smaller at high redshift due to increased
pressure.
For the \bd galaxies, we find in Section~\ref{sect:gradients}
that the least extended CO and SFR discs relative to $R_{25}$
are those of the \hi-rich galaxies ID1, ID15, and ID17. These
three galaxies are grouped in the category of discs with \hi
excess between $R_{25}$ and R1(\hi) by \cite{wang-2014}.
Perhaps this surrounding \hi gas will not be transported further 
in the disc and will participate in the growth of the CO disc from
the outskirts.

\section{Summary}
\label{sect:summary}
To shed light on the role of gas accretion in the evolution and
growth of galaxies, the \bd survey has investigated optical and
radio properties of 50 nearby disc galaxies. Among those, 25
have enhanced \hi content and bluer discs than average for
their stellar mass \citep{wang-2013,wang-2014,carton-2015}.
The objective of this paper is to establish whether the \hi-rich
galaxies also have different molecular gas and star-formation
properties than the normal, control galaxies.
We have observed the molecular gas tracer CO in 26 galaxies
of the \bd survey and resolved emission in 15 of them, 11 of which
are \hi-rich galaxies and 4 of which are control galaxies. 

Despite the excess of \hi gas in the \hi-rich galaxies, we find
that the \hi-rich and control galaxies have similar total molecular
masses, $\sim3\times10^9$\,\msun. \hi-rich galaxies
show marginally shorter molecular depletion times and
longer gas (atomic and molecular) depletion times. 
The large atomic gas reservoir of the \hi-rich galaxies may play
a role in sustaining their SFR but does not imply a significant
or systematic increase in the molecular gas mass or SFR.

We detect CO in radial profiles out to $\sim0.7\times R_{25}$. 
We compare those to radial profiles of \hi, SFR, and metallicity.
Besides higher molecular SFE and lower \htwo/\hi ratios
in the centres of the \hi-rich galaxies, the main differences between
\hi-rich and control galaxies lie in the outer parts of the disc:
more atomic gas, less molecular gas, and lower metallicities.
CO discs also appear slightly smaller in the \hi-rich galaxies. 

Signatures from atomic gas accretion on the inner disc and
on the molecular gas properties are found almost inexistent.
We conjecture that the excess \hi gas of the \hi-rich galaxies will
be consumed and participate to the growth of the molecular disc.
Comparing the relevant time-scales, we conclude that the conversion
of this gas into \htwo and then into stars is a rather slow process
($\sim$5\,Gyr). A secular evolution of the \hi-rich galaxies into
control galaxies is therefore expected.

\section*{Acknowledgements}
We are grateful to Mei-Ling Huang for her help with the
\bd star-formation rate data.
We would like to thank Barbara Catinella and Sacha Hony
for careful reading of the manuscript and helpful comments,
as well as David Elbaz and Adam Leroy for interesting discussions.
We also thank the referee for a very constructive report.
DC and FB acknowledge support from DFG grant BI 1546/1-1. 
JP acknowledges support from the CNRS program ``Physique et
Chimie du Milieu Interstellaire'' (PCMI).
JMvdH acknowledges support from the European Research Council
under the European Union's Seventh Framework Programme
(FP/2007-2013)/ ERC Grant Agreement nr. 291531.
This work is based on observations carried out with the IRAM 
30-m Telescope. IRAM is supported by INSU/CNRS (France), 
MPG (Germany) and IGN (Spain). 
This publication makes use of data products from the Wide-field
Infrared Survey Explorer, which is a joint project of the University
of California, Los Angeles, and the Jet Propulsion Laboratory/California
Institute of Technology, funded by the National Aeronautics and
Space Administration.

\bibliographystyle{mnras}
\bibliography{mn-16-1944-arxiv.bib}

\appendix{
%
\section{IRAM data of the \bd galaxies}
\label{append:hera}
We present SDSS images and the IRAM data of the \bd galaxies 
in Fig.~\ref{fig:hera} for the galaxies mapped with HERA and 
in Fig.~\ref{fig:emir} for the galaxies observed with EMIR. 
SDSS images were downloaded from the DR7 Data Archive Server 
(\protect\url{http://das.sdss.org/www/html/}). 
Images are 1.6\arcm$\times$1.6\arcm in size. In Fig.~\ref{fig:hera}, 
the HERA coverage (1.2\arcm$\times$1.2\arcm) 
and beam size (11\arcs at 230\,GHz) are drawn on top. 
In Fig.~\ref{fig:emir}, the EMIR beam at 115\,GHz (22\arcs) 
is drawn on top of the image.
All IRAM spectra have spectral resolution 15.6\,\kms. 
The vertical lines show the optical centre (plain red), the CO 
centre (plain blue), and the width of the CO line (dashed blue). 
Note that the width indicated for the HERA spectra is only
shown for comparison, it is computed for each pixel individually
in these resolved maps.
Spectra are labeled (HI) for the \hi-rich glaxies, (C) for the control
galaxies, (E) for excluded galaxies, and (P) for poor quality.

\begin{figure*}
\centering
 \includegraphics[clip,width=2.8cm]{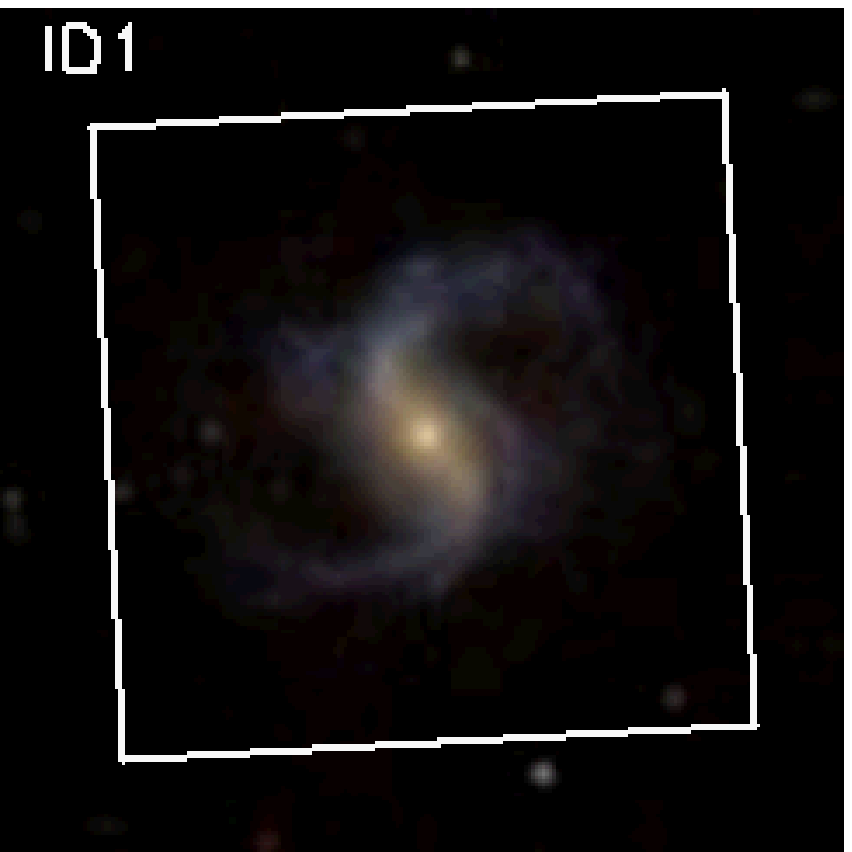}\hspace{-1mm}
 \includegraphics[clip,width=2.8cm]{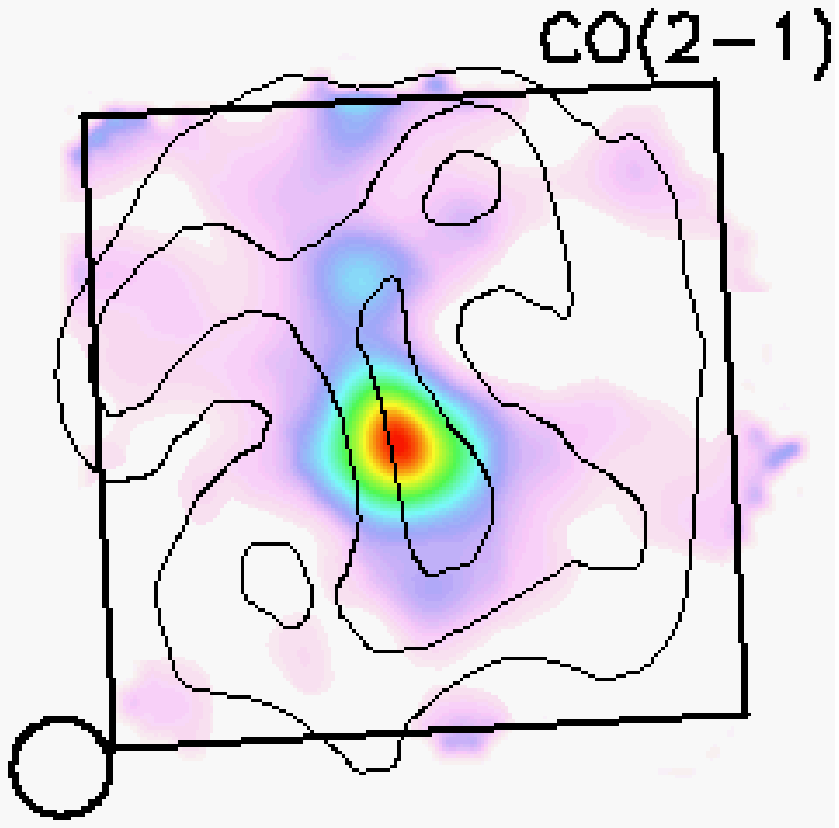}
 \includegraphics[clip,width=3cm]{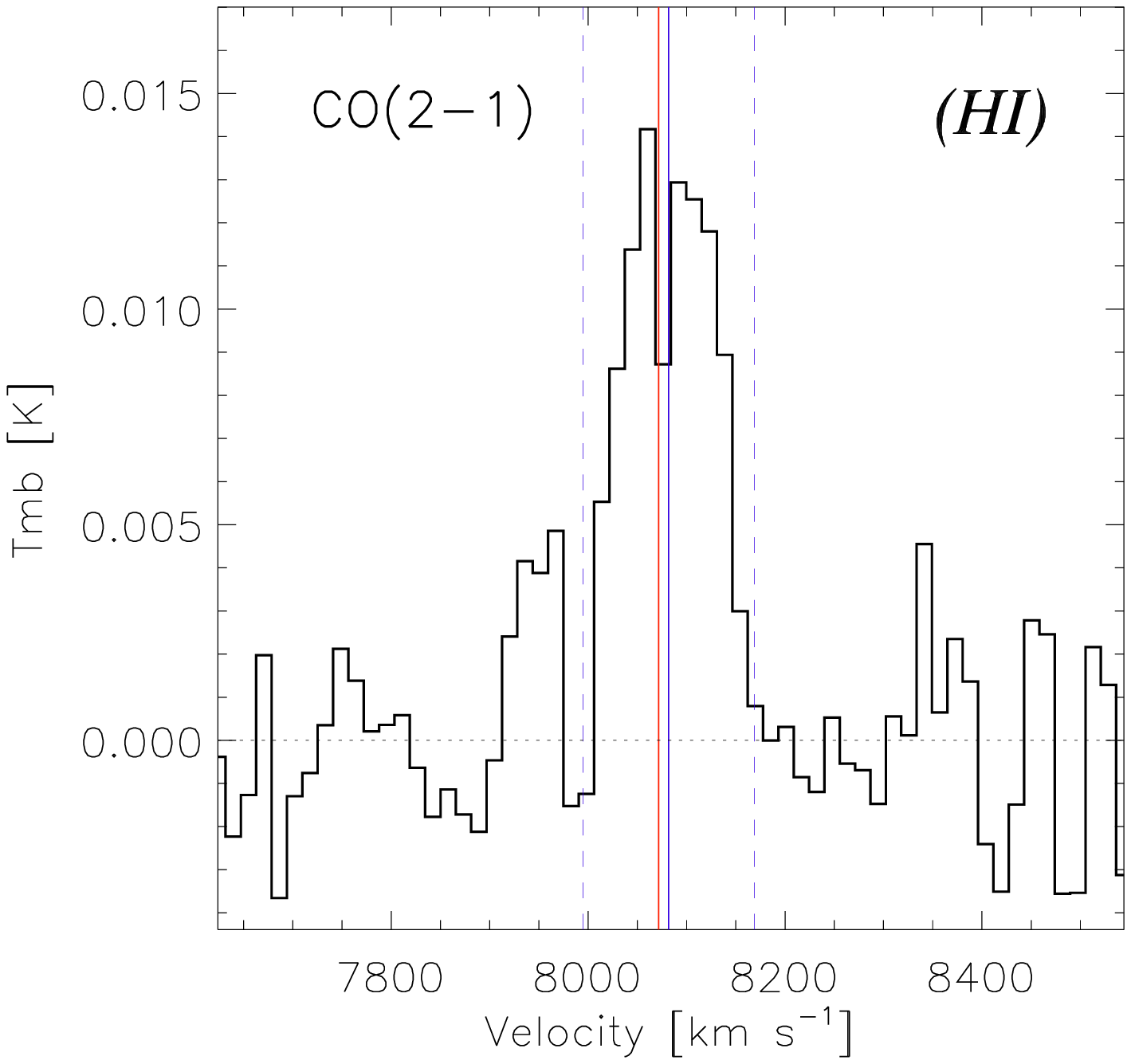}\hfill
 \includegraphics[clip,width=2.8cm]{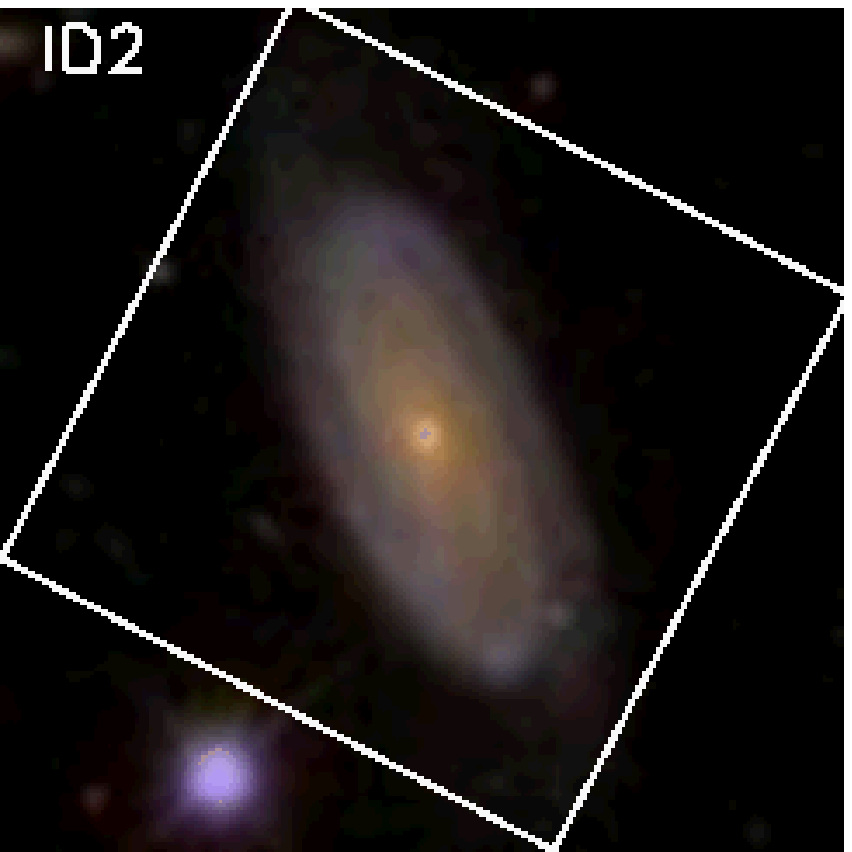}\hspace{-1mm}
 \includegraphics[clip,width=2.8cm]{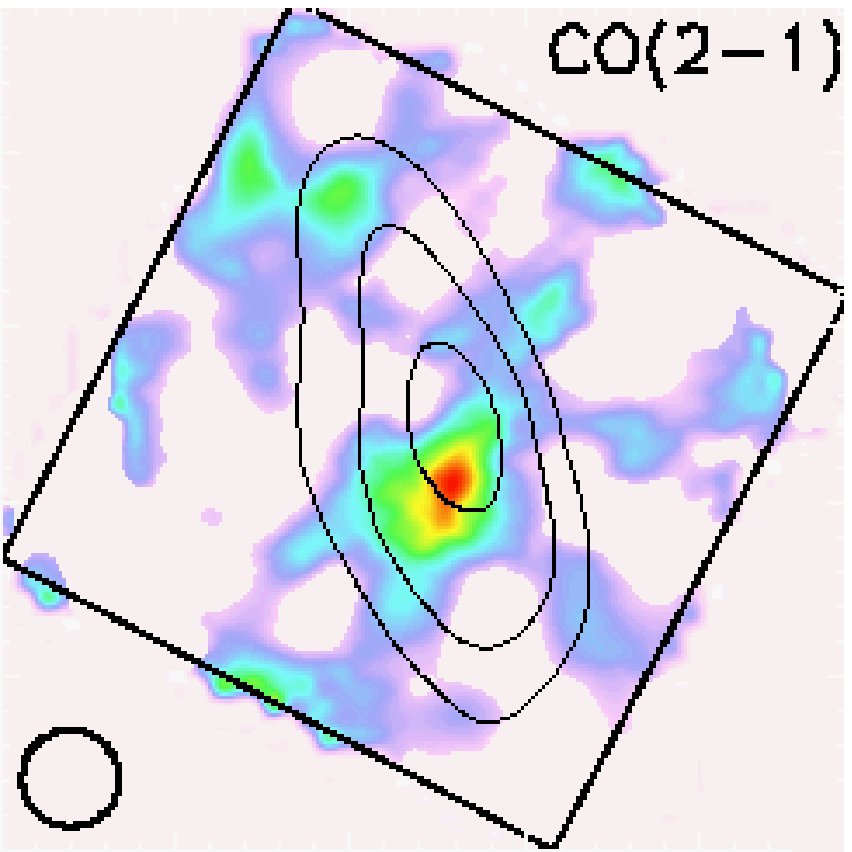}
 \includegraphics[clip,width=3cm]{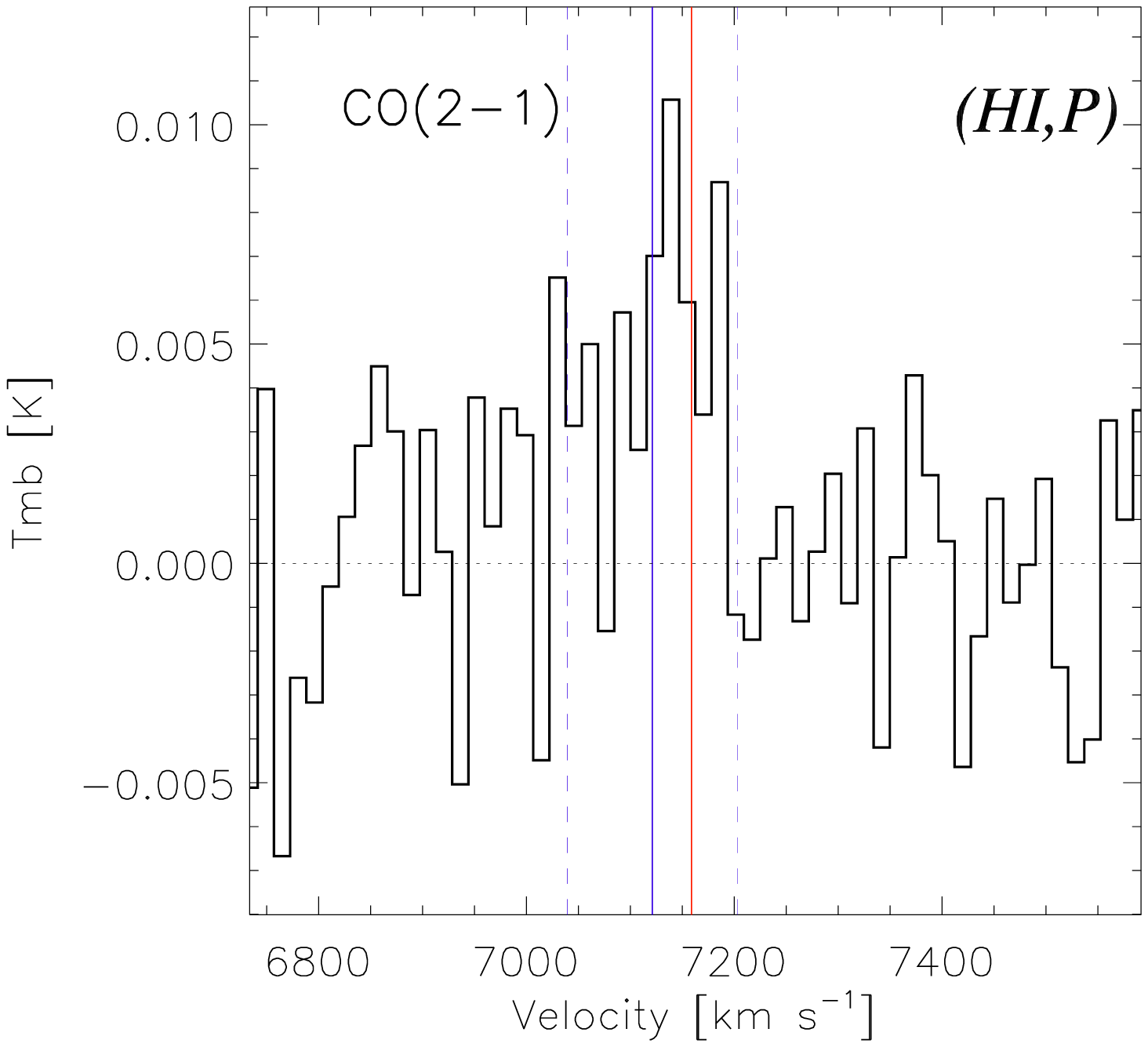}
 \includegraphics[clip,width=2.8cm]{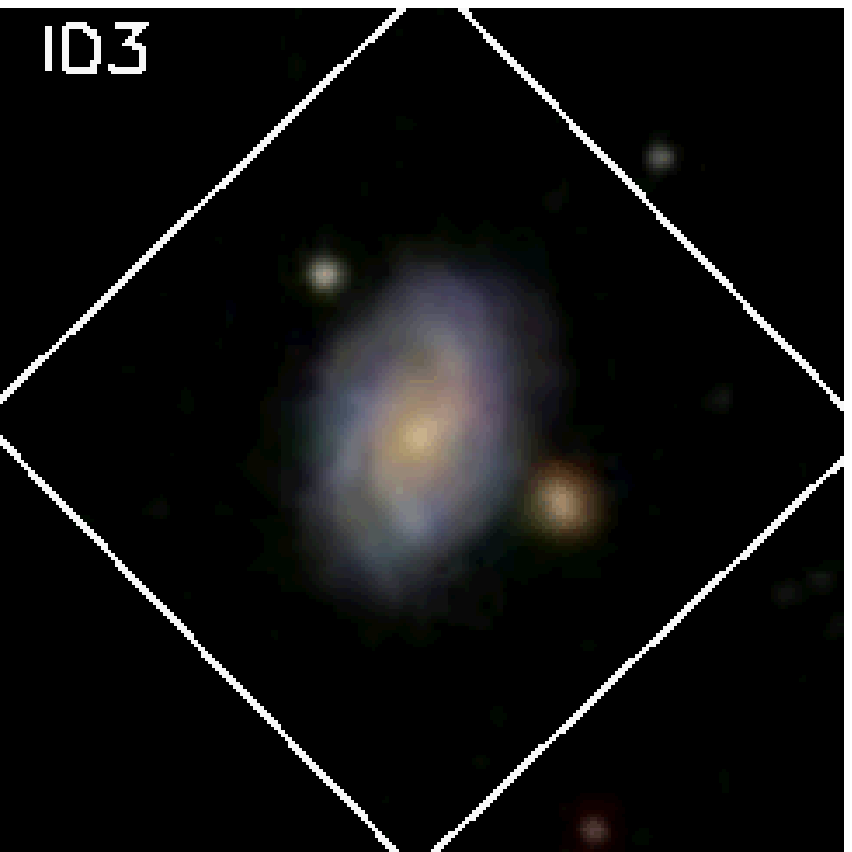}\hspace{-1mm}
 \includegraphics[clip,width=2.8cm]{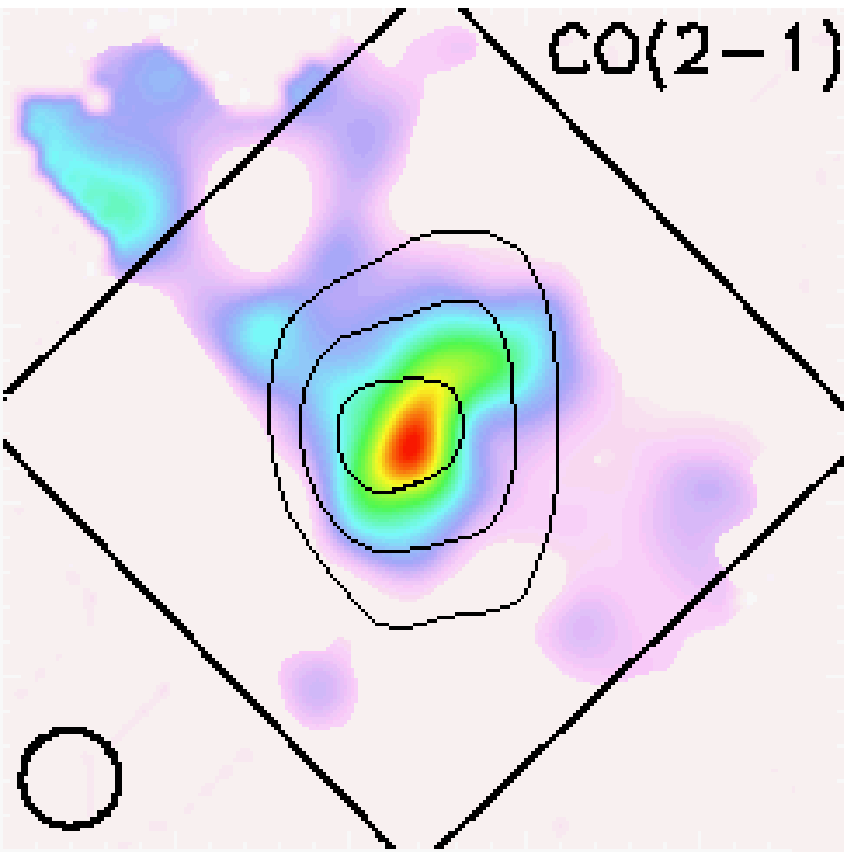}
 \includegraphics[clip,width=3cm]{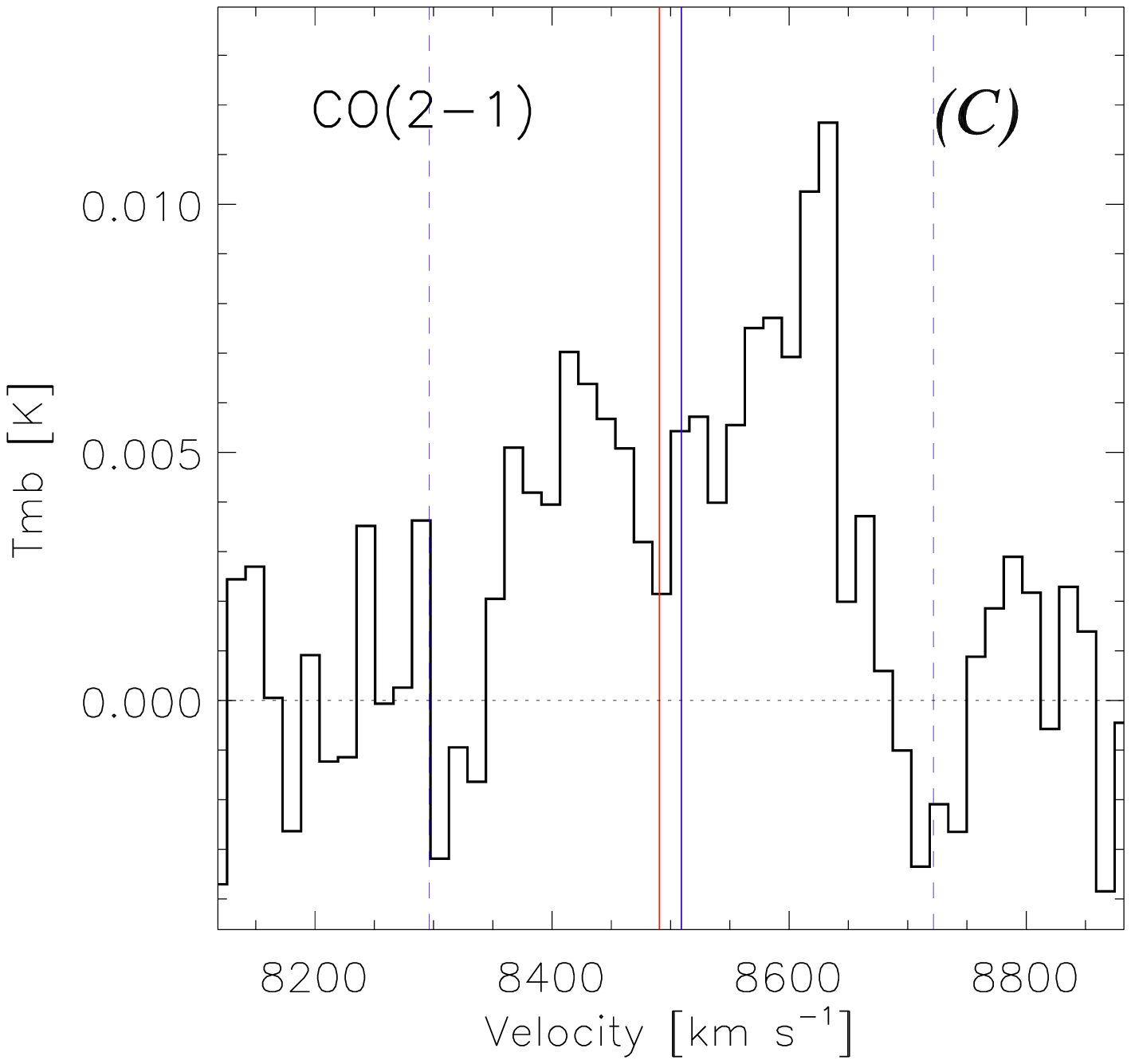}\hfill
 \includegraphics[clip,width=2.8cm]{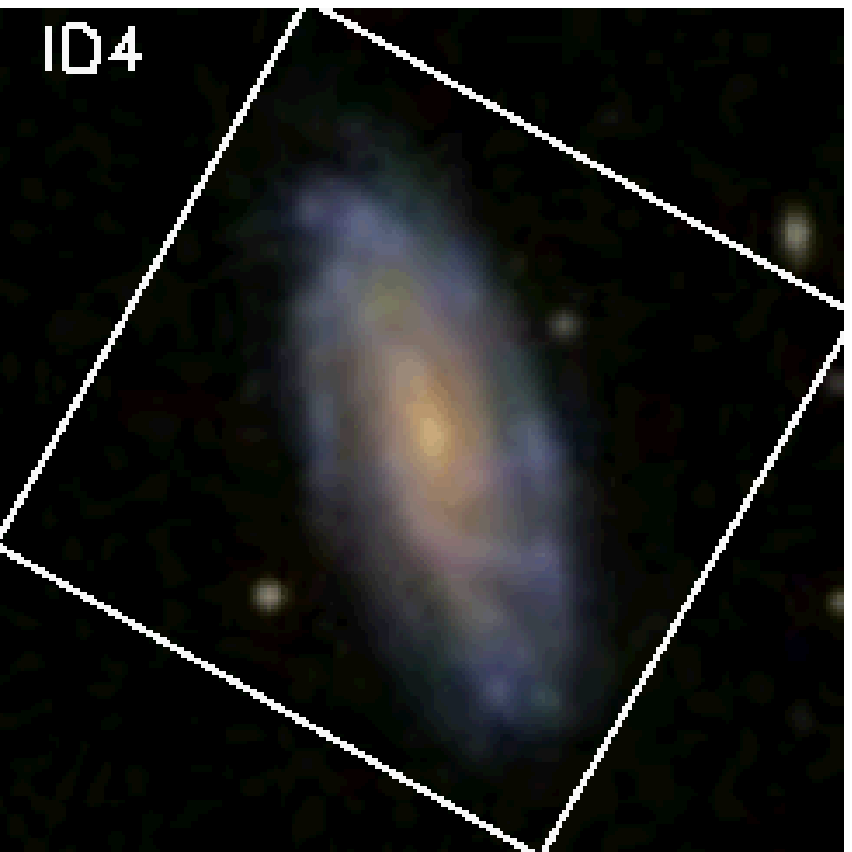}\hspace{-1mm}
 \includegraphics[clip,width=2.8cm]{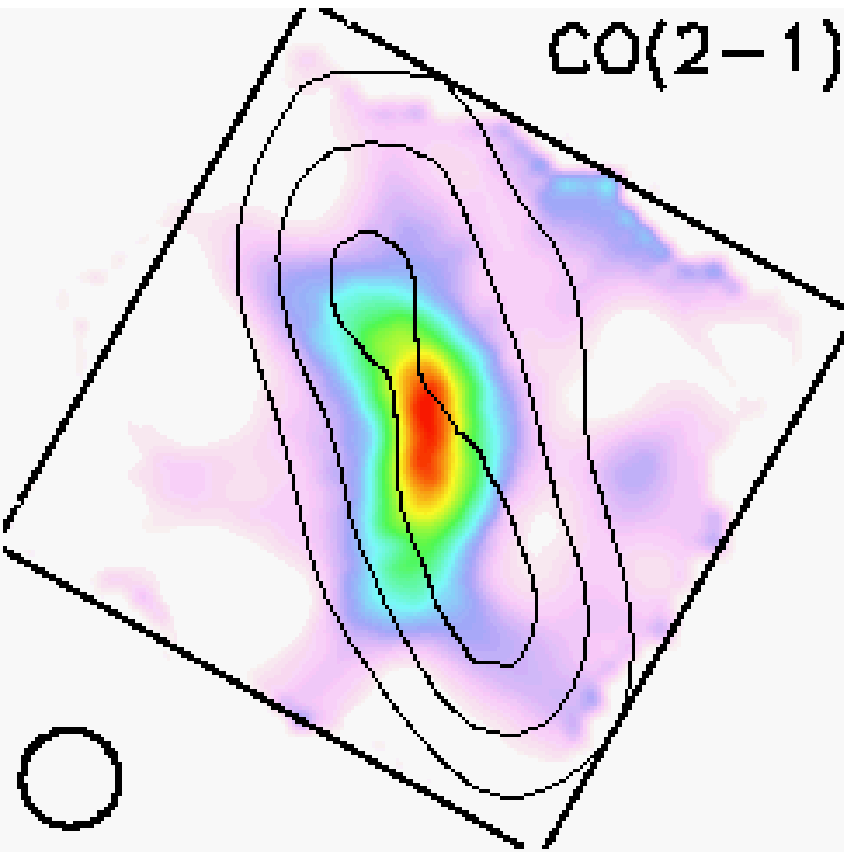}
 \includegraphics[clip,width=3cm]{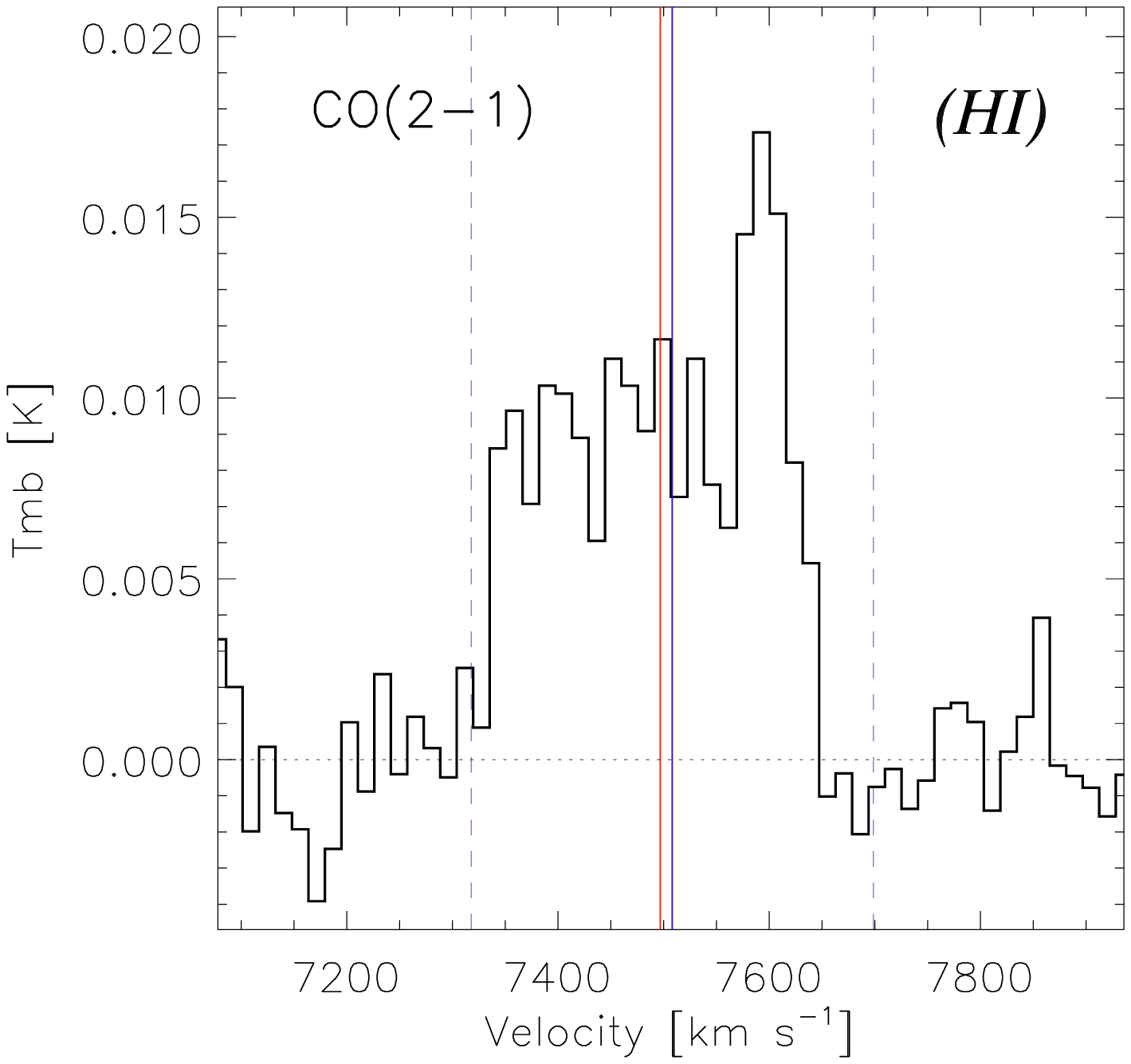}
 \includegraphics[clip,width=2.8cm]{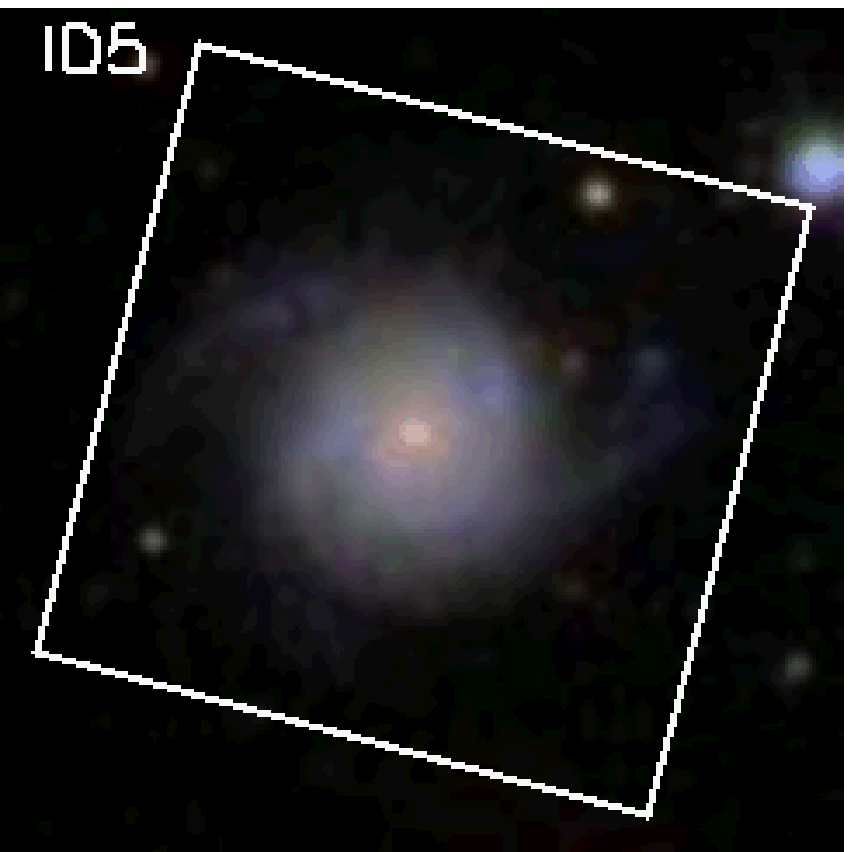}\hspace{-1mm}
 \includegraphics[clip,width=2.8cm]{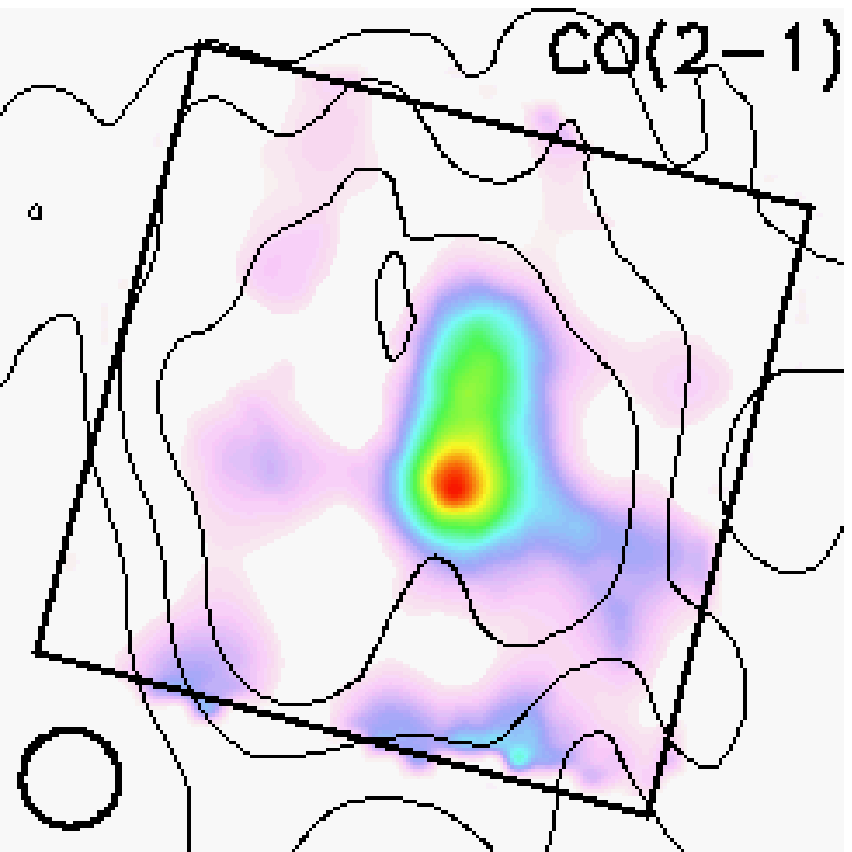}
 \includegraphics[clip,width=3cm]{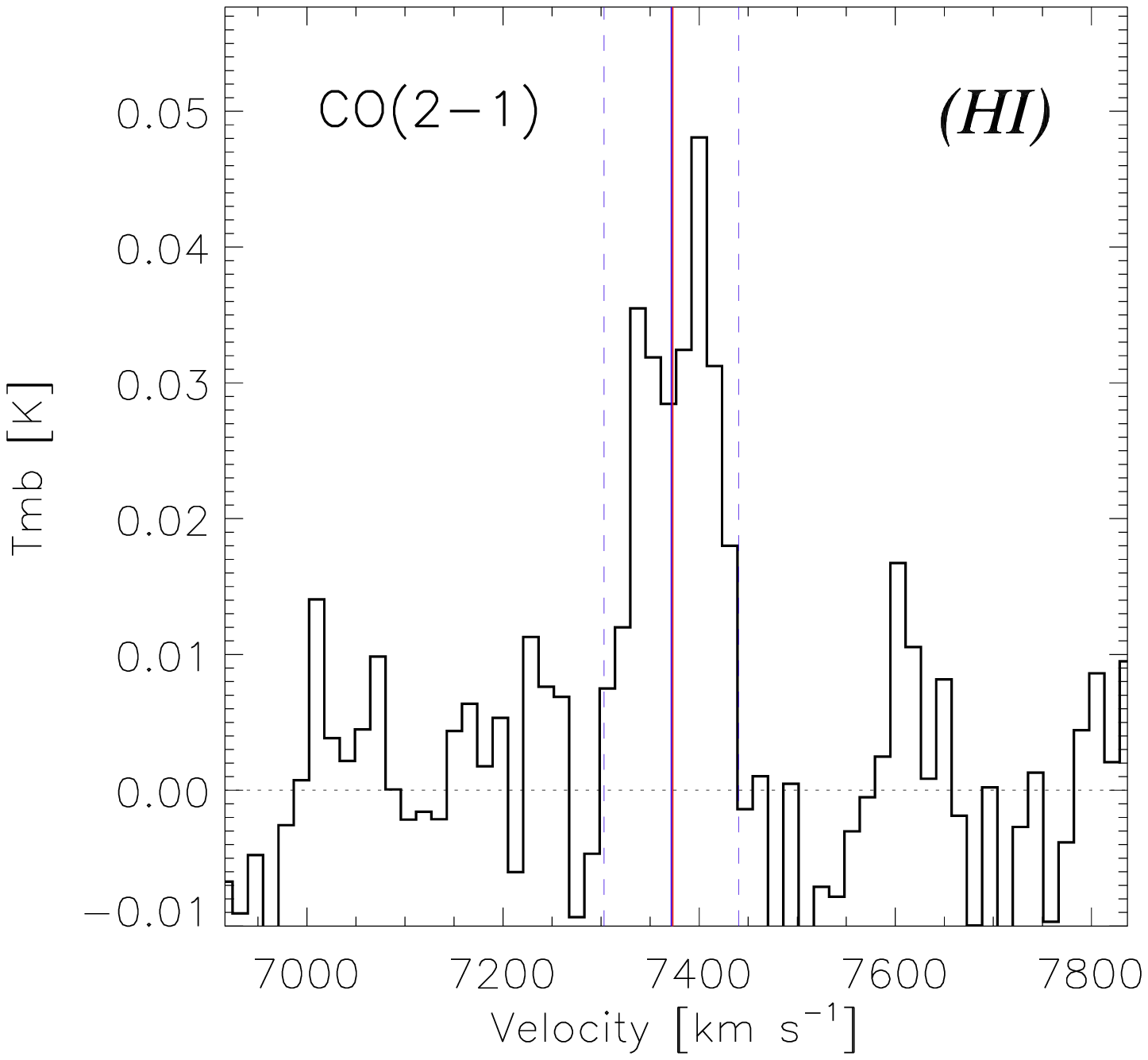}\hfill
 \includegraphics[clip,width=2.8cm]{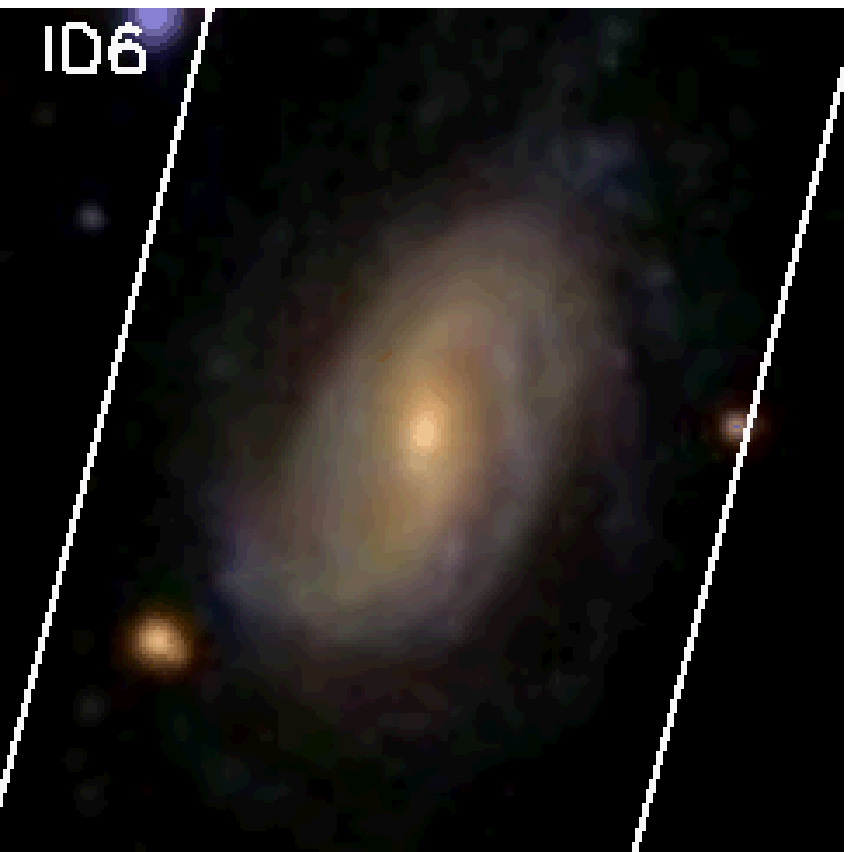}\hspace{-1mm}
 \includegraphics[clip,width=2.8cm]{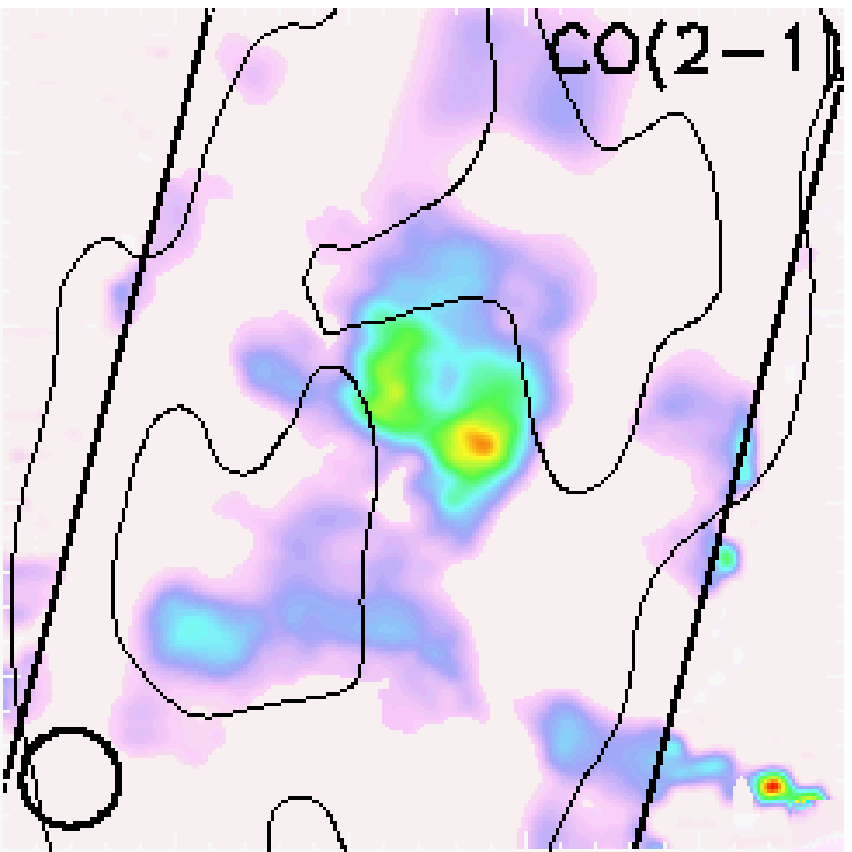}
 \includegraphics[clip,width=3cm]{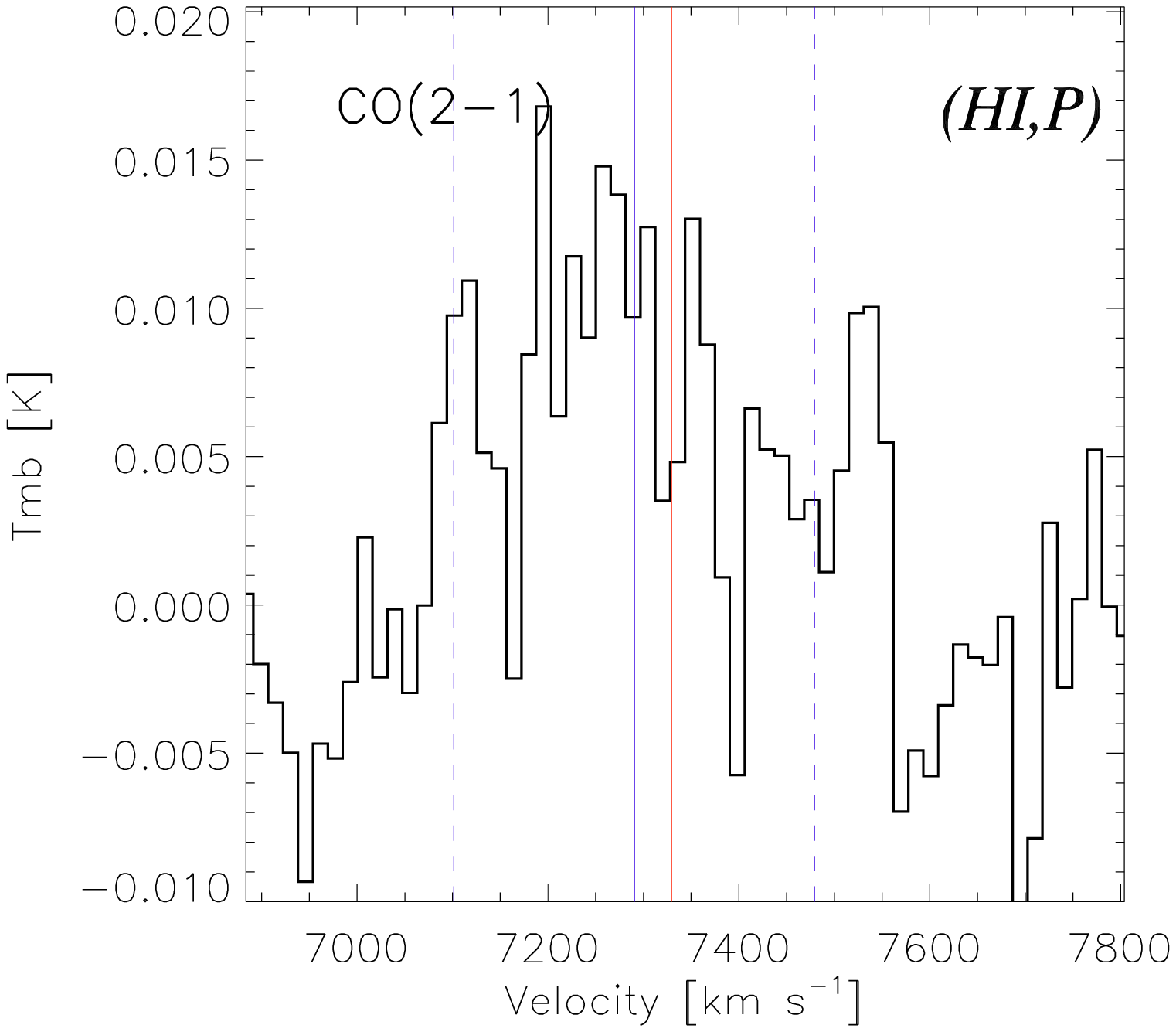}
 \includegraphics[clip,width=2.8cm]{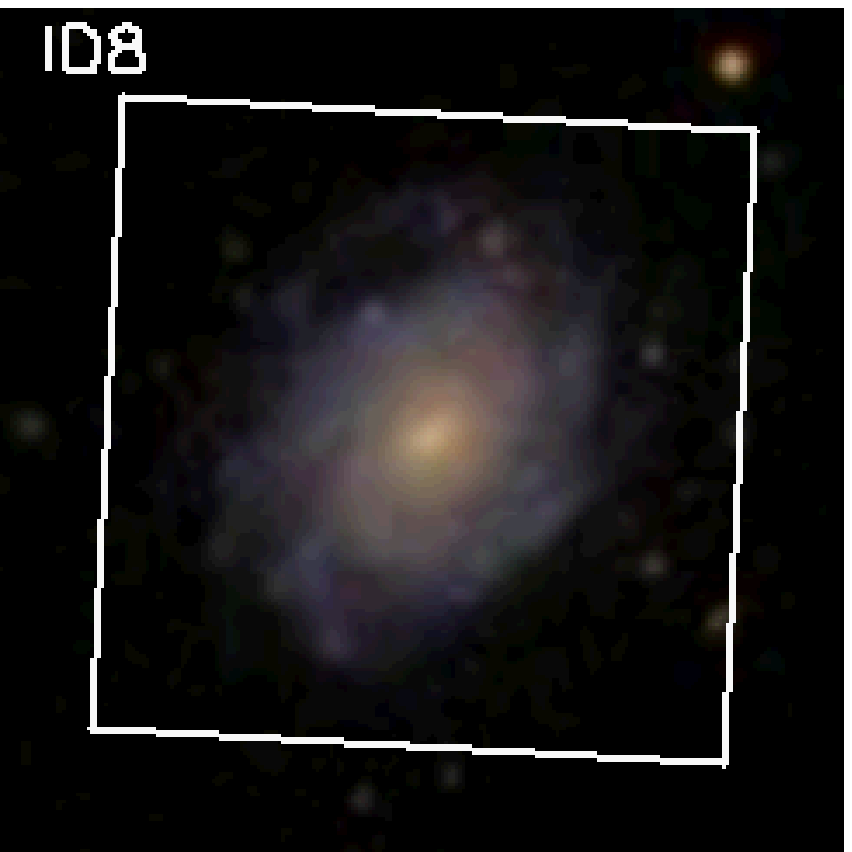}\hspace{-1mm}
 \includegraphics[clip,width=2.8cm]{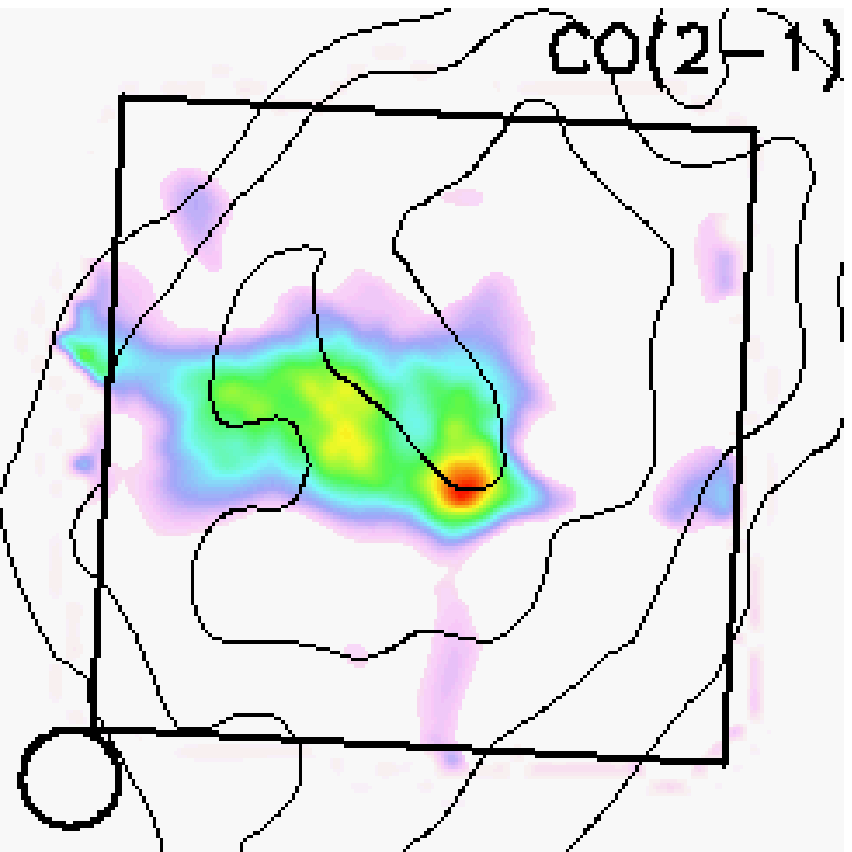}
 \includegraphics[clip,width=3cm]{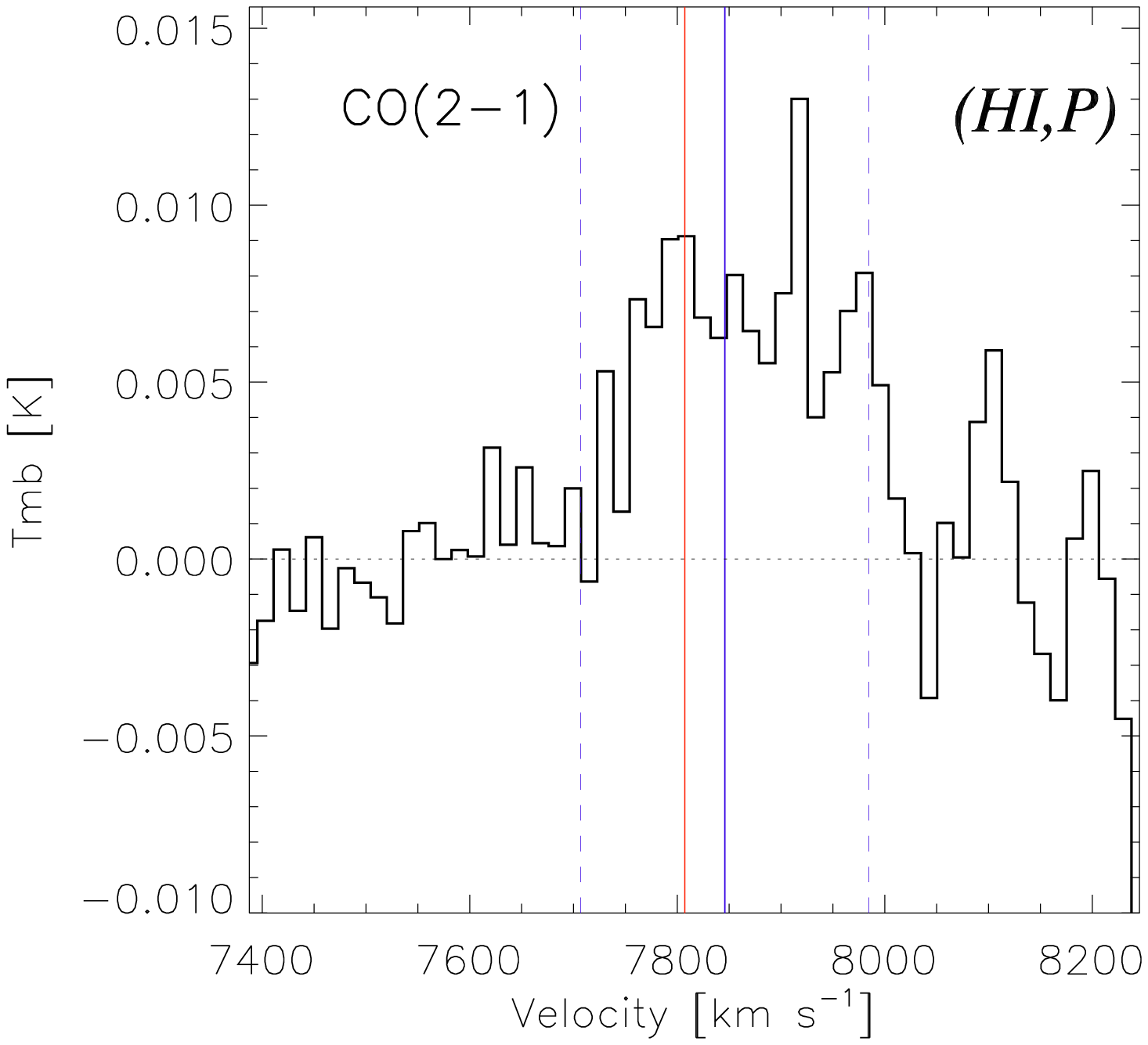}\hfill
 \includegraphics[clip,width=2.8cm]{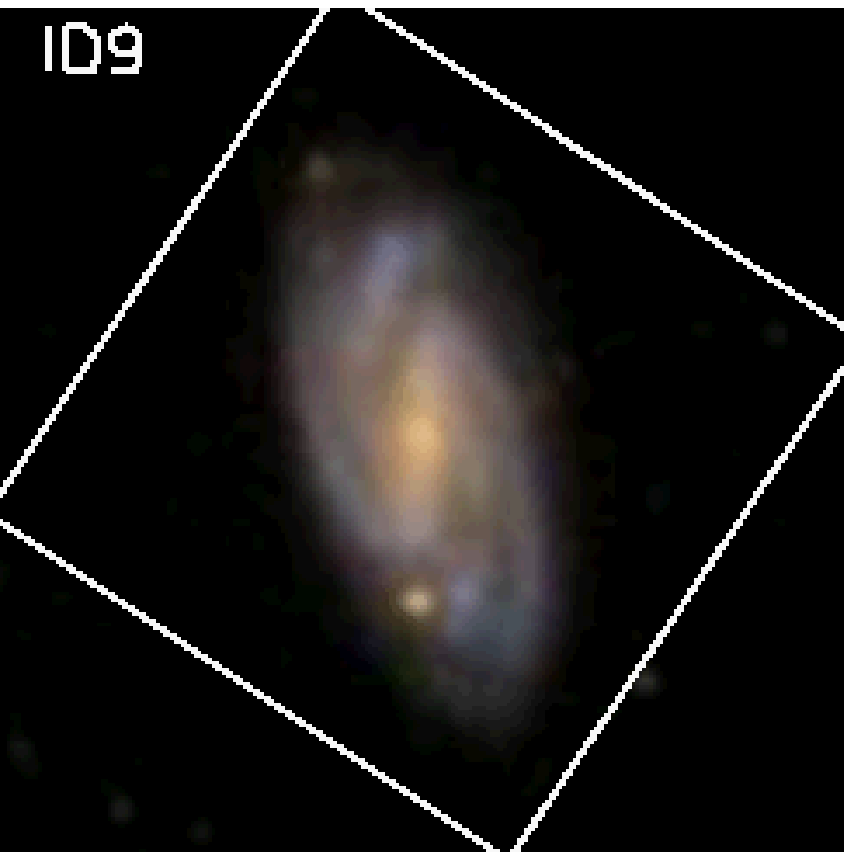}\hspace{-1mm}
 \includegraphics[clip,width=2.8cm]{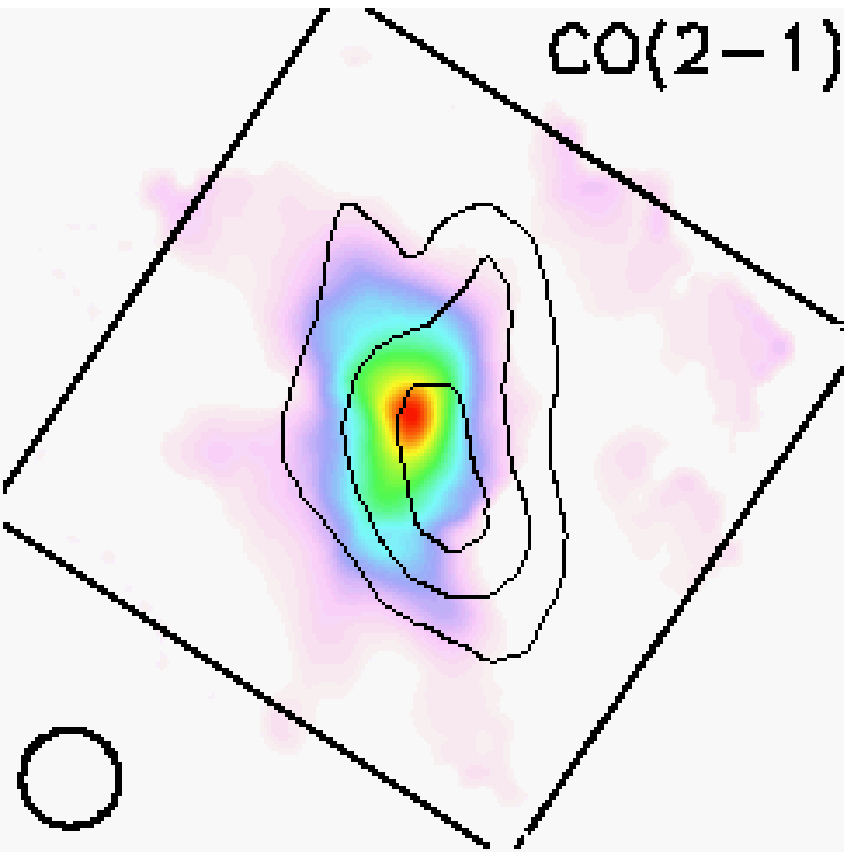}
 \includegraphics[clip,width=3cm]{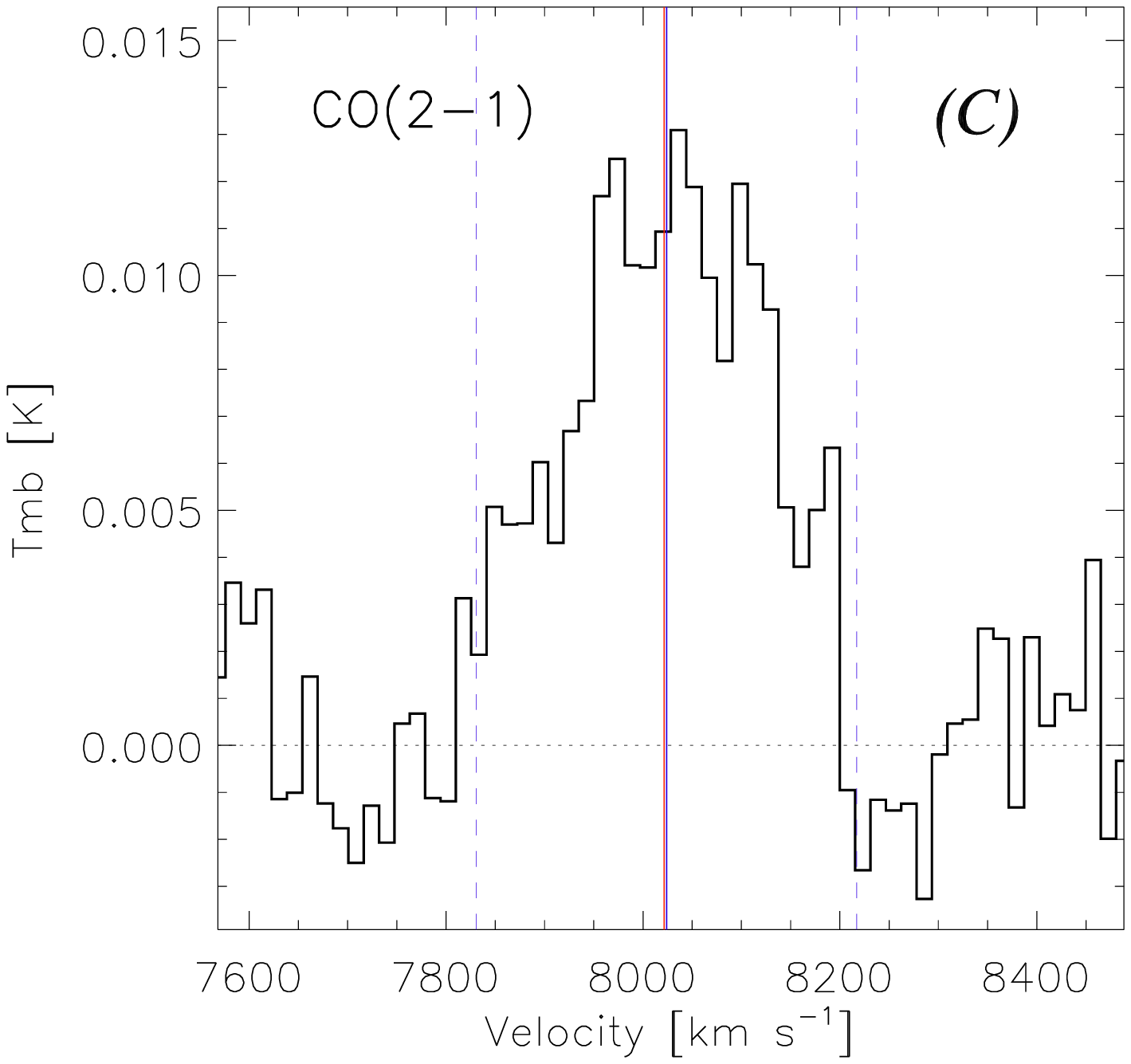}
 \includegraphics[clip,width=2.8cm]{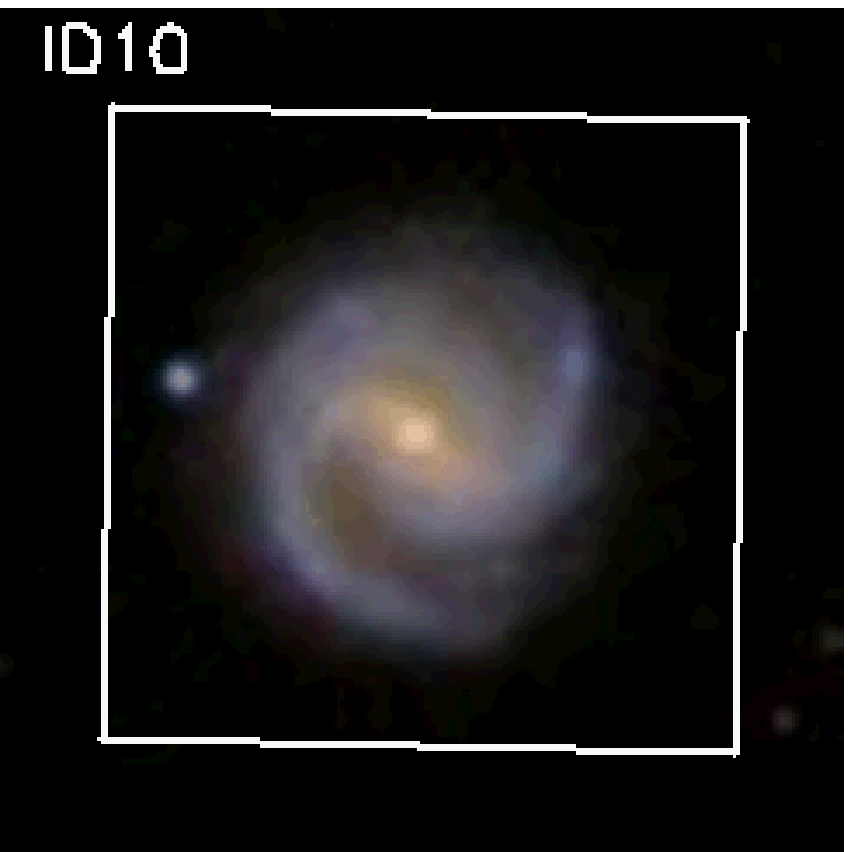}\hspace{-1mm}
 \includegraphics[clip,width=2.8cm]{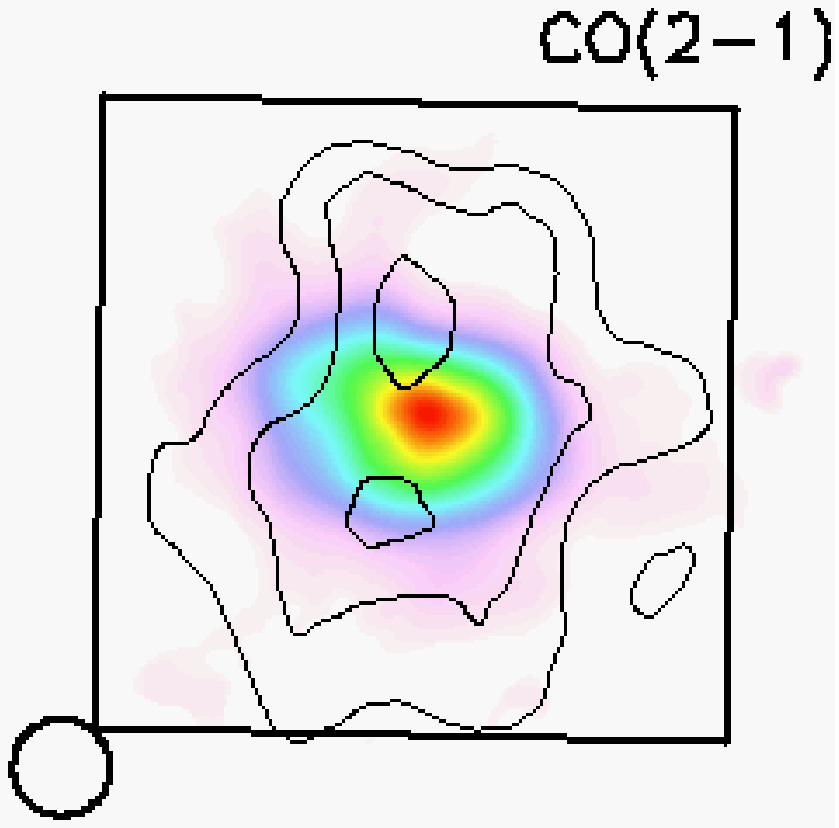}
 \includegraphics[clip,width=3cm]{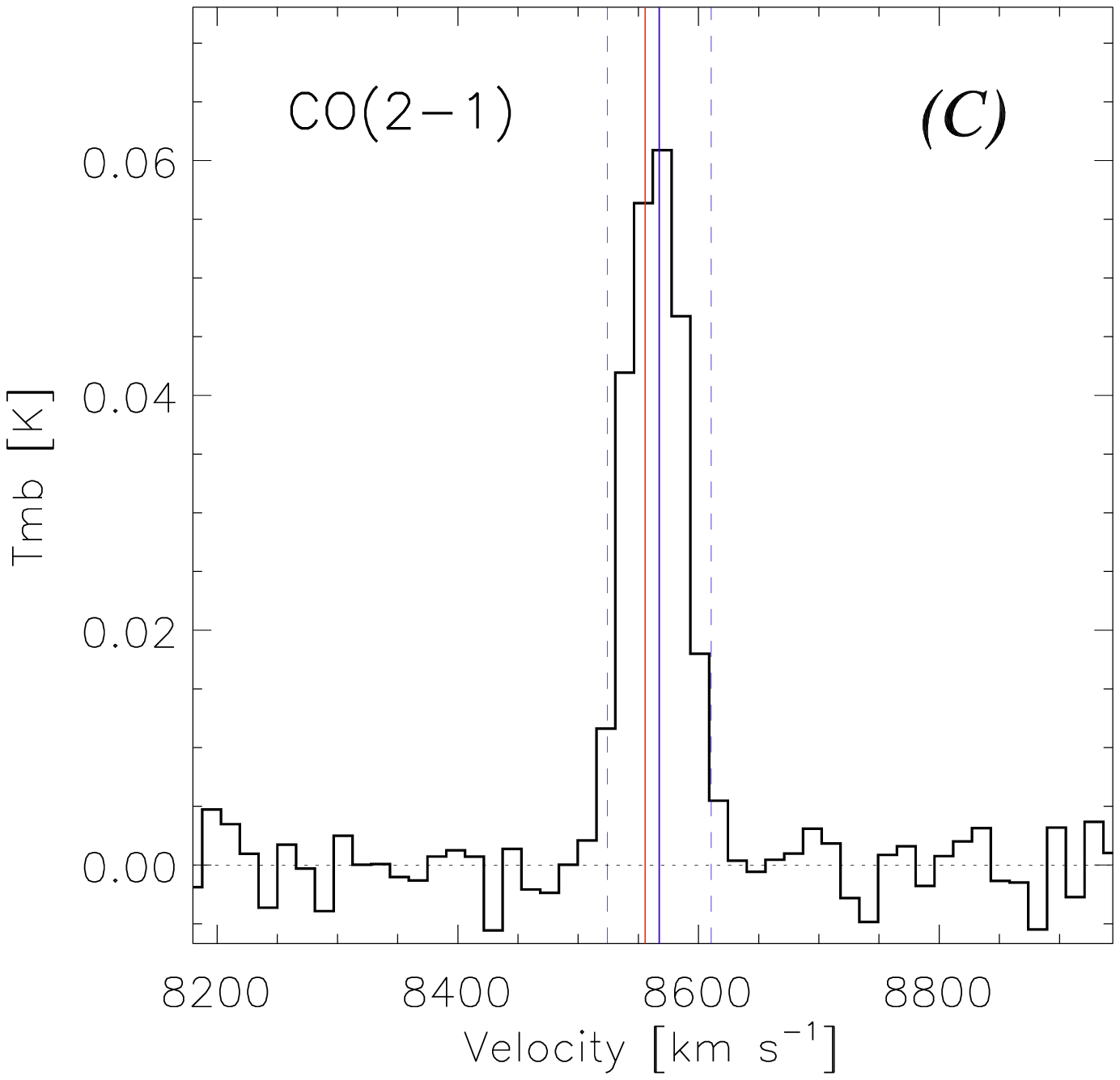}\hfill
 \includegraphics[clip,width=2.8cm]{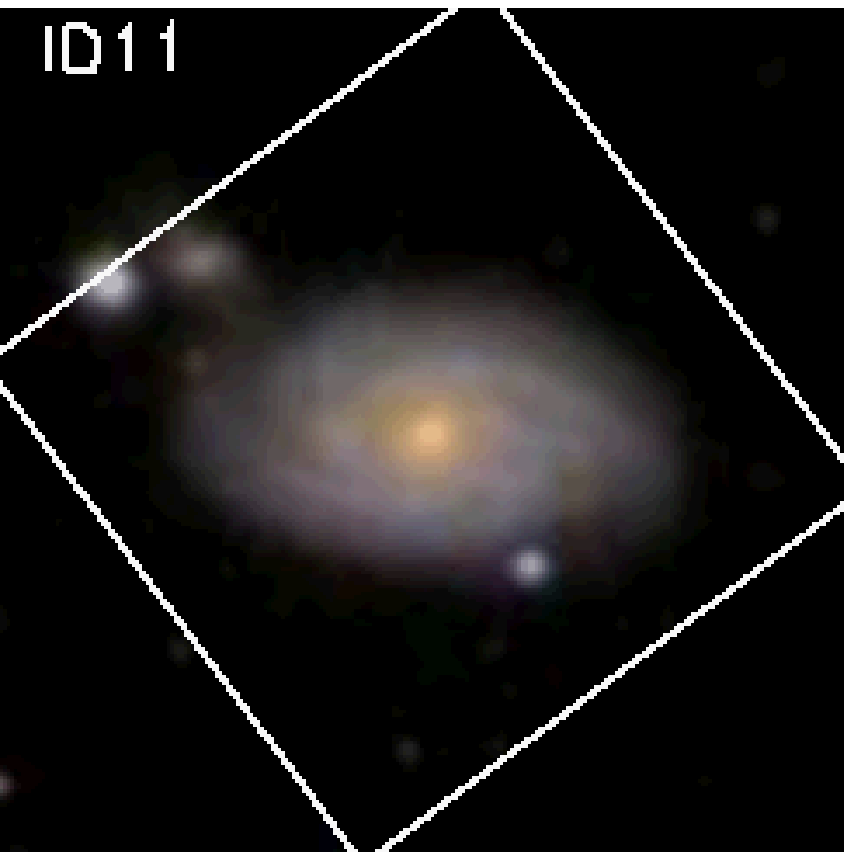}\hspace{-1mm}
 \includegraphics[clip,width=2.8cm]{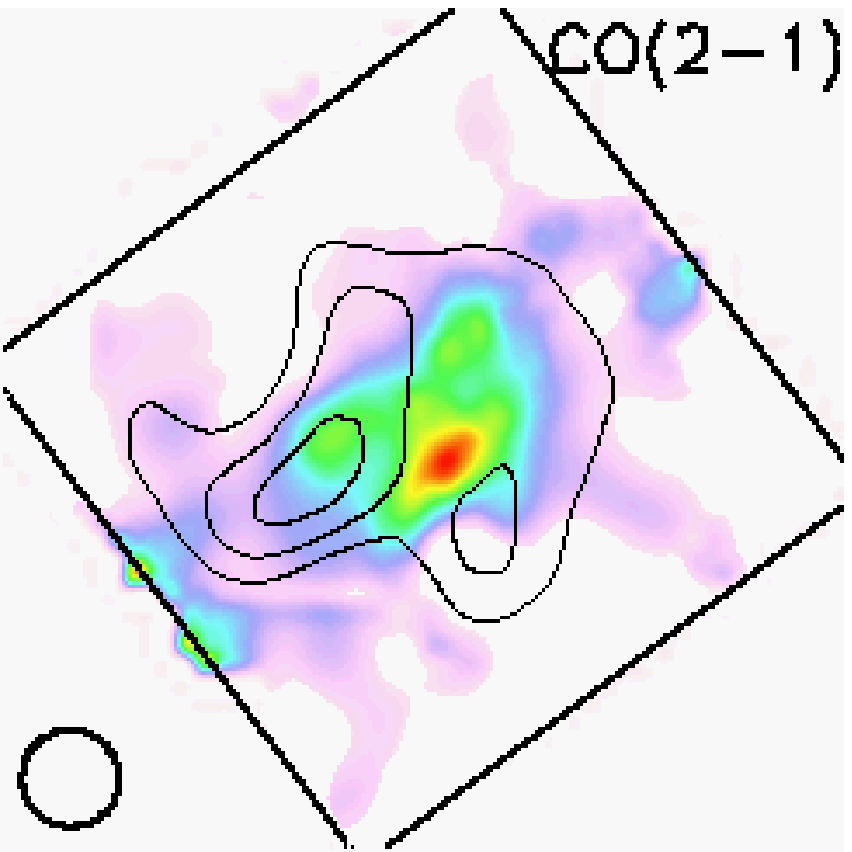}
 \includegraphics[clip,width=3cm]{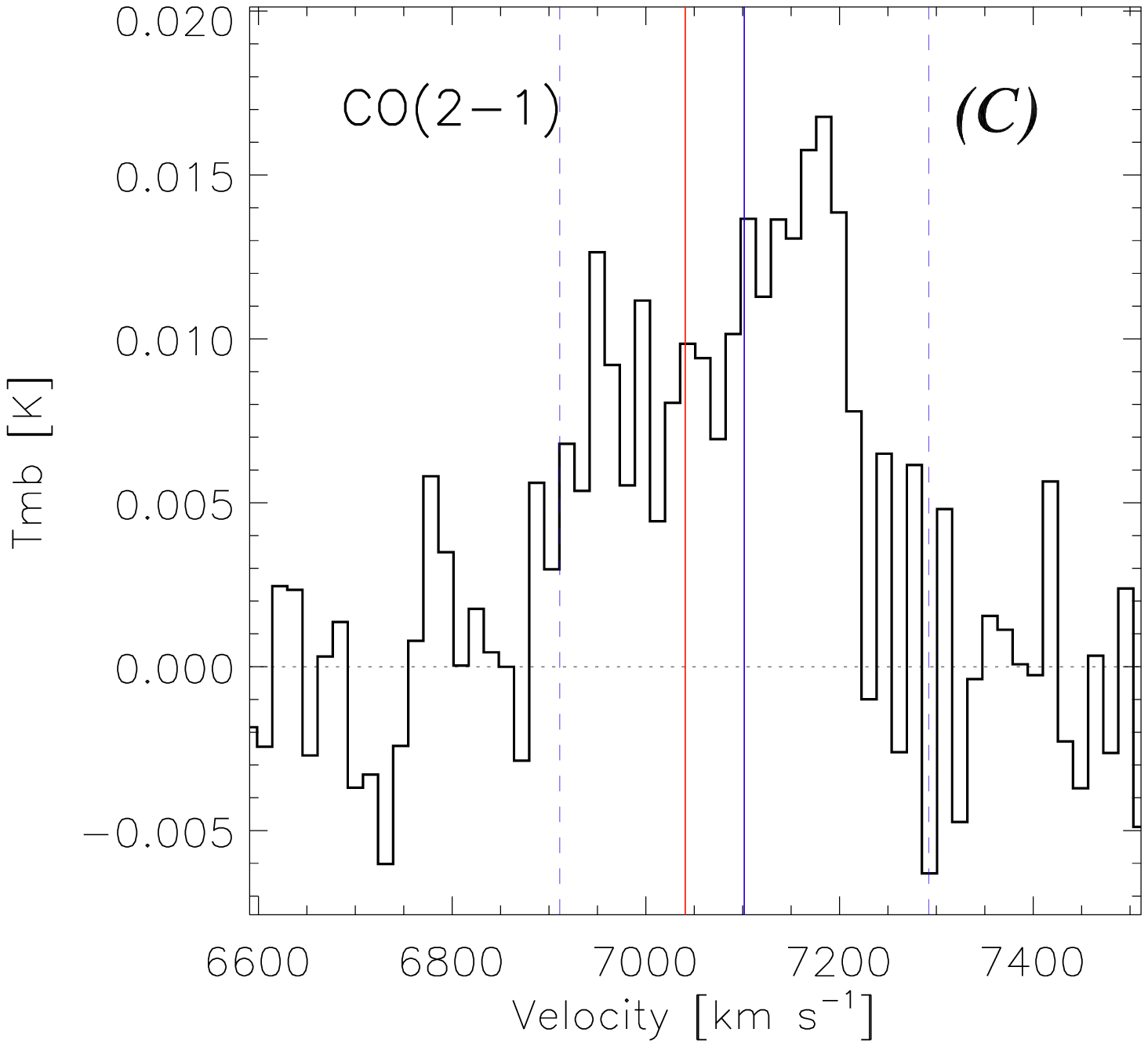}
 \includegraphics[clip,width=2.8cm]{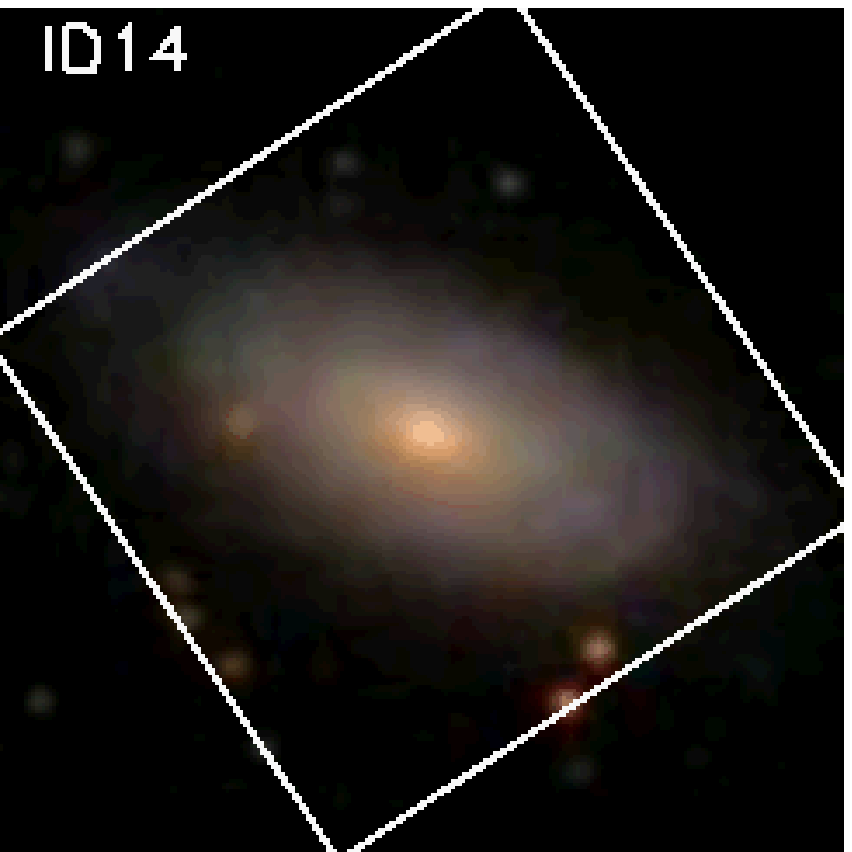}\hspace{-1mm}
 \includegraphics[clip,width=2.8cm]{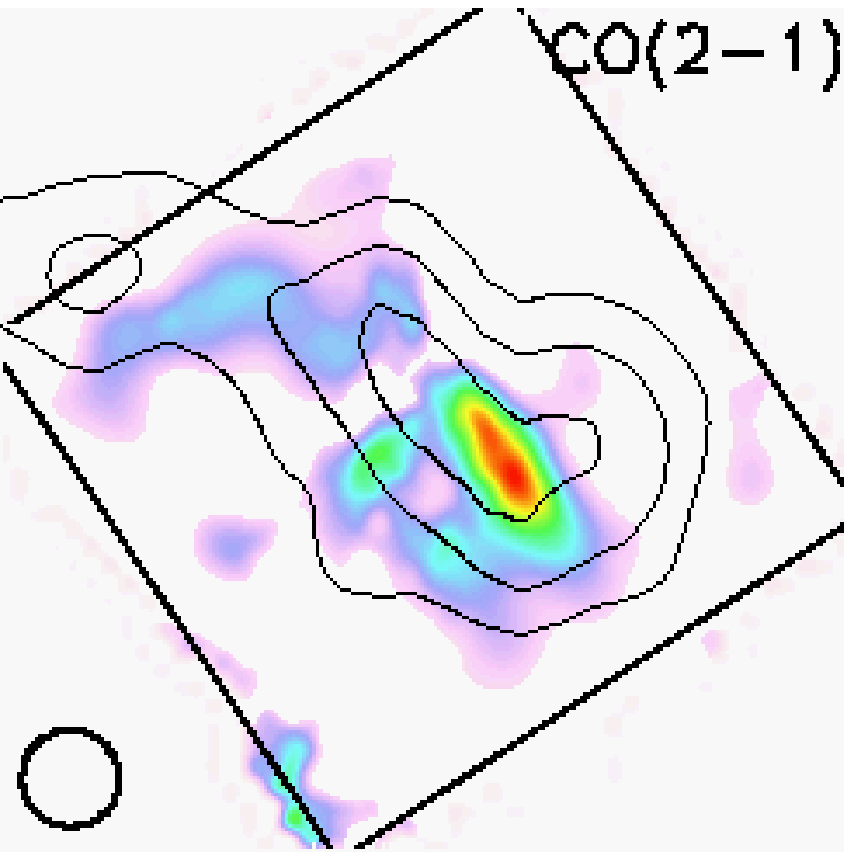}
 \includegraphics[clip,width=3cm]{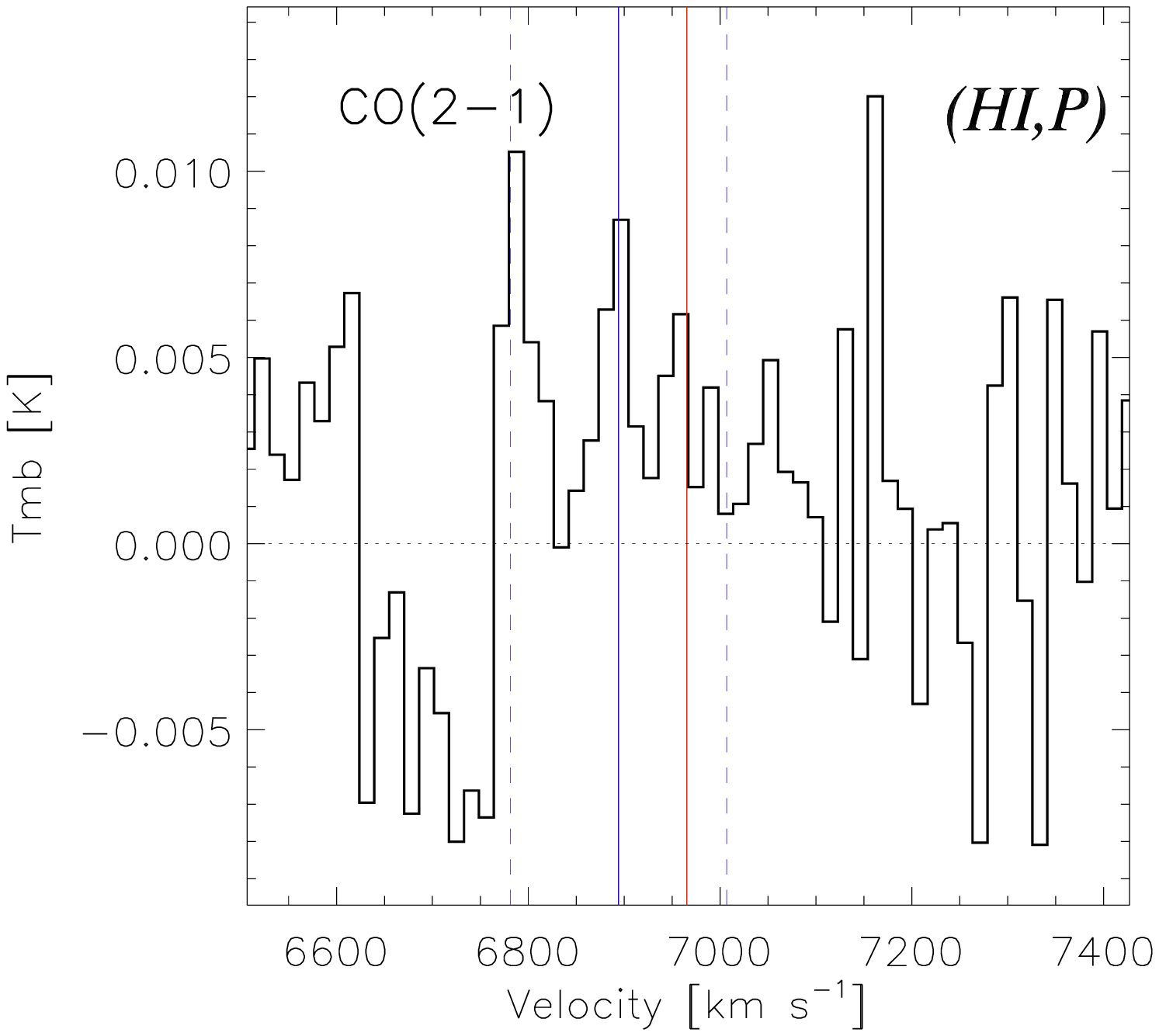}\hfill
 \includegraphics[clip,width=2.8cm]{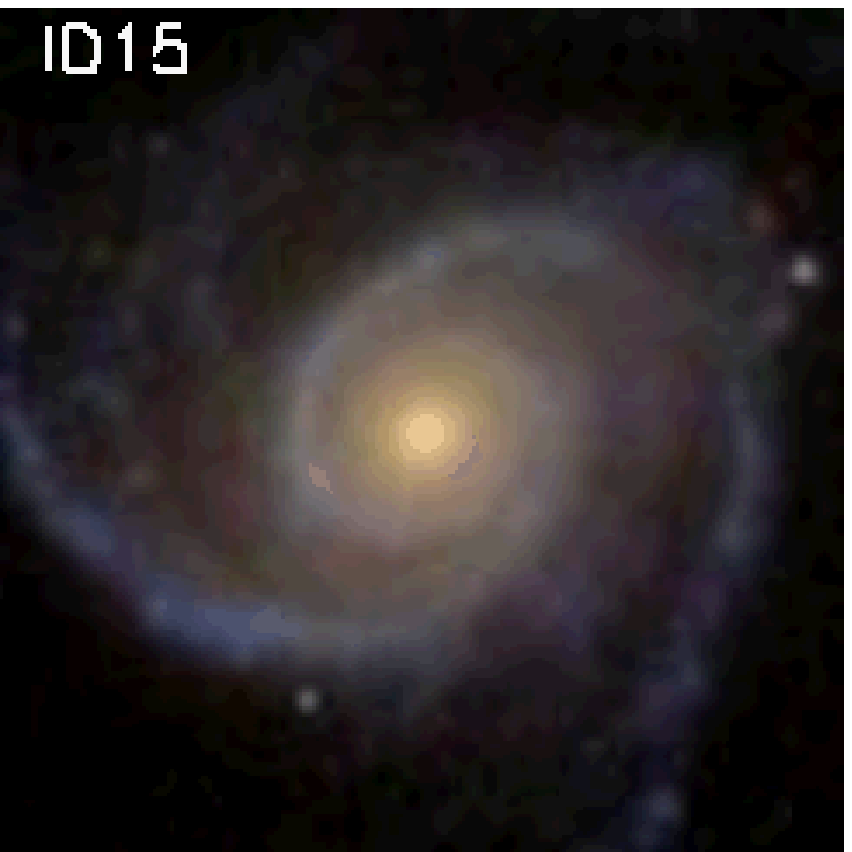}\hspace{-1mm}
 \includegraphics[clip,width=2.8cm]{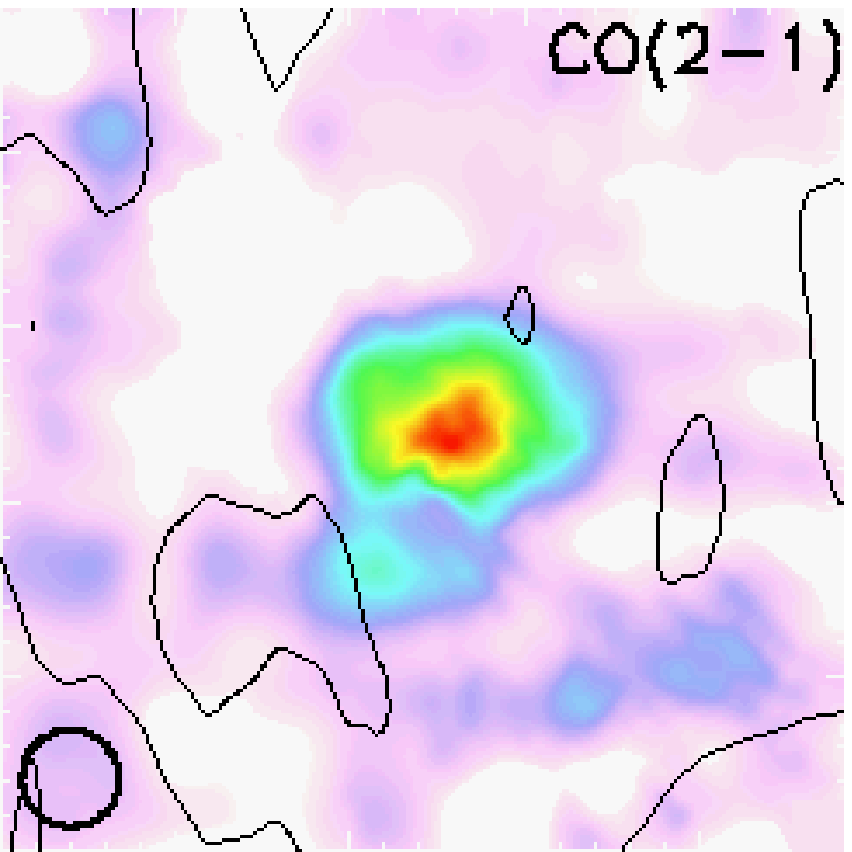}
 \includegraphics[clip,width=3cm]{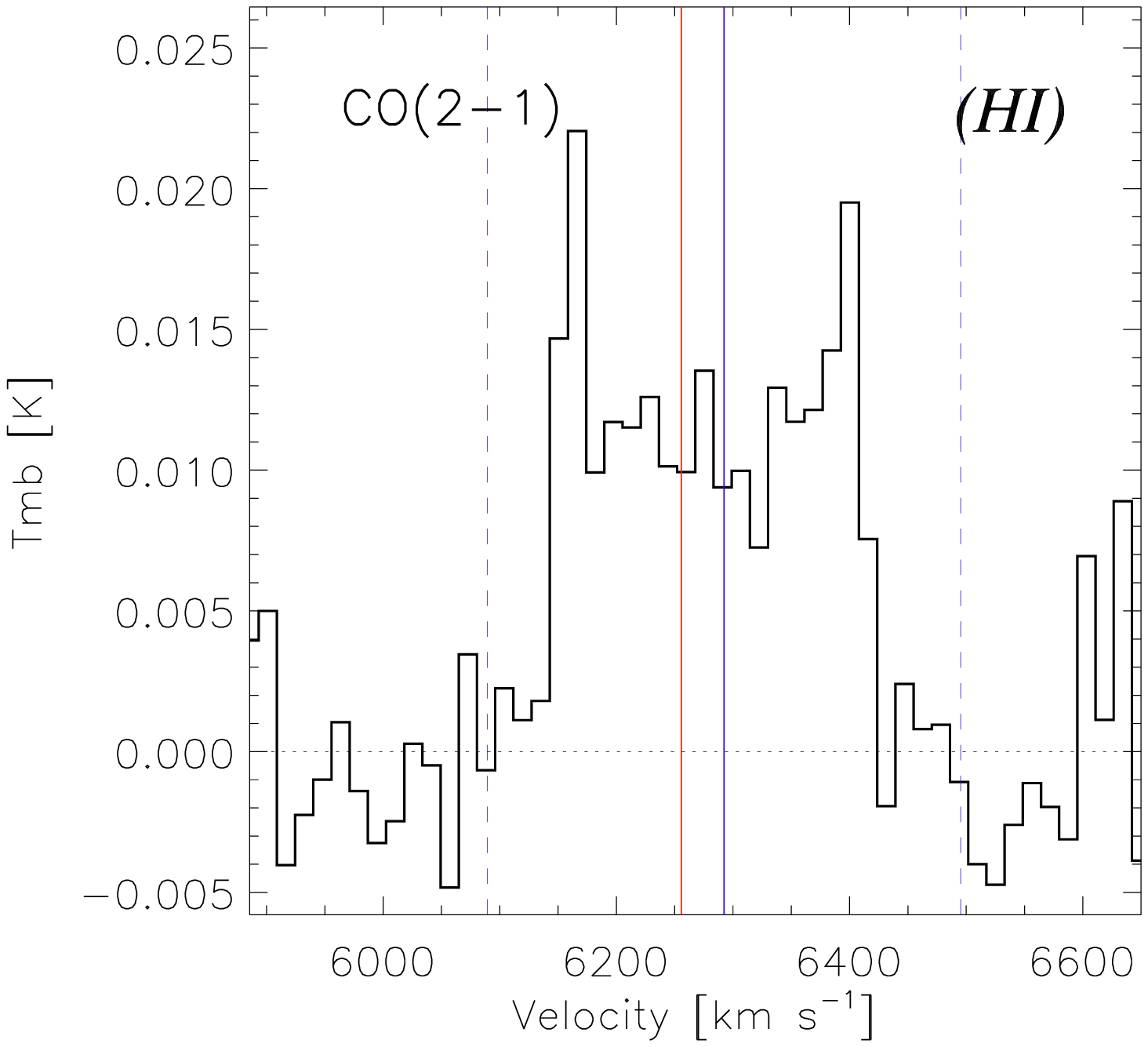}
 \includegraphics[clip,width=2.8cm]{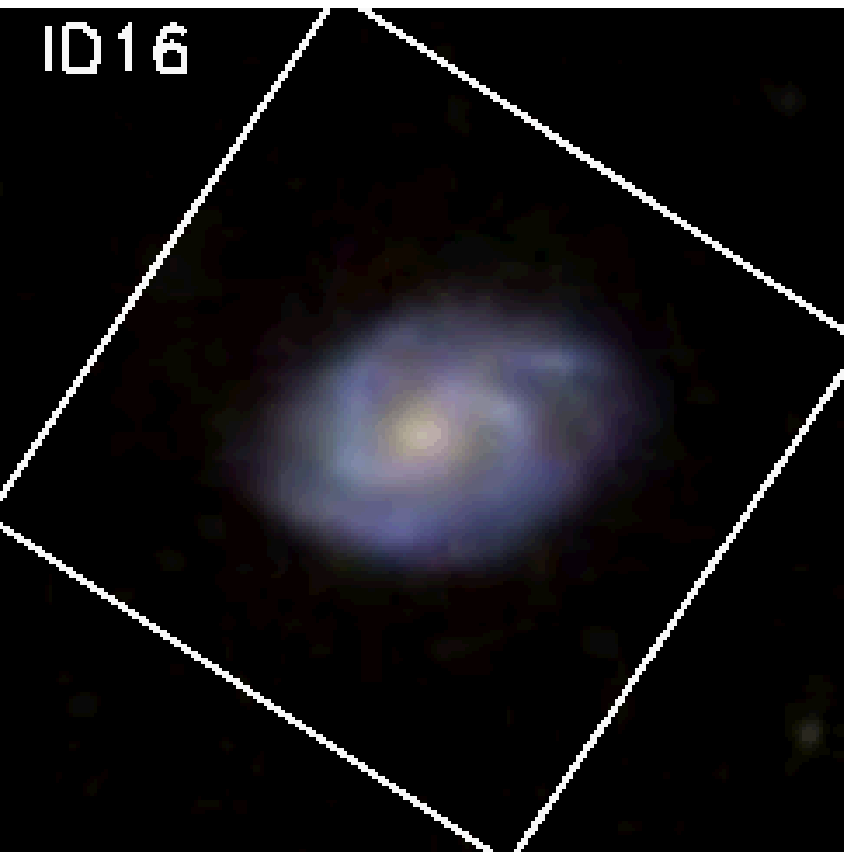}\hspace{-1mm}
 \includegraphics[clip,width=2.8cm]{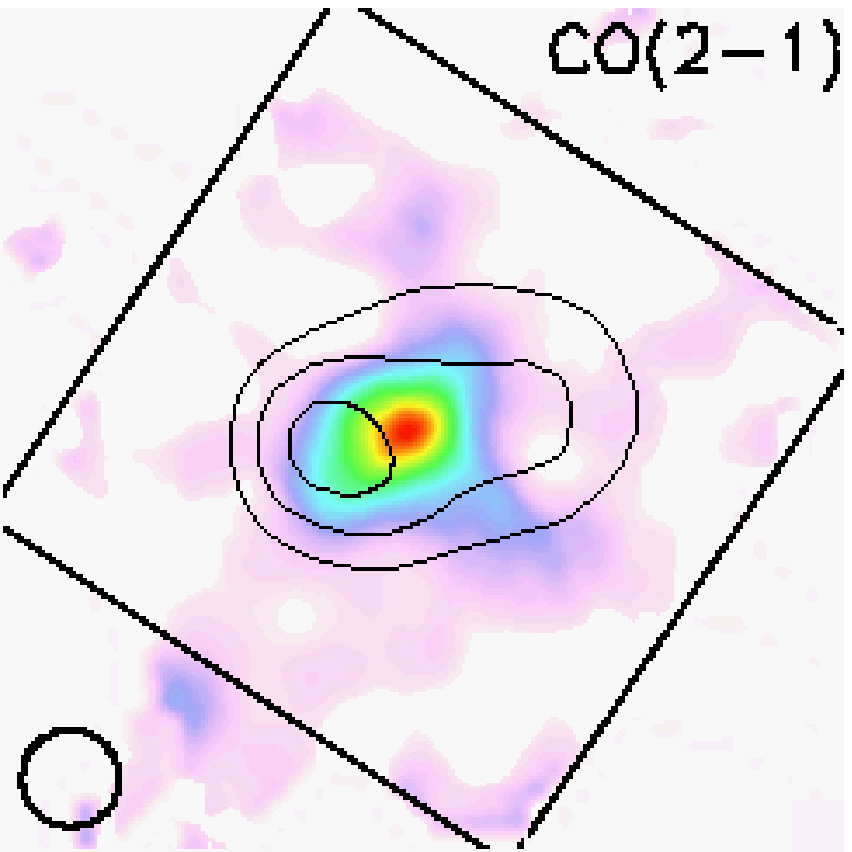}
 \includegraphics[clip,width=3cm]{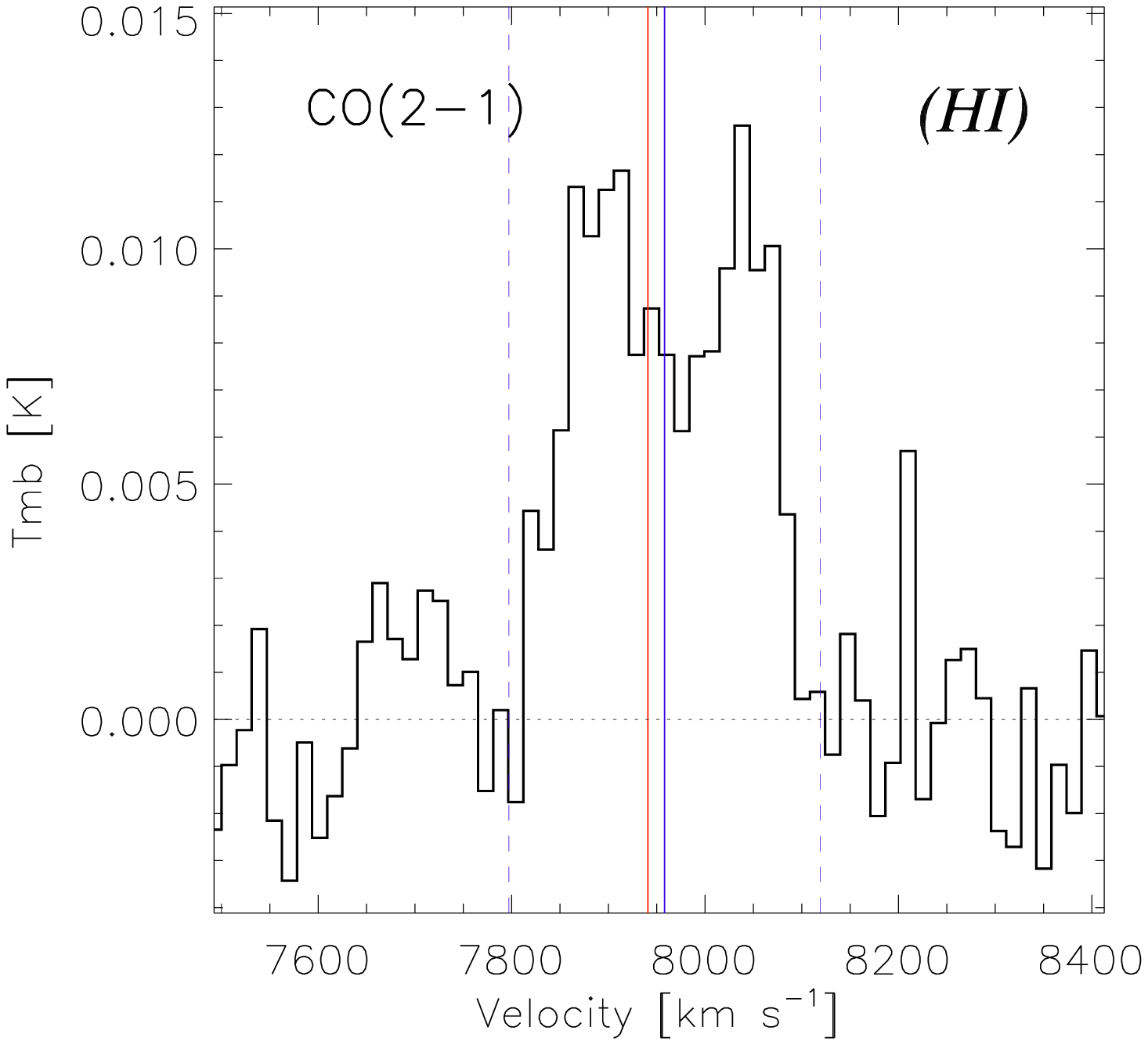}\hfill
 \includegraphics[clip,width=2.8cm]{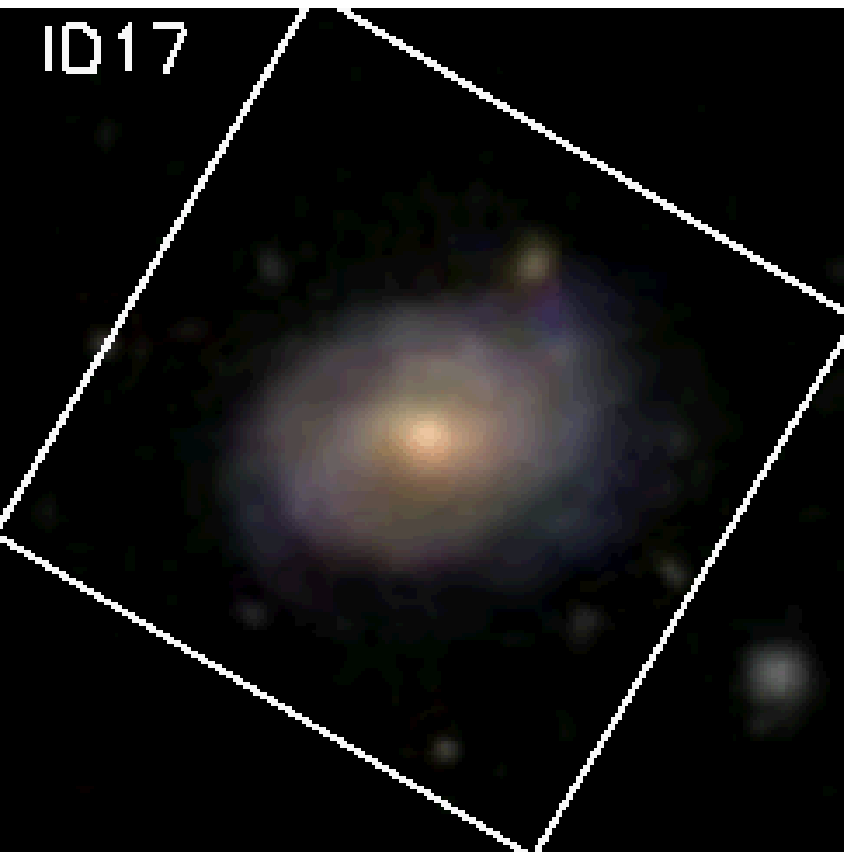}\hspace{-1mm}
 \includegraphics[clip,width=2.8cm]{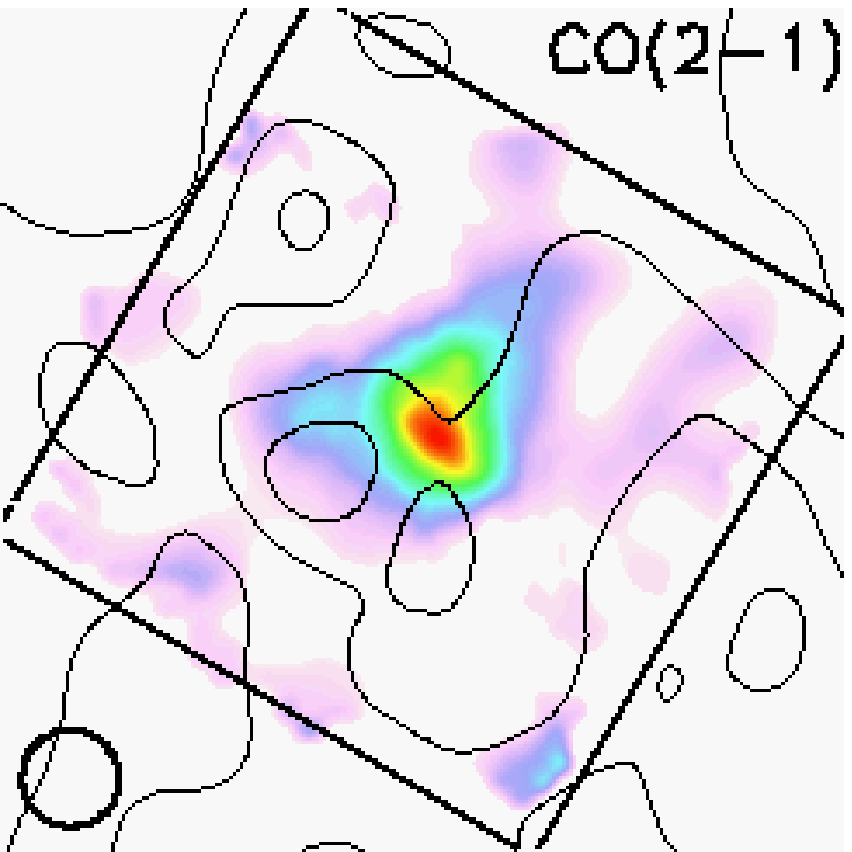}
 \includegraphics[clip,width=3cm]{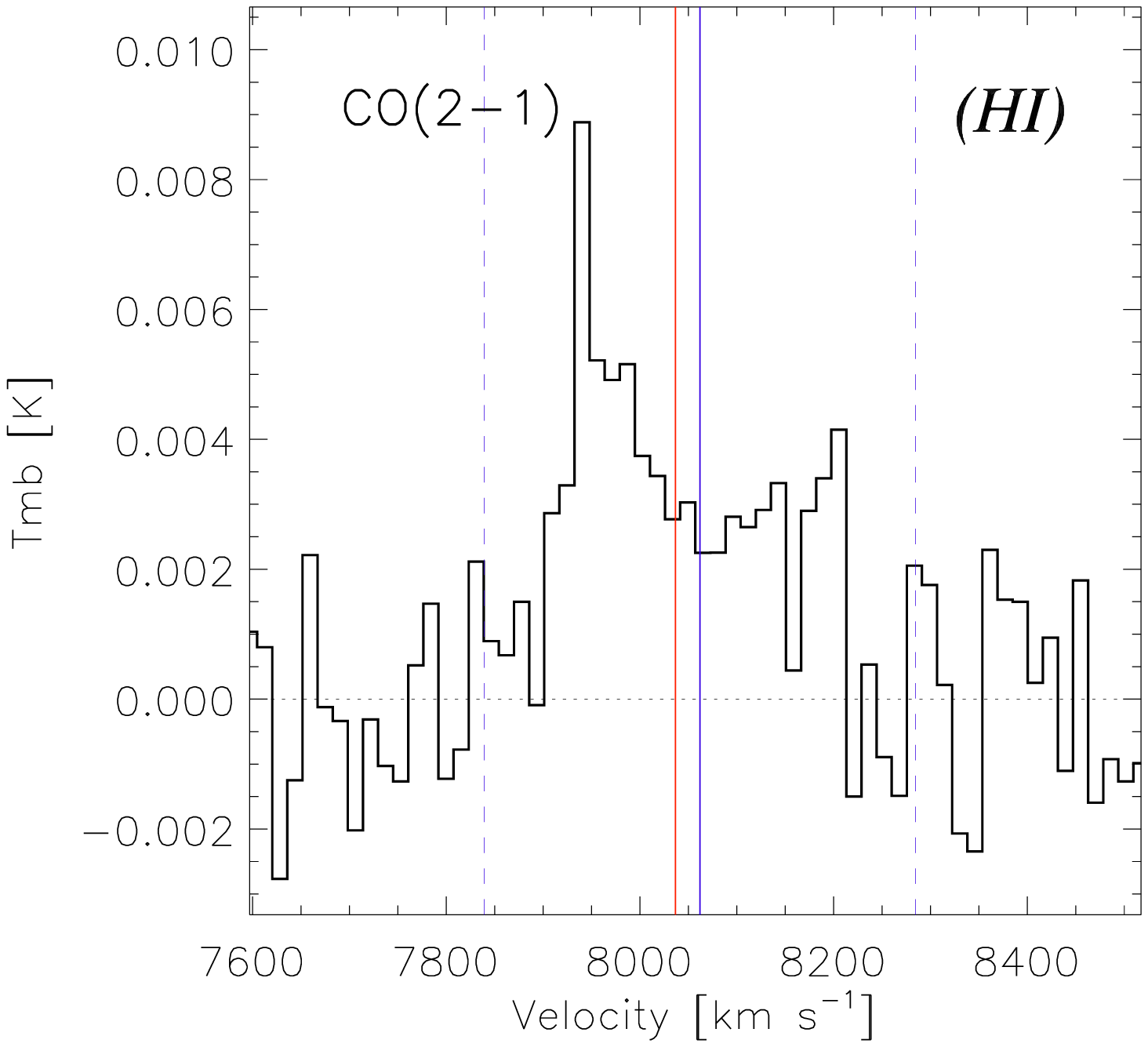}
 \includegraphics[clip,width=2.8cm]{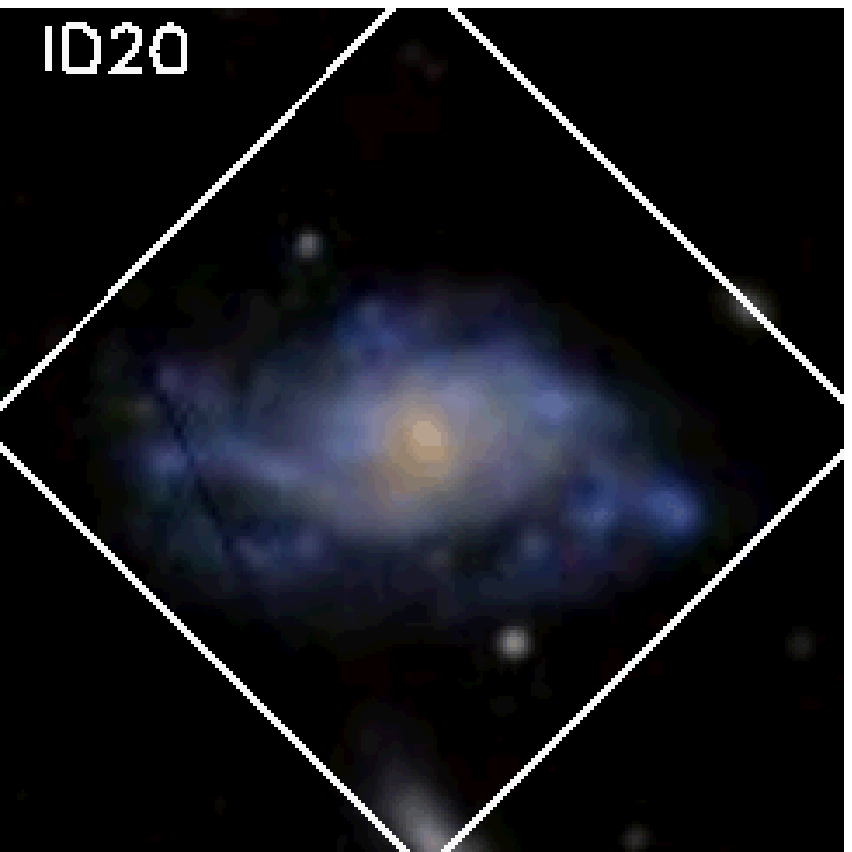}\hspace{-1mm}
 \includegraphics[clip,width=2.8cm]{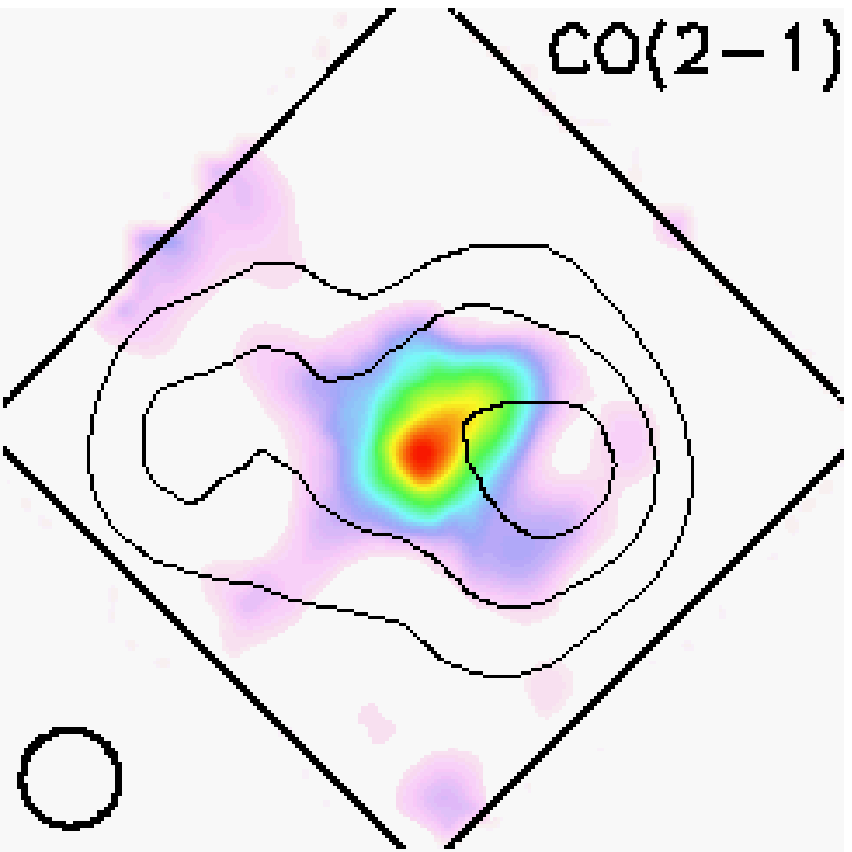}
 \includegraphics[clip,width=3cm]{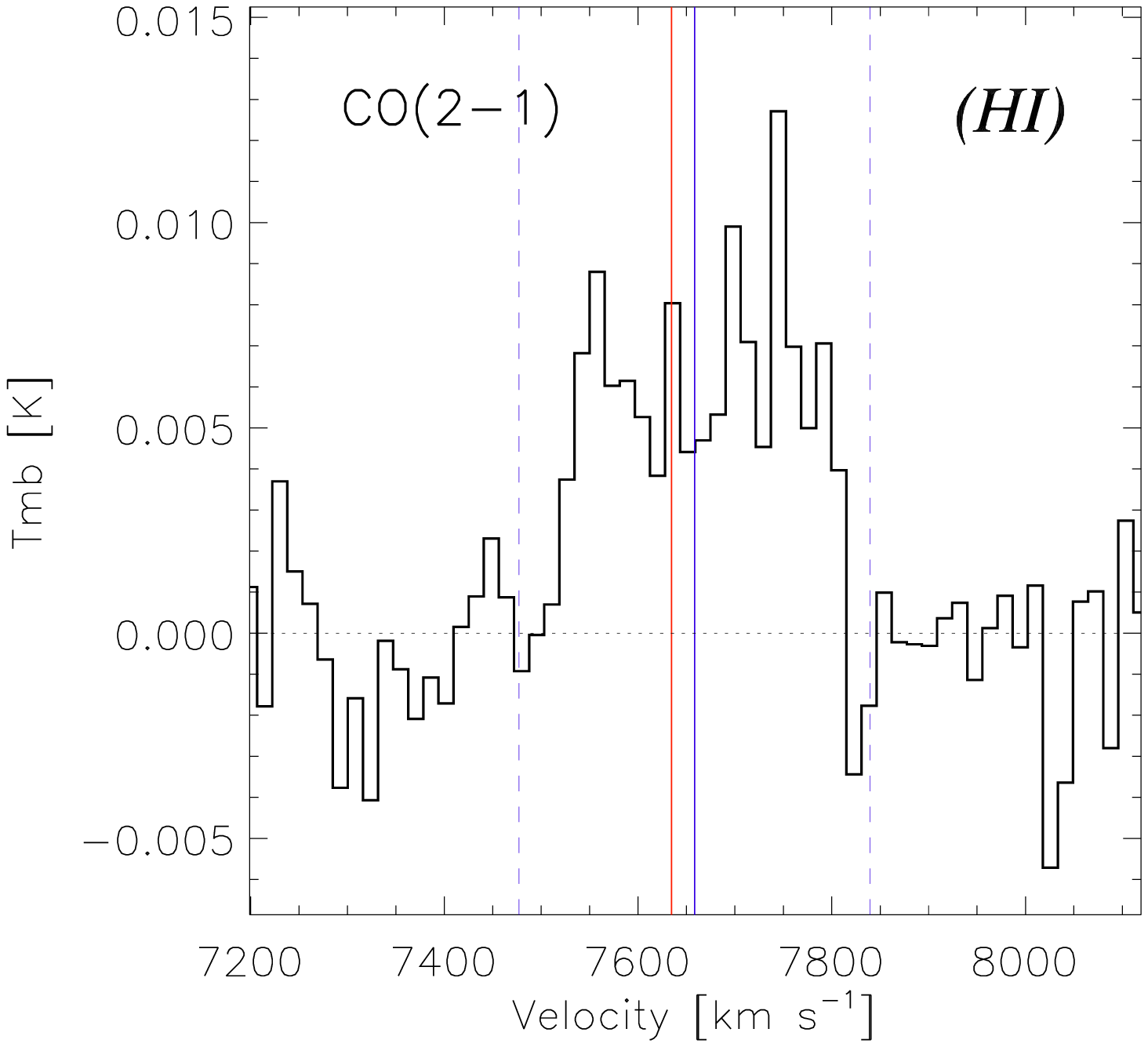}\hfill\hspace{8cm}
\caption{ 
 SDSS 3-colour images ($r$, $g$, $i$ bands), 
 HERA CO(2-1) intensity map with \hi contours (levels at roughly 
 50\%, 70\% and 90\% of the emission peak), and spectrum within 
 a central aperture of diameter 22\arcs (EMIR beam size at
 the observed frequency of 112\,GHz). 
}
\label{fig:hera}
\end{figure*}

\begin{figure*}
\centering
 \includegraphics[clip,width=2.9cm]{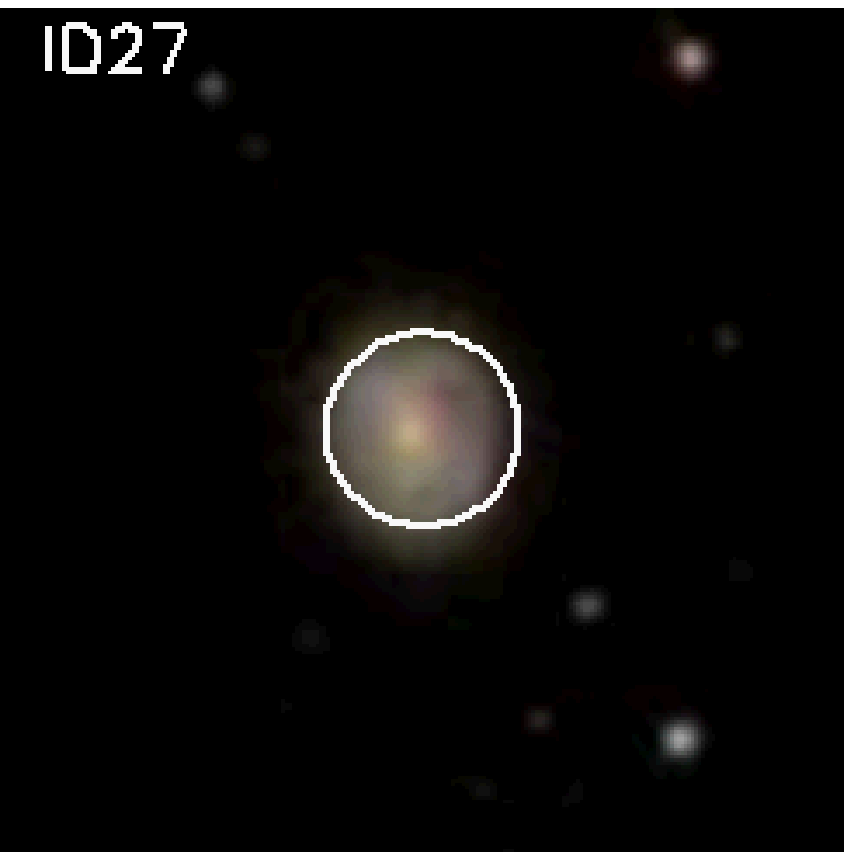}
 \includegraphics[clip,width=5.8cm]{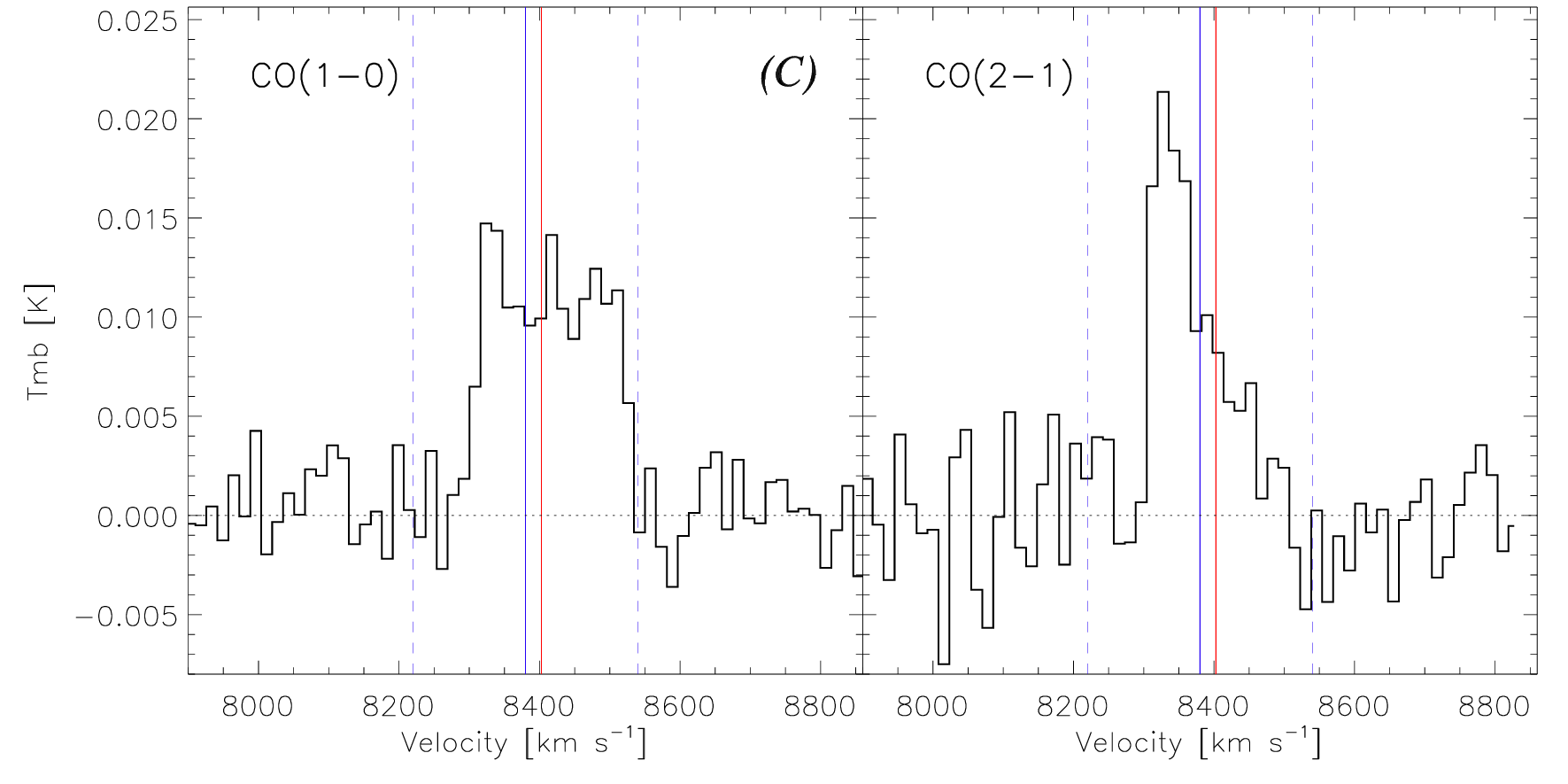}\hfill
 \includegraphics[clip,width=2.9cm]{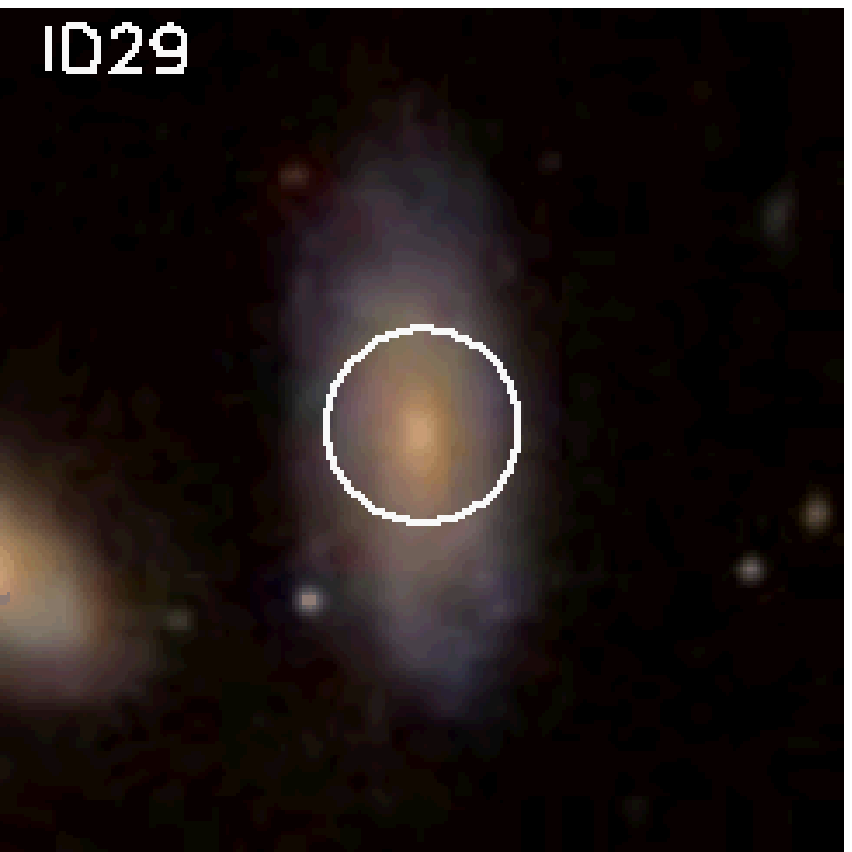}
 \includegraphics[clip,width=5.8cm]{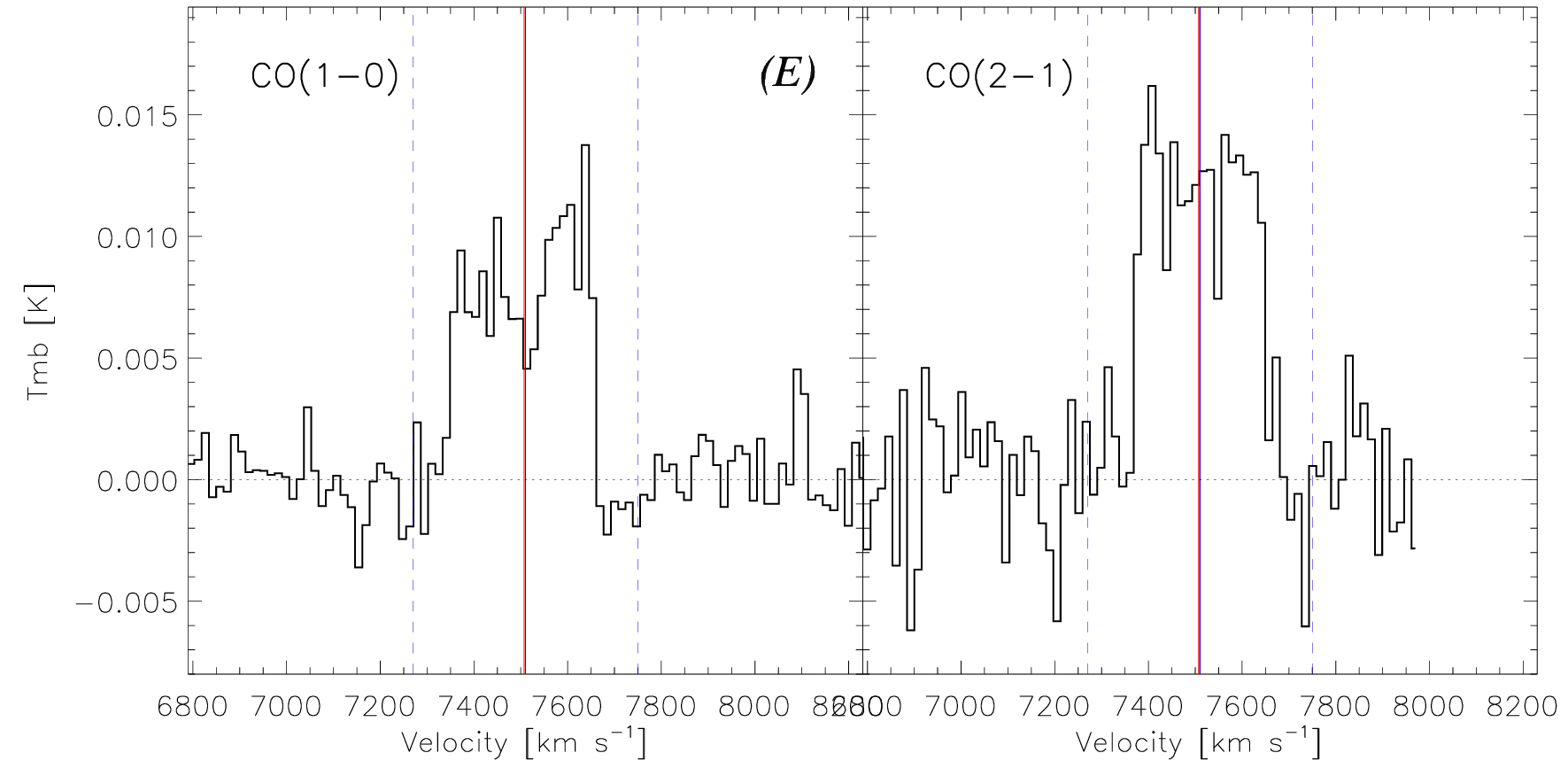}
 \includegraphics[clip,width=2.9cm]{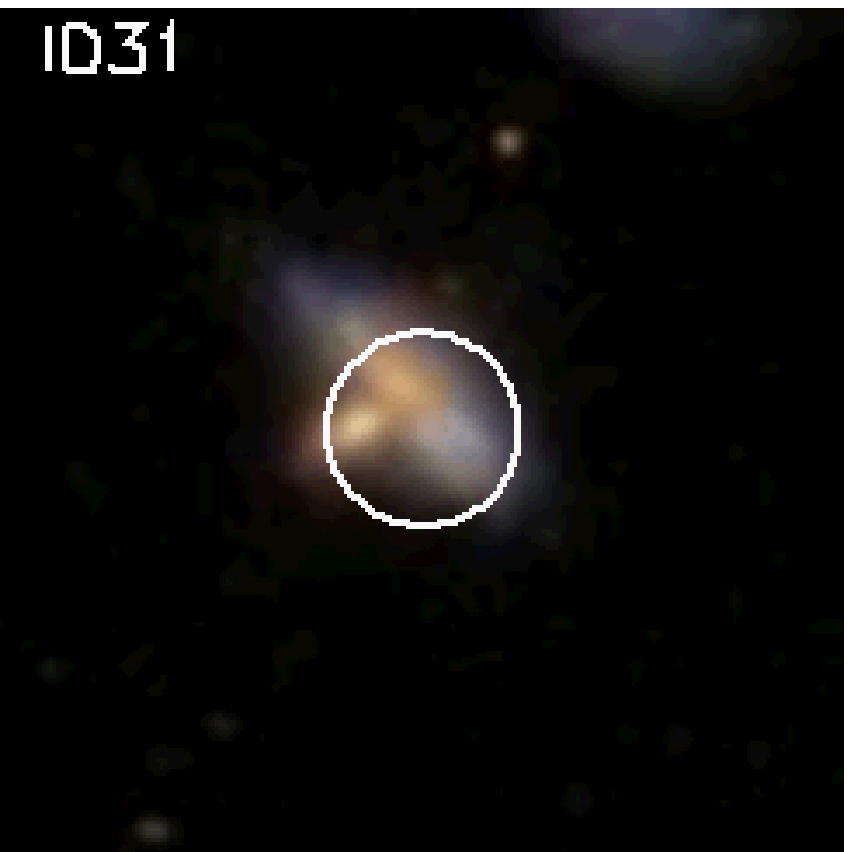}
 \includegraphics[clip,width=5.8cm]{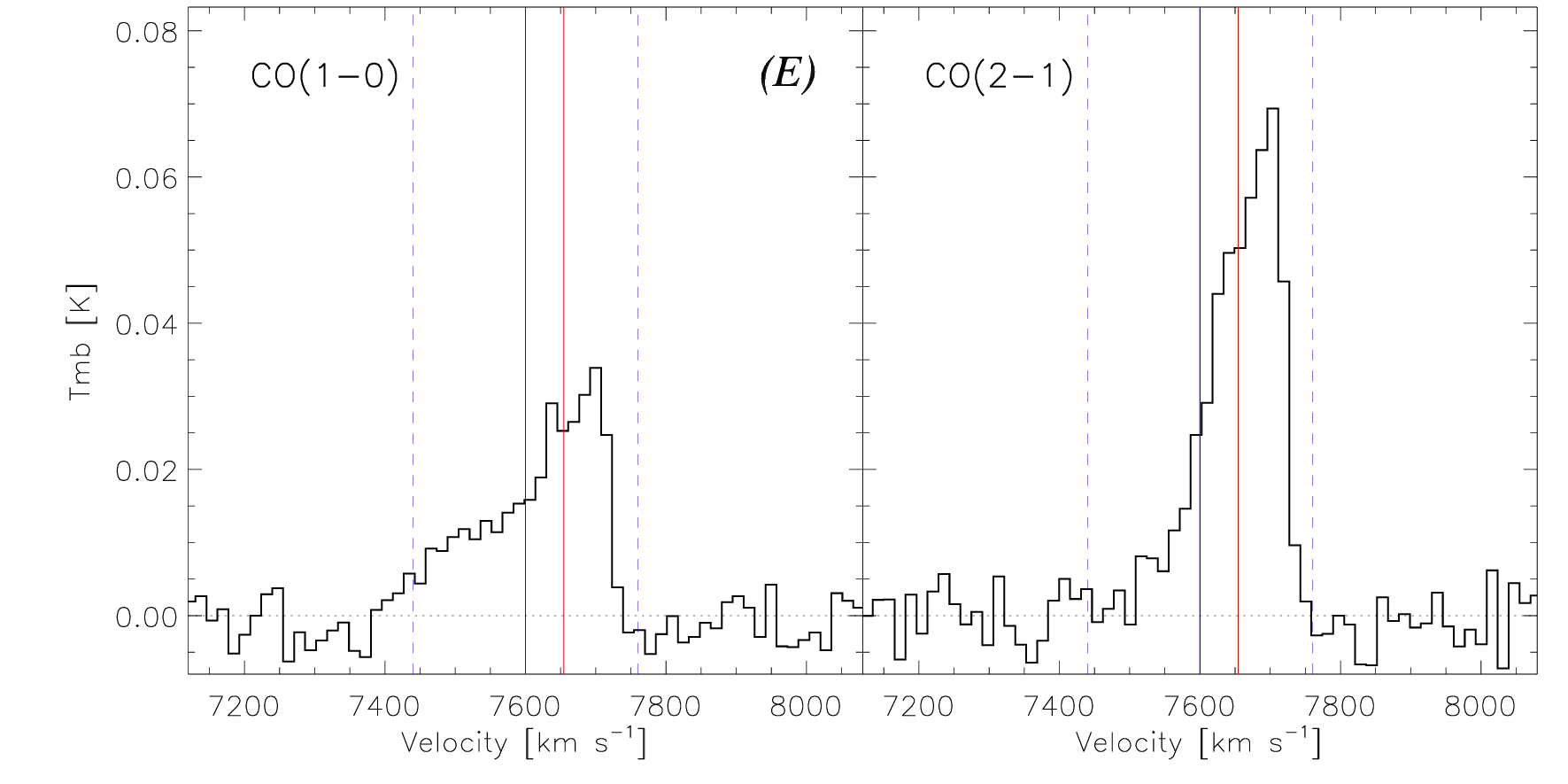}\hfill
 \includegraphics[clip,width=2.9cm]{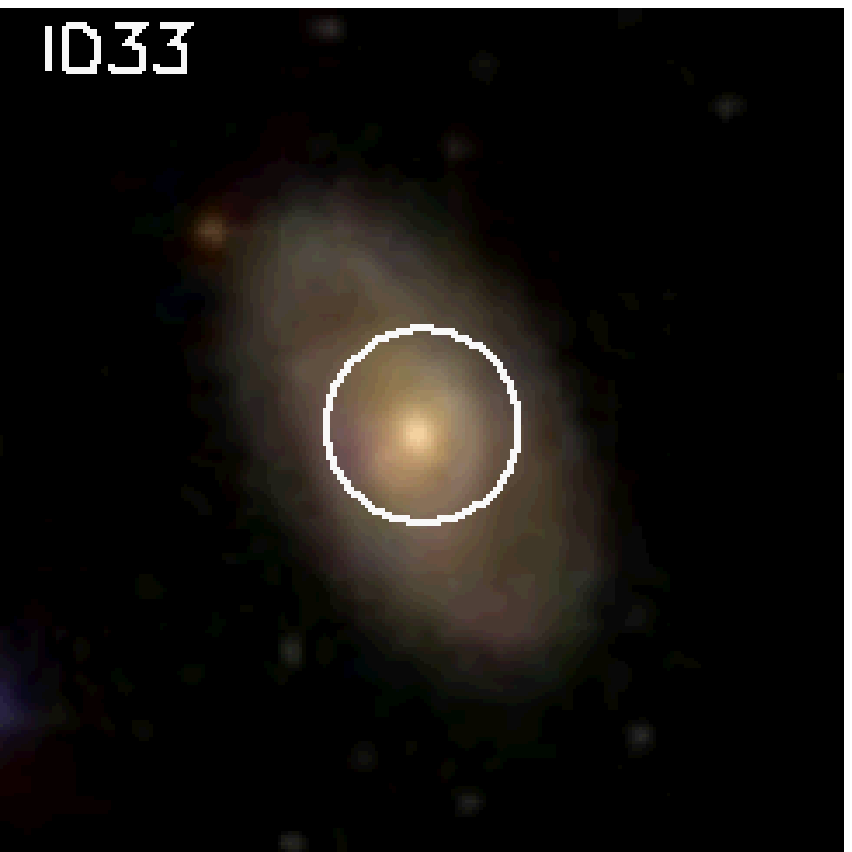}
 \includegraphics[clip,width=5.8cm]{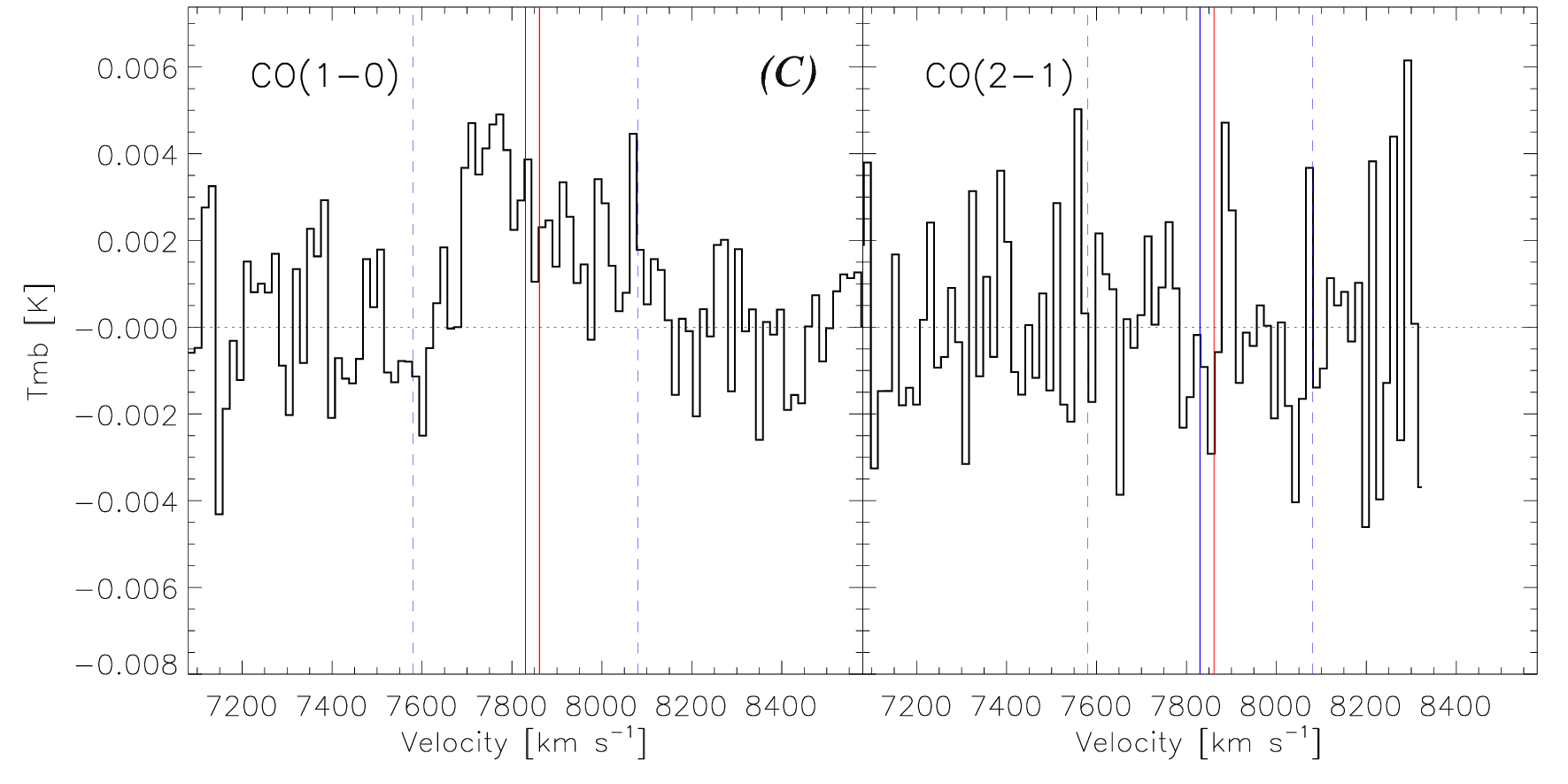}
 \includegraphics[clip,width=2.9cm]{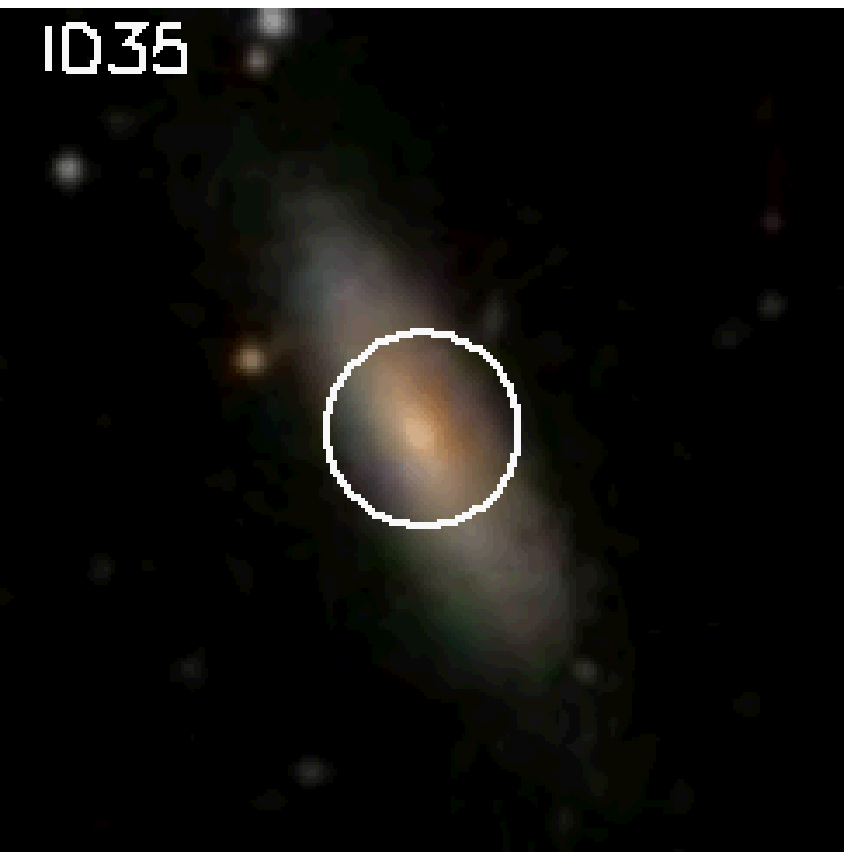}
 \includegraphics[clip,width=5.8cm]{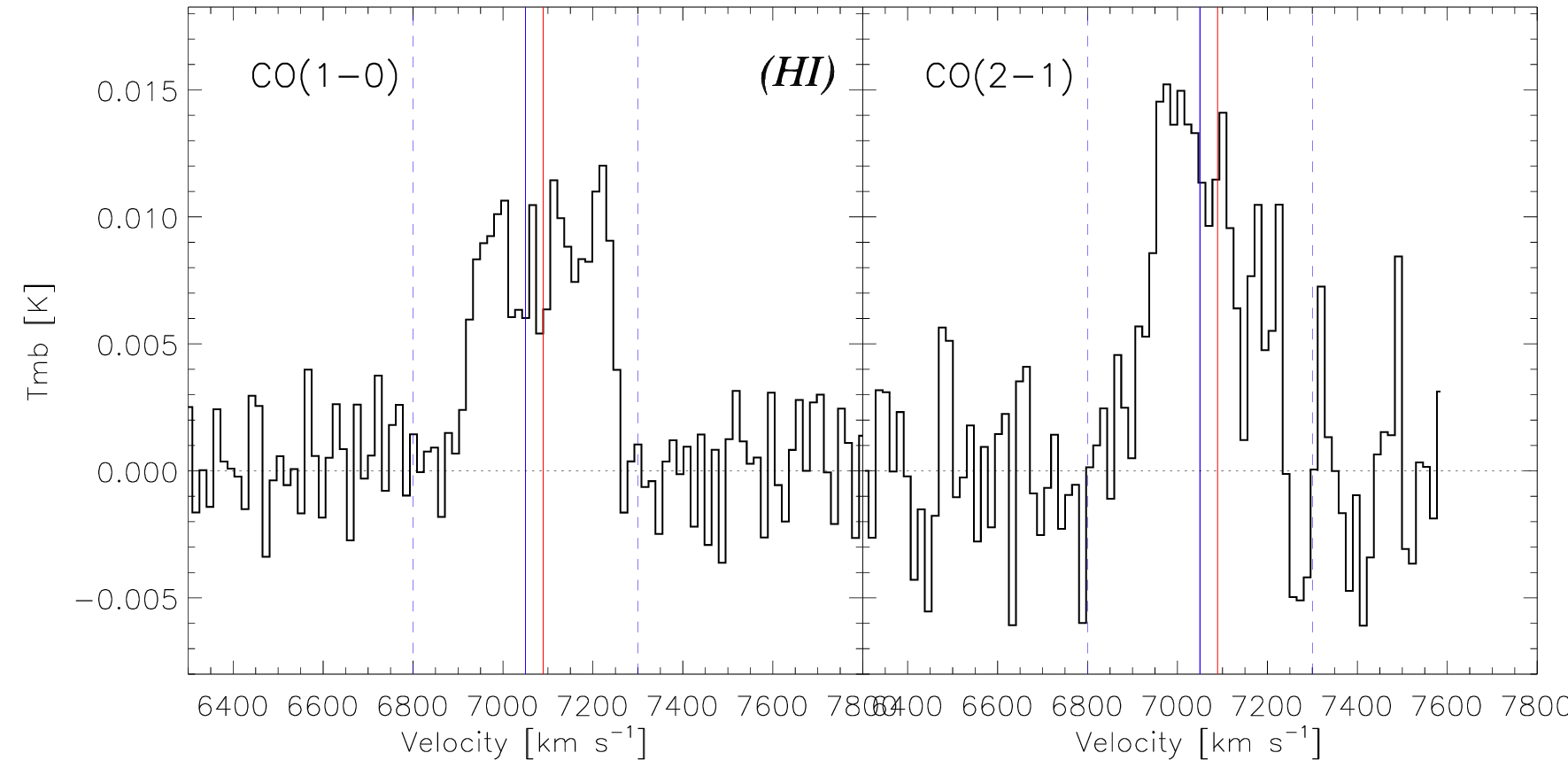}\hfill
 \includegraphics[clip,width=2.9cm]{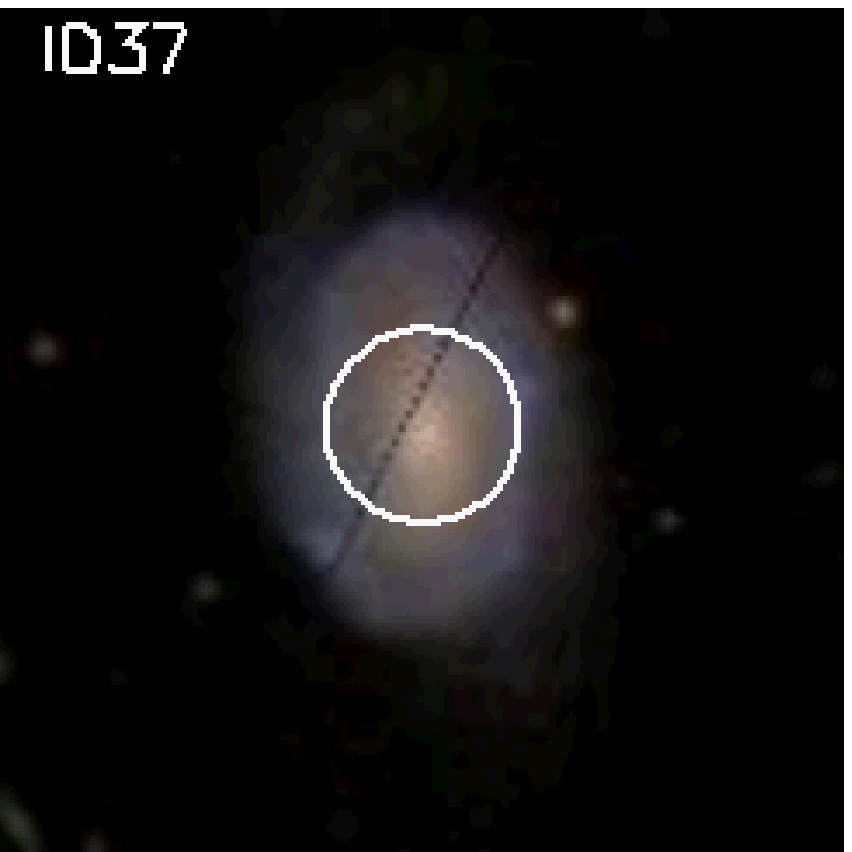}
 \includegraphics[clip,width=5.8cm]{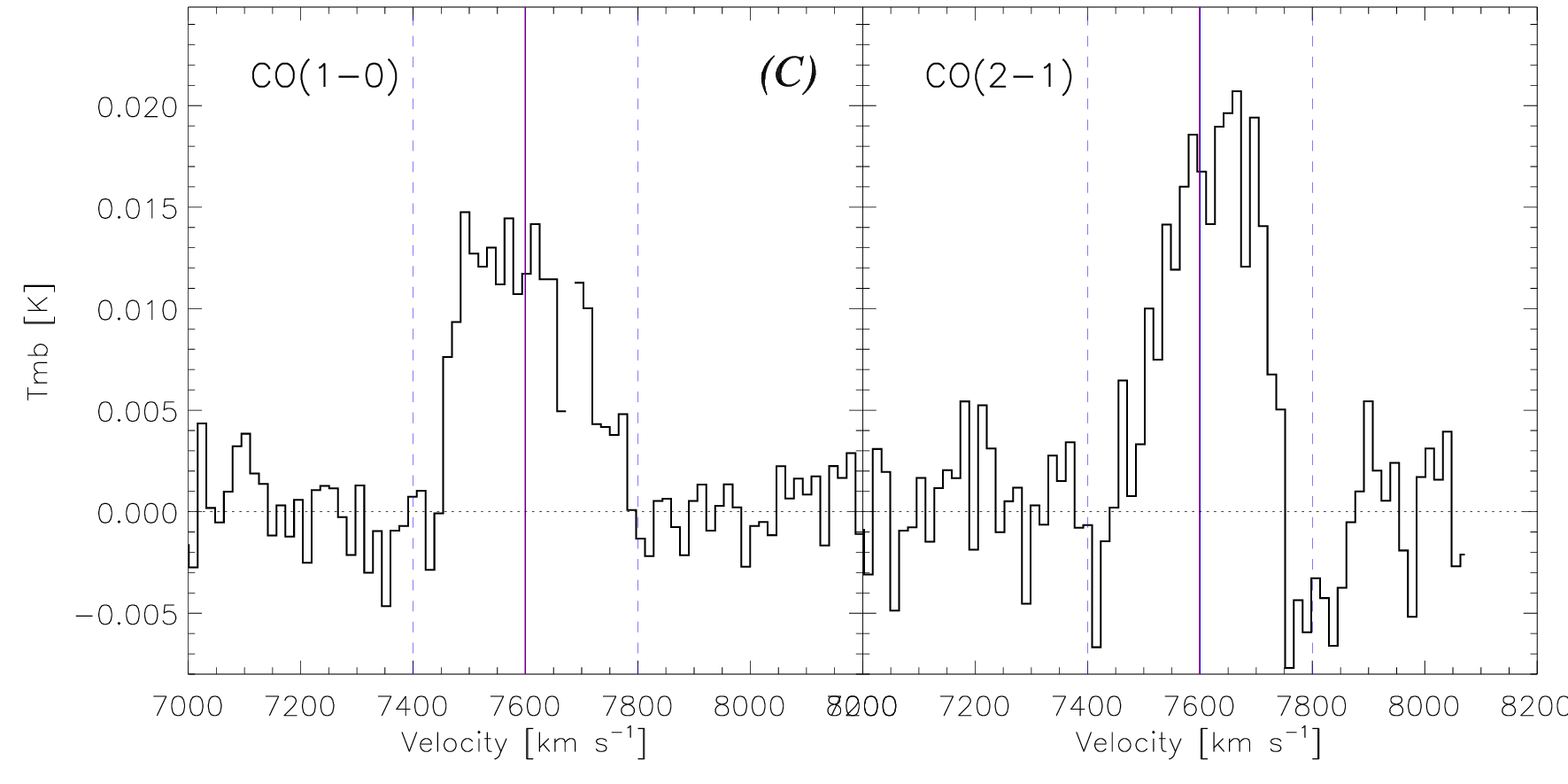}
 \includegraphics[clip,width=2.9cm]{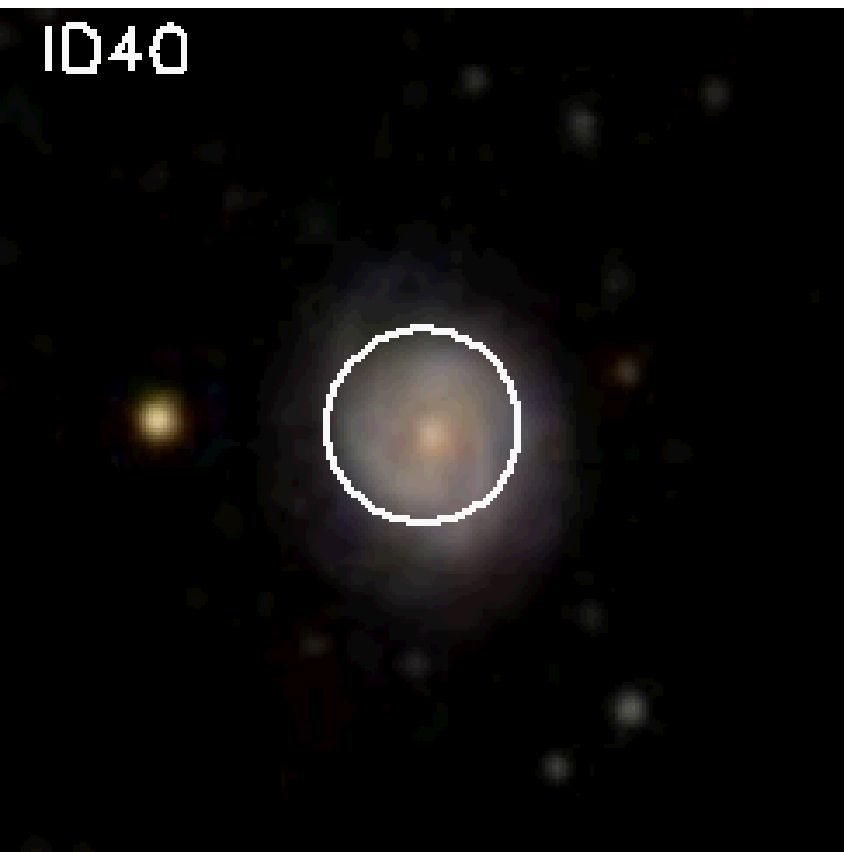}
 \includegraphics[clip,width=5.8cm]{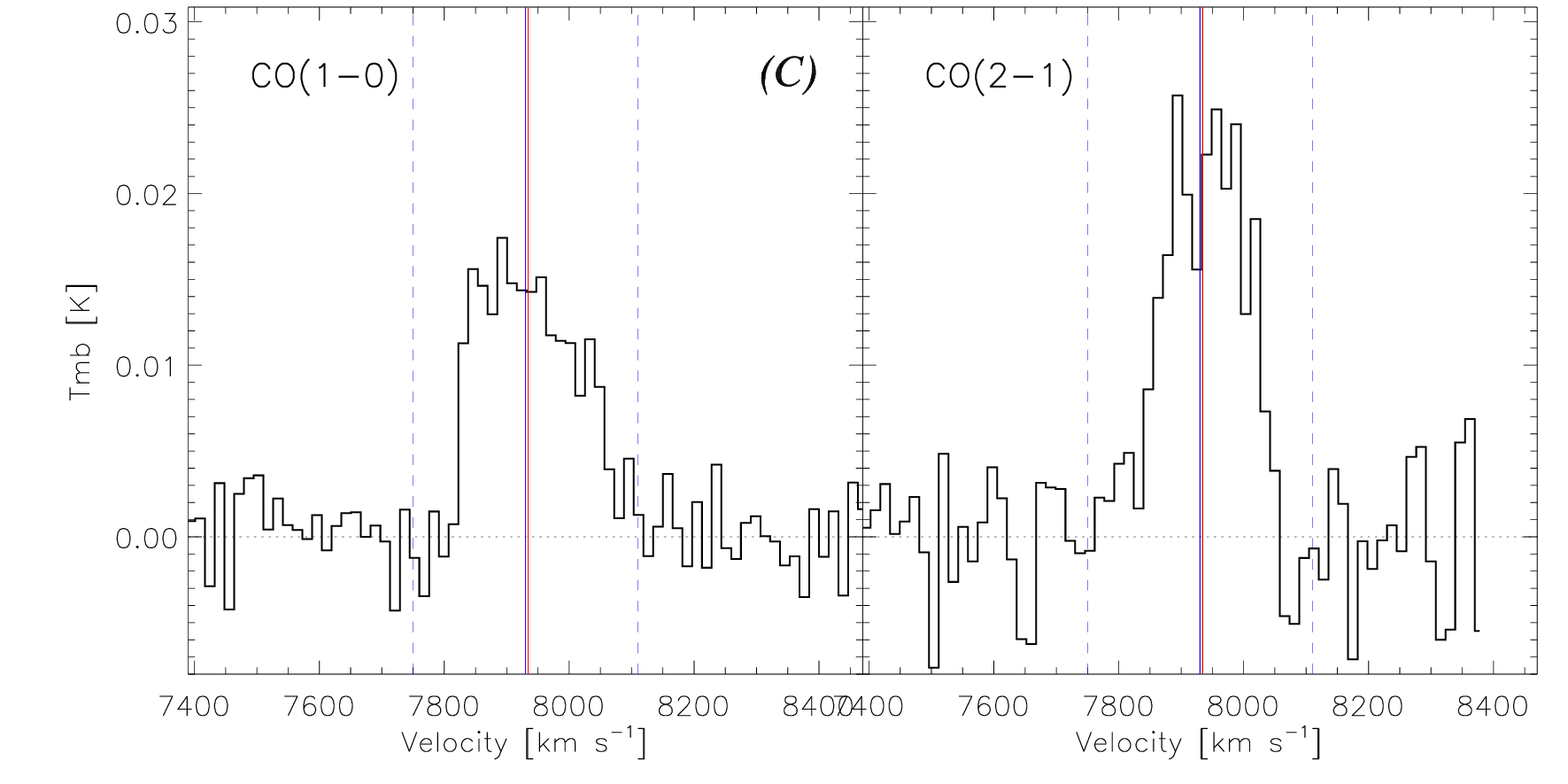}\hfill
 \includegraphics[clip,width=2.9cm]{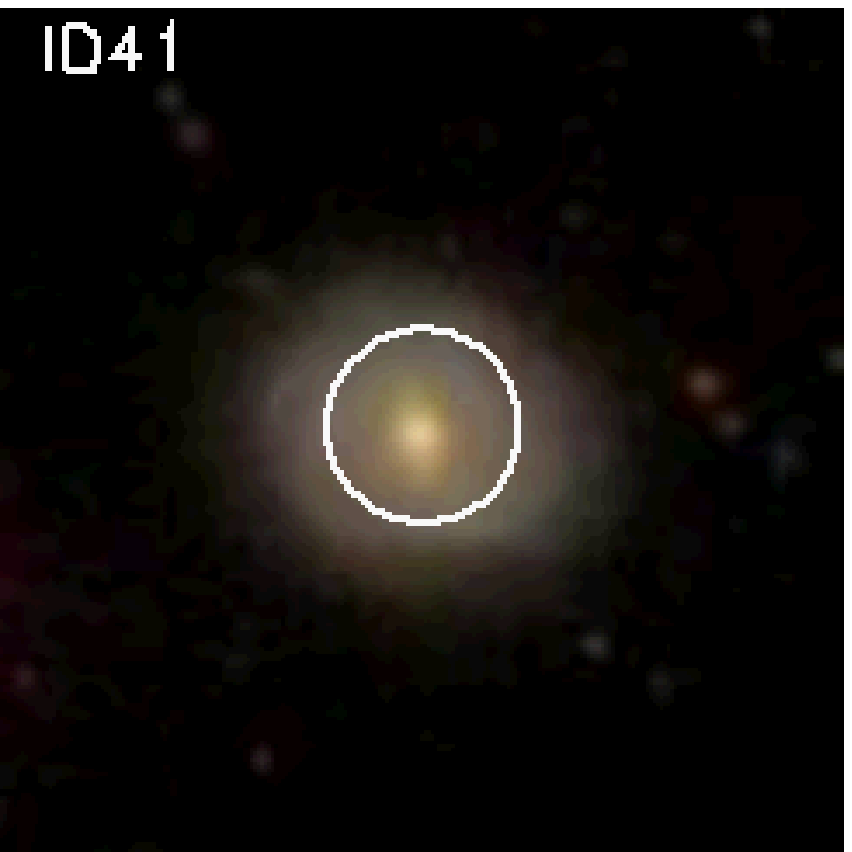}
 \includegraphics[clip,width=5.8cm]{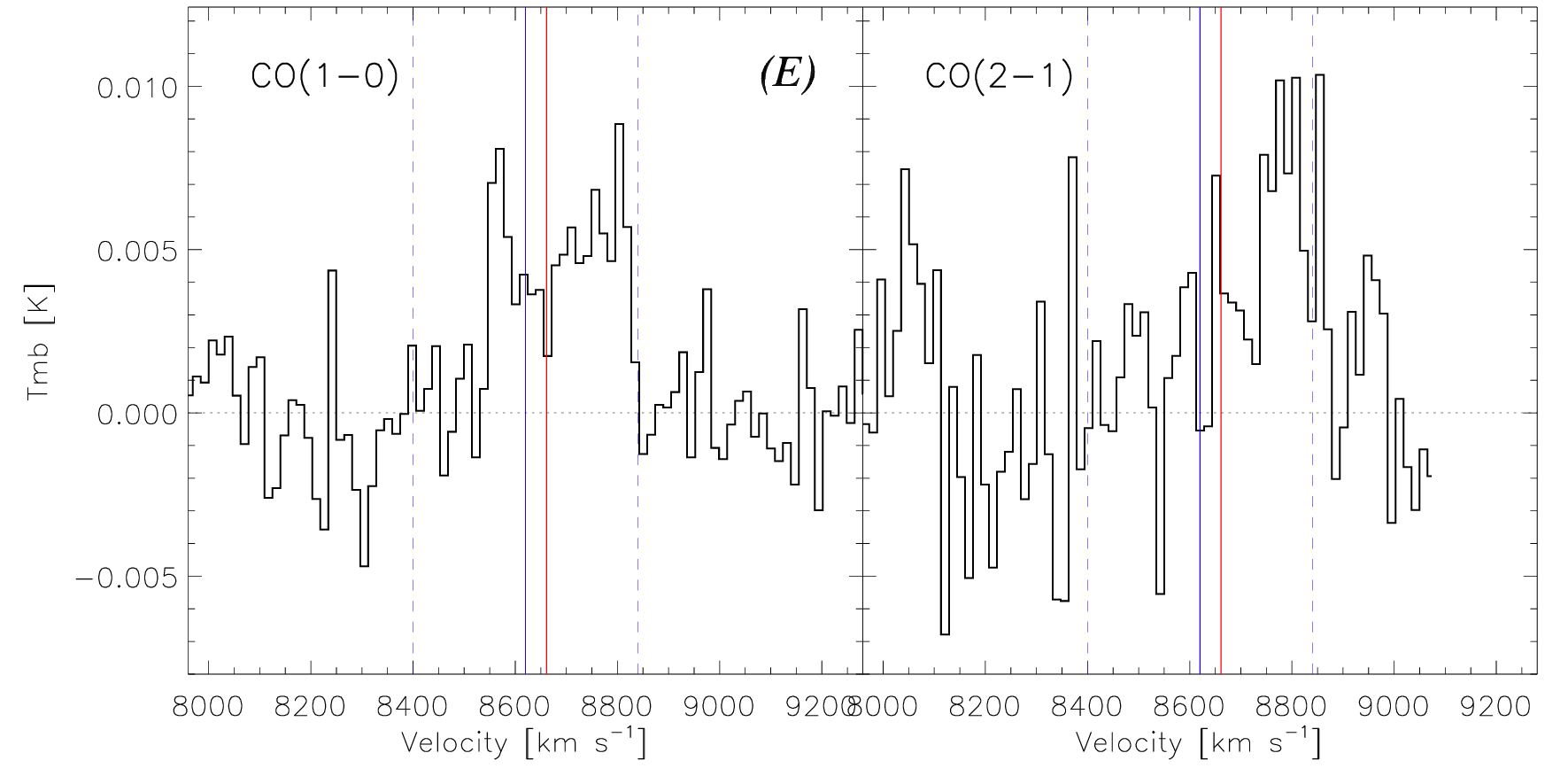}
 \includegraphics[clip,width=2.9cm]{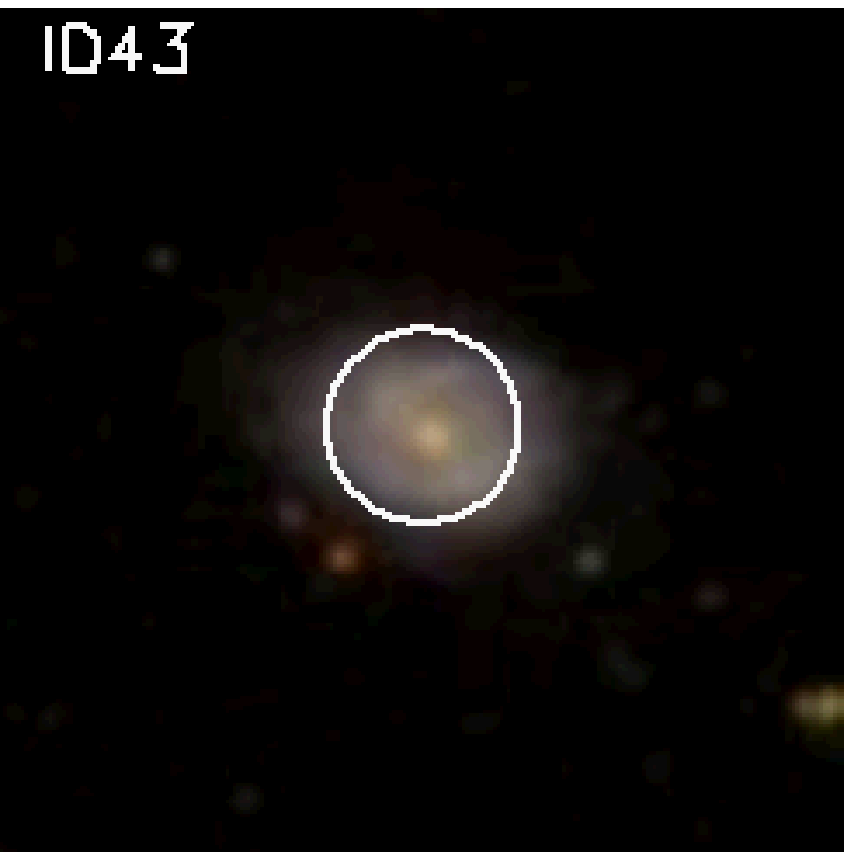}
 \includegraphics[clip,width=5.8cm]{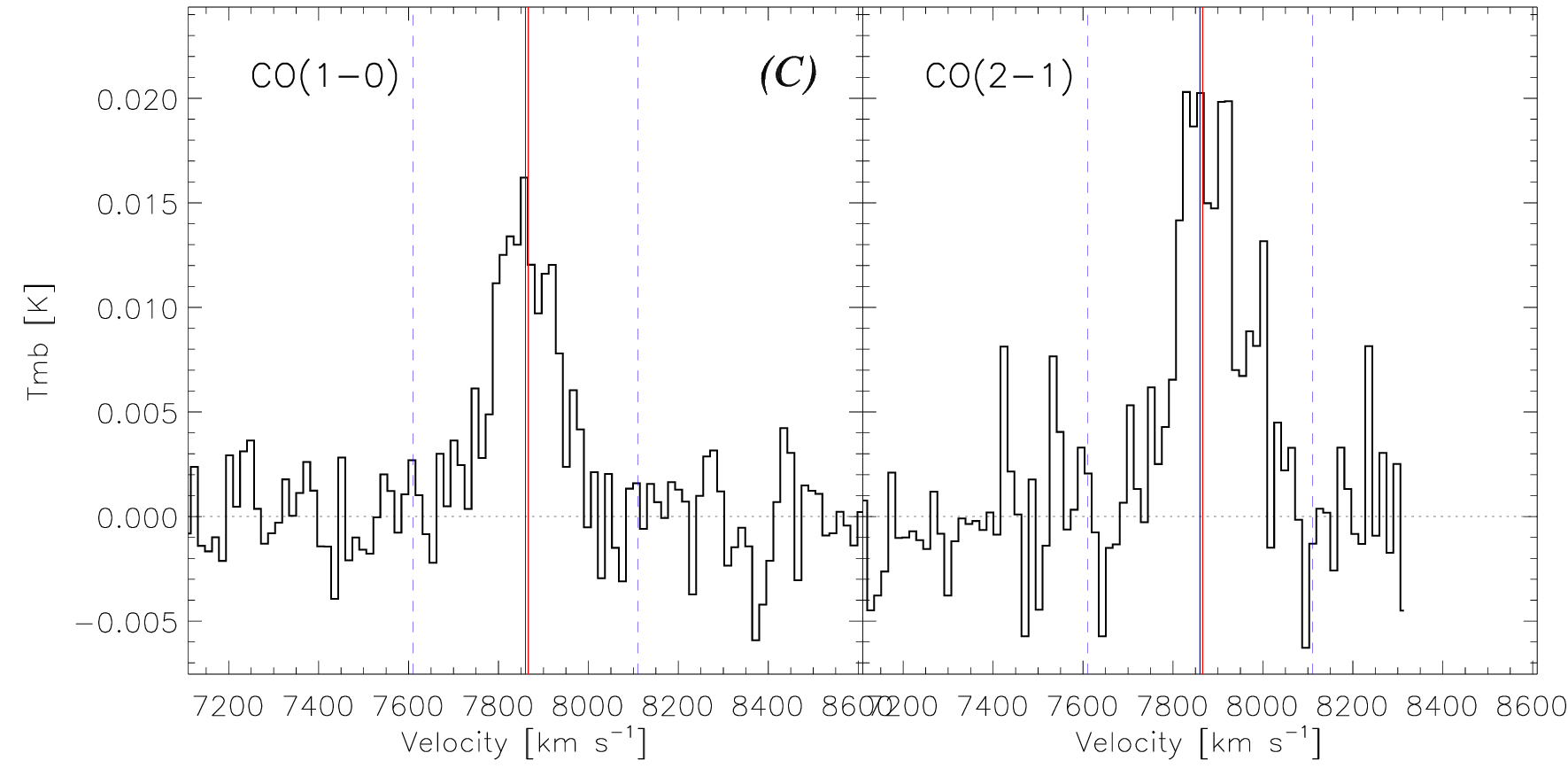}\hfill
 \includegraphics[clip,width=2.9cm]{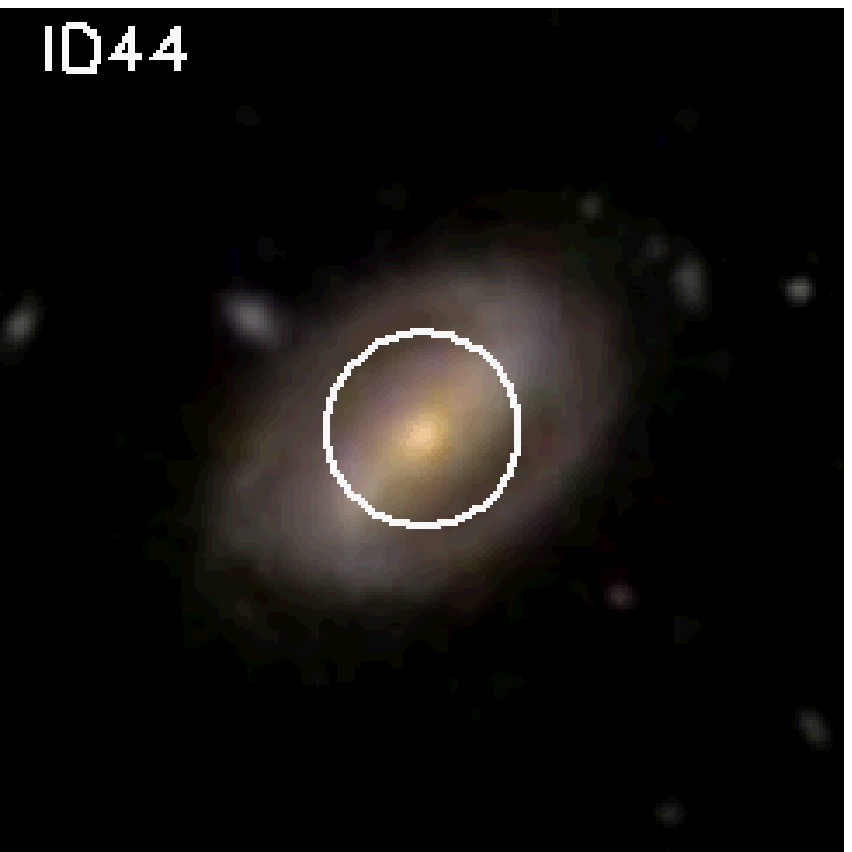}
 \includegraphics[clip,width=5.8cm]{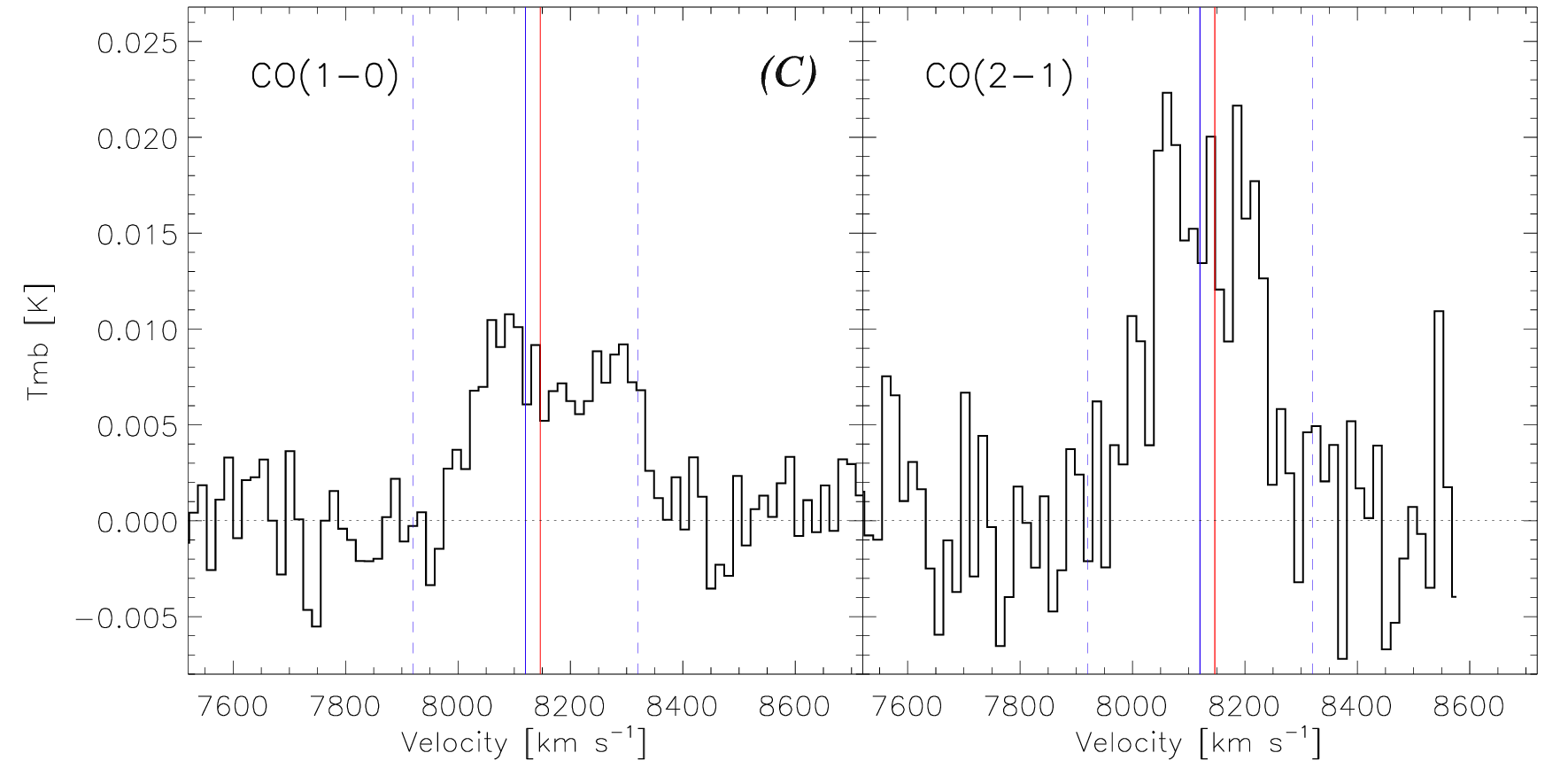}
 \includegraphics[clip,width=2.9cm]{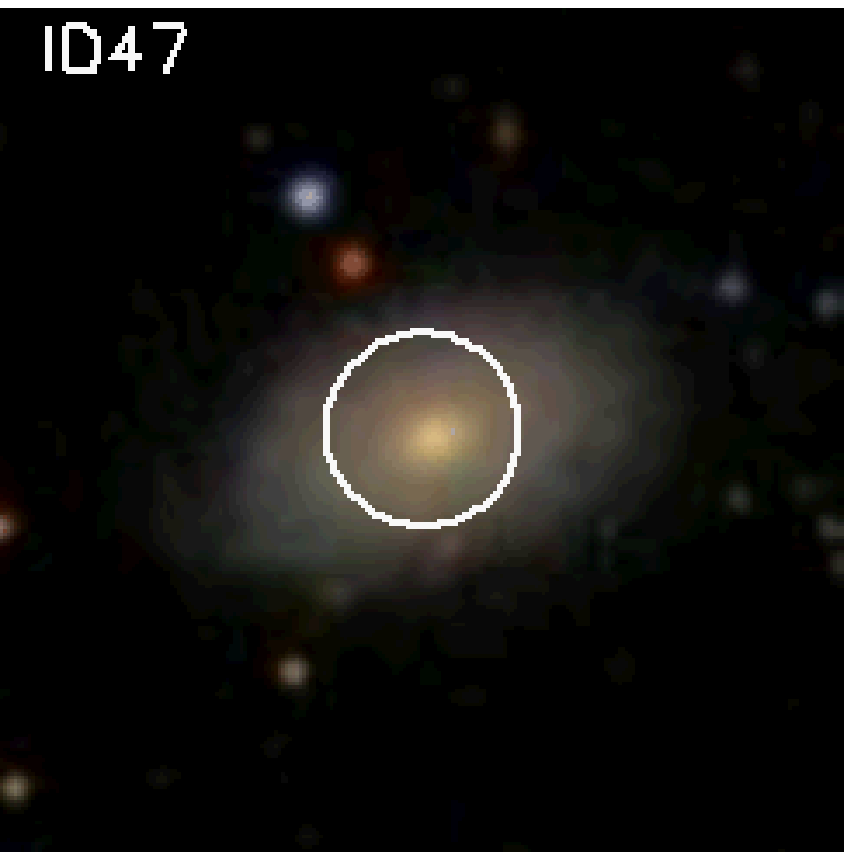}
 \includegraphics[clip,width=5.8cm]{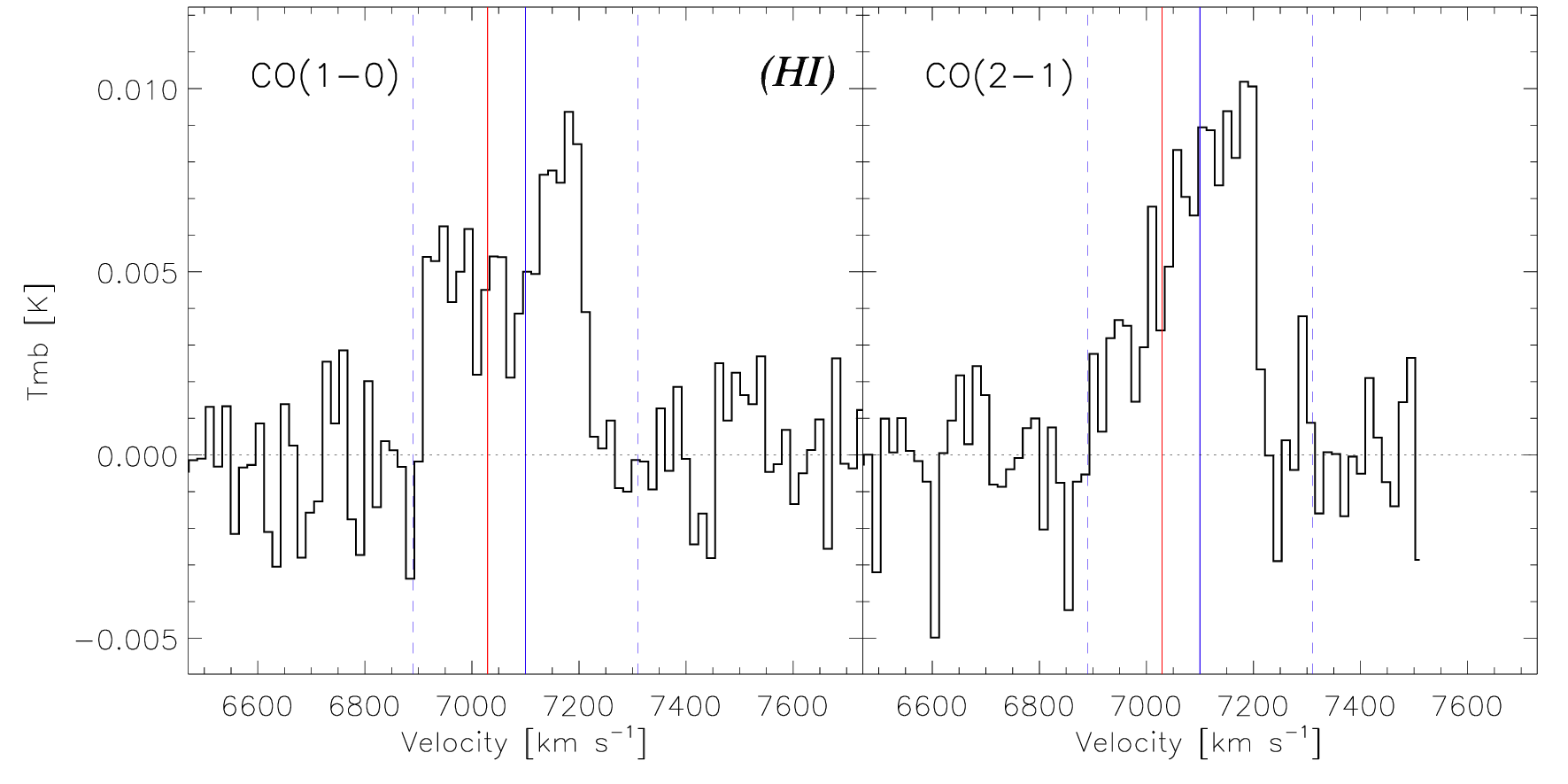}\hfill\hspace{8cm}
\caption{ 
 SDSS 3-colour images ($r$, $g$, $i$ bands) of the control galaxies 
 and their EMIR spectra of CO(1-0) and CO(2-1). 
}
\label{fig:emir}
\end{figure*}

}

\label{lastpage}

\end{document}